\documentstyle[tighten,preprint,prb,aps,epsf]{revtex}

\def\plotfiddle#1#2#3#4#5#6#7{\centering \leavevmode
    \vbox to#2{\rule{0pt}{#2}}
    \includegraphics{#1}}

\def\plotfidtwo#1#2#3#4#5#6#7#8#9{\centering \leavevmode
    \vbox to#3{\rule{0pt}{#3}}
    \includegraphics{#1}
    \includegraphics{#2}}
\epsfverbosetrue

\def\plotport#1#2#3#4#5#6#7#8{\centering \leavevmode
     \vbox to#3{\rule{0pt}{#3}}
     \includegraphics{#2}
     \vbox to 0 pt{\hfill $\smash{\lower
          0.9 true in\hbox{Fig.\ #1\hskip -0.6 true in}}$}}
\def\plotland#1#2#3#4#5#6#7#8{\centering \leavevmode
     \vbox to#3{\rule{0pt}{#3}}
     \includegraphics{#2}
     \vbox to 0 pt{\hfill $\smash{\hbox{\rotatebox{90}{$\smash{\lower
          0.6 true in\hbox{{\hskip -0.9 true in}Fig.\ #1}}$}}}$}}

\def\etal{{\it et al.}}

\def\cp{Chem.\ Phys.\ }
\def\cpl{Chem.\ Phys. Lett.\ }

\def\jpc{J.\ Phys.\ Chem.\ }

\def\movup#1{{\vphantom{#1}\smash{\raise1pt\hbox{$\scriptstyle #1$}}}} 
\def\movdn#1{{\vphantom{#1}\smash{\lower1pt\hbox{$\scriptstyle #1$}}}} 
\def\argoversim#1#2{\lower2pt\vbox{\baselineskip0.5pt \lineskip0.5pt
       \ialign{$\mathsurround=0pt #1\hfil##\hfil$\crcr#2\crcr\sim\crcr}}} 

\def\har{\hbox{$E_h$}}
\def\millih{\hbox{$m \mskip -1 mu E_h$}}

\def\microh{\hbox{$\mu \mskip -1 mu E_h$}}

\def\bohr{\hbox{$a_{\hbox{$\scriptscriptstyle 0$}}$}}

\def\phn{\hbox{$\phantom{0}$}}
\def\phnn{\hbox{$\phantom{00}$}}
\def\phnnn{\hbox{$\phantom{000}$}}

\def\fad{f_{\!Ad}}    

\def\Crsq{C_{\!\hbox{$\scriptscriptstyle R$}}^\movup{2}} 
\def\m{\!\raise 2 pt\hbox{$\scriptscriptstyle -$}} 

\def\hehh{\hbox{He-H$_2$}}
\def\hephh{\hbox{$\rm He+H_2$}}

\def\abinitio{{\it ab~initio\/}}

\def\etal{{\it et~al.}}

\begin{document}

\draft

\title{An accurate analytic He-H$_2$ potential energy surface
 from a greatly expanded set of ab initio energies}

\author{Arnold I. Boothroyd and
Peter G. Martin}
\address{Canadian Institute for Theoretical Astrophysics, University of
Toronto,\\ Toronto, Ontario, M5S 3H8, Canada}

\author{Michael R. Peterson}
\address{Department of Computing and Networking
Services, University of Toronto,\\ Toronto, Ontario, M5S 3J1, Canada.}

%

\date{\today: J.~Chem.~Phys., in press; preprint CITA-2003-19}

\maketitle

\begin{abstract}

The interaction potential energy surface~(PES) of~\hehh\ is of great
importance for quantum chemistry, as the simplest test case for
interactions between a molecule and a closed-shell atom.
It is also required for a detailed understanding of certain astrophysical
processes, namely collisional excitation and dissociation of~H$_2$ in
molecular clouds, at densities too low to be accessible experimentally.
A new set of $23\,703$ \abinitio\ energies was computed, for
\hehh\ geometries where the interaction energy was expected to
be non-negligible.
These
have an estimated rms ``random'' error
of~$\sim 0.2$~millihartree
and a systematic error
of~$\sim 0.6$~millihartree (0.4~kcal/mol).
A new analytic \hehh\ PES, with 112 parameters,
was fitted to $20\,203$ of these new \abinitio\ energies
(and to an additional 4862 points generated at large separations). This
yielded an improvement by better than an order of magnitude in the fit to
the interaction region, relative to the best previous surfaces
(which were accurate only for near-equilibrium H$_2$ molecule sizes).
This new PES has an rms error of
0.95~millihartree (0.60~kcal/mole)
relative to the the $14\,585$ \abinitio\ energies that lie below
twice the H$_2$ dissociation energy, and
2.97~millihartree (1.87 kcal/mole)
relative to the full set of $20\,203$ \abinitio\ energies (the fitting
procedure used a reduced weight for high energies, yielding a weighted
rms error of
1.42~millihartree, i.e., 0.89~kcal/mole).
These rms errors are comparable to the estimated error in
the \abinitio\ energies themselves;
the conical intersection between the ground
state and the first excited state is the largest source of error in the
PES\hbox{}.

\end{abstract}



\section{Introduction}
\label{sec:intro}

The interaction potential energy surface~(PES) of~\hehh\
(i.e.,~two hydrogen atoms plus one helium atom) is of great importance
for quantum chemistry.  \hehh\ is the simplest test case for
theories of the interaction of a molecule with a closed-shell atom.
It is highly desirable to have a PES for \hehh\ that attains or at least
approaches the ``chemical accuracy'' (of order one millihartree, i.e.,
better than one kcal/mol) required for reaction dynamics ---
cf.\ available \abinitio\ potential energy surfaces
for~H$_3$ (molecule plus open-shell atom)\cite{th78,vbmtb87,bkmp,bkmp2,wu99}
or H$_4$ (two molecules)\cite{ke92,ag94,bmkp02,bmkp02epaps}.  Previous
analytic \hehh\ surfaces\cite{Rob63,kk64,km65,gs70,gh72,tk73,gwa75,%
wkb74,tt77,tt78,dr78,rg82,sk85,thccl92,mr94}
were based on only a few dozen \abinitio\ energies,
which provided highly inadequate coverage of all but a very restricted
region in the conformation space of \hbox{He--H$_2$}.  This paper
reports a greatly-expanded set of \abinitio\ energies, and a greatly
improved \hehh~PES fitted to these energies.

The \hehh~PES is also of particular astrophysical interest for
studying \hephh\ interactions in physical conditions not accessible
to experiment, namely the low densities characteristic of giant molecular
clouds in the interstellar medium, where star formation occurs.  Helium is
the second-most-common collision partner for H$_2$ in these molecular clouds,
which contain roughly one helium atom for every ten hydrogen atoms.
Heating of these clouds by strong shock waves
causes rotational and vibrational excitation of the H$_2$~molecules, and
can lead to collision-induced dissociation of~H$_2$ into free H~atoms.
The collision rates in molecular clouds can be so low that the (observed)
forbidden (quadrupole) infrared emission of excited H$_2$~molecules
can induce highly non-thermal distributions over the internal states
of~H$_2$
(see, e.g., Refs.\ \onlinecite{drcm87},  \onlinecite{ch91},
and~\onlinecite{msm96}).
Because the mean free paths of molecules
are thousands of kilometers at such low densities, these processes will
remain inaccessible to laboratory experiment.
Computer simulation is therefore a {\it sine qua non\/} in the study of the
physics and chemistry of star forming regions.

\subsection{Notation}
\label{ssec:notation}

Atomic units are used in this paper unless otherwise specified, i.e.,
distances are in bohrs~($\bohr$) and energies are in hartrees~($\har$),
millihartrees~($\millih$), or microhartrees~($\microh$).  Recall that
$1\;\bohr = 0.529177\;$\AA, while $1\;\millih = 0.0272114\;$eV${} =
0.62751\;$kcal/mol.

Unless otherwise stated, energy values are measured relative to the energy
of two isolated hydrogen atoms plus an isolated helium atom.  Thus
an energy $E = 0.0 \; \har$ corresponds to the H$_2$ dissociation energy,
while an isolated ground-state H$_2$ molecule (plus an isolated helium atom)
lies at $E = -164.6\;\millih = -103.3\;$kcal/mol.  The minimum of the
H$_2$ potential energy curve of Schwenke\cite{sh88}
is $E = -174.496\;\millih = -109.50\;$kcal/mol; the lowest possible energy
on the \hehh\ PES lies about $0.05\;\millih$
($0.03\;$kcal/mol) below this, due to the \hephh\ van der Waals well
(discussed in \S~\ref{sssec:generateheh2}).

Figure~\ref{fig:notation} illustrates the notation used in this paper
for the interatomic distances and angles.  The $(r,R,\gamma)$ coordinate
system is the natural one to use when considering low-energy interactions
of a He~atom with a near-equilibrium H$_2$~molecule, e.g., the van der
Waals well.  It is less suitable for higher-energy interactions, where
the He~atom may be relatively close to one of the H~atoms; \hehh\ surfaces
that attempt to fit this interaction region typically use the
$(r,R_A,R_B)$ coordinate system.

\subsection{Previous surfaces}
\label{ssec:prevsurf}

Many of the early \hehh\ surfaces were intended only to describe
the \hephh\ van der Waals potential well and the outer (low-energy)
part of the \hehh\ repulsive
wall\cite{Rob63,kk64,km65,gs70,gh72,tk73,gwa75,tt77,tt78,thccl92},
and were valid only for equilibrium H$_2$ molecules.  However,
several analytic \hehh\ surfaces were intended to be valid in
the higher-energy interaction region.

In 1974,
Wilson, Kapral, \& Burns\cite{wkb74} presented a more general \hehh~PES,
albeit with an extremely simple functional form (comprising the three
pair-wise potentials plus an interaction term with only four free
parameters).  Shortly thereafter, Dove \& Raynor\cite{dr78} used a
modified form of this surface (with somewhat improved pair-wise potentials)
to investigate the collisional dissociation of H$_2$ by~He.  Both of these
surfaces\cite{wkb74,dr78} contain an unphysical ``hole'' at relatively
short distances where the PES dropped abruptly to large negative values.
For these surfaces, the lowest point on the rim of this hole lies
at~$53 \; \millih$ (33~kcal/mol) and $87 \; \millih$ (55~kcal/mol),
respectively (where zero energy corresponds to separated {\it atoms}:
see \S~\ref{ssec:notation} above);
i.e., unphysical effects due to this hole could be encountered
at less than twice the dissociation energy relative to
equilibrium H$_2 + {}$He.  In addition, neither of
these surfaces\cite{wkb74,dr78} included any \hephh\ van der Waals well.

The 1982 surface of Russek \& Garcia\cite{rg82} was intended to fit only
the repulsive wall; their \abinitio\ energies had been calculated for
relatively small \hbox{$\rm He - H_2$} separations for near-equilibrium
H$_2$ molecules (of size $r = 1.2$, 1.4, and~$1.6 \; \bohr$).

The 1985 rigid-rotor \hehh\ PES
of Schaefer \& K\"ohler\cite{sk85} was fitted to the relatively
accurate \abinitio\ energies of Meyer, Hariharan, \& Kutzelnigg\cite{mhk80};
these had been calculated for relatively large \hbox{$\rm He - H_2$}
separations for near-equilibrium H$_2$ molecules ($r = 1.28$, 1.449,
and~$1.618 \; \bohr$), supplemented by less-accurate calculations over
a wider range ($0.9 \; \bohr \le r \le 2.0 \; \bohr$).
The PES
fitted the van der Waals well and the \hephh\ repulsive wall, but was
defined only for a limited range of H$_2$ molecule sizes
(differing by no more than 40\% from equilibrium~H$_2$),
and thus could not be used for cases where the H$_2$ molecule might
be dissociated or highly excited.

In 1994, Muchnick \& Russek\cite{mr94} presented a general \hehh\ PES,
with 19 fitted parameters.  This surface was fitted to a
combination of the Meyer, Hariharan, \& Kutzelnigg\cite{mhk80}
\abinitio\ energies (the less-comprehensive \abinitio\ energies of
Senff \& Burton\cite{sb85} were used as comparison values but not
actually fitted), and the somewhat less accurate \abinitio\ energies
of Russek \& Garcia\cite{rg82}.
This Muchnick \& Russek \hehh\ PES was a great improvement over previous
surfaces: it was designed to be accurate in the van der Waals well and
in the repulsive wall, and to behave reasonably in regions not constrained
by \abinitio\ data.

Recently, Tao\cite{Tao94} presented improved \abinitio\ \hehh\ energies
in the van der Waals well and repulsive wall, for near-equilibrium
H$_2$ molecules ($r = 1.28$, 1.449, and~$1.618 \; \bohr$).
These \abinitio\ energies should
be significantly more accurate than the earlier energies of
Meyer, Hariharan, \& Kutzelnigg\cite{mhk80} or Senff \&
Burton\cite{sb85} (and in fact show a van der Waals well that is
several percent deeper than indicated by the earlier energies).  However,
Tao\cite{Tao94} did not fit an analytic surface to the \abinitio\ energies.

\subsection{Goals of the present work}
\label{ssec:goals}

An analytic general \hehh\ PES was desired that would not only accurately
represent the van der Waals well, but also fit the interaction region
at the ``chemical accuracy'' level required for reaction dynamics (i.e.,
of order one millihartree).  In order to do this, $20\,203$ new
\abinitio\ energies were computed in the interaction region.  Both the
ground state energy and the first few excited state energies were
computed, and the conical intersection of the ground state with the
first excited state was mapped out approximately.  A new
analytic \hehh\ PES was fitted to these $20\,203$ ground
state \abinitio\ energies (and to an additional 4862 points generated to
constrain the fit at large separations),
yielding an order-of-magnitude improvement over previous \hehh\ surfaces.
These \abinitio\ energies and a Fortran program for this analytic PES
(including analytic first derivatives)
are available from EPAPS\cite{epaps} or from the authors\cite{bmp}.

\section{Methods}
\label{sec:methods}

\subsection{The grid of conformations to be fitted}
\label{ssec:grid}

\subsubsection{The main grid}
\label{sssec:maingrid}

\begin{figure}   
%
%
\plotfiddle{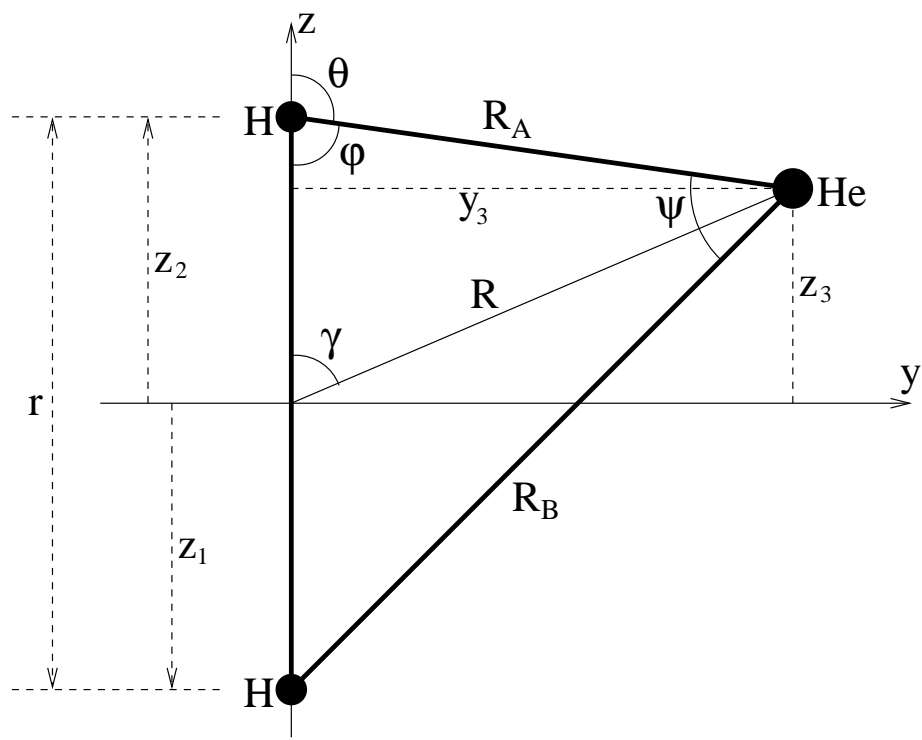}{3.7 true in}{-90}{50}{50}{-200}{295}
%
%
\caption[fig:notation]{Notation used in this paper for interatomic
distances and angles of \hehh.  By definition,
$\vert z_1 \vert = z_2 = r / 2$, with atom-3 being the helium atom.
}
\label{fig:notation}
\end{figure}

\begin{figure}   
%
%
\plotfidtwo{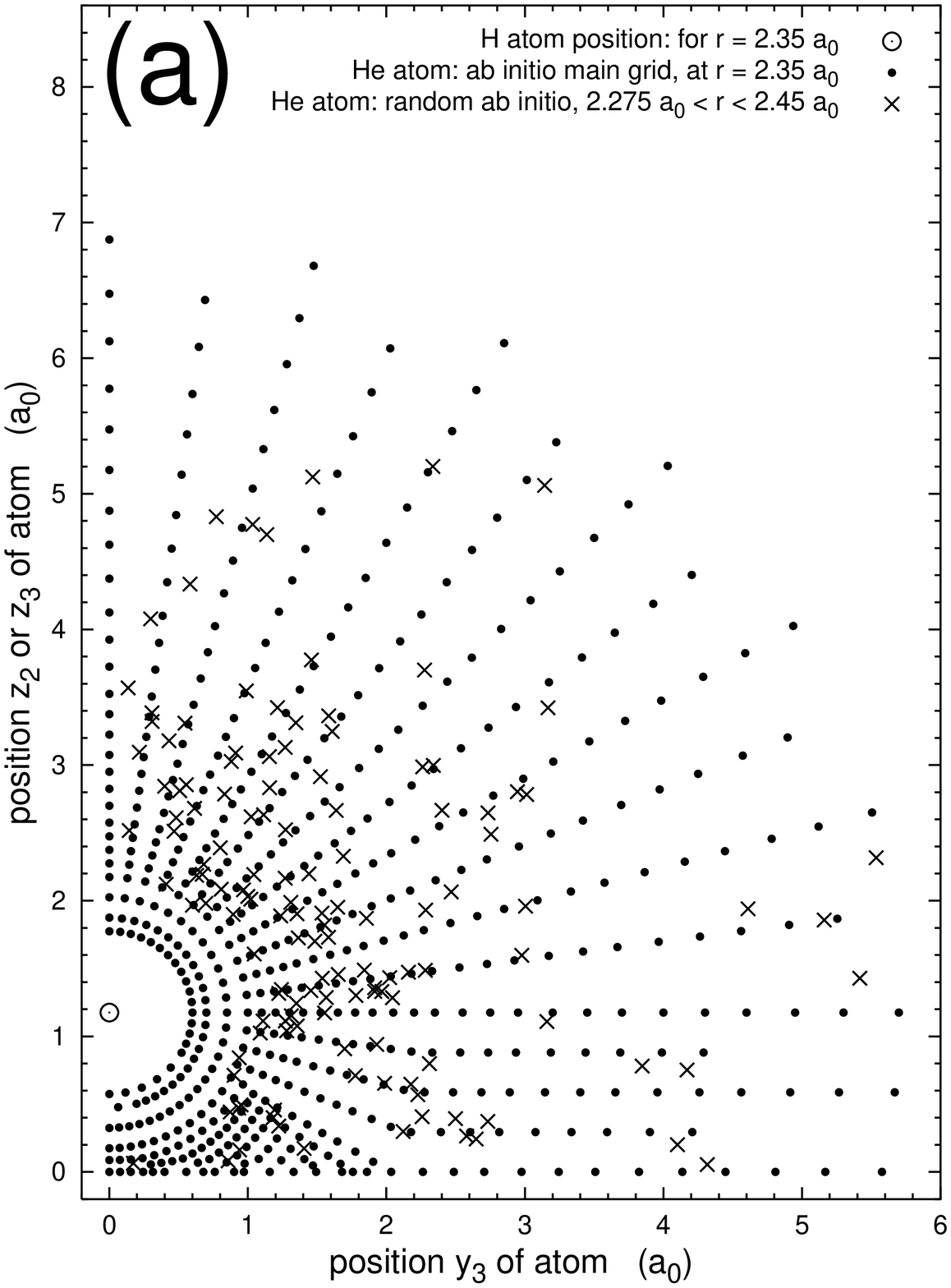}{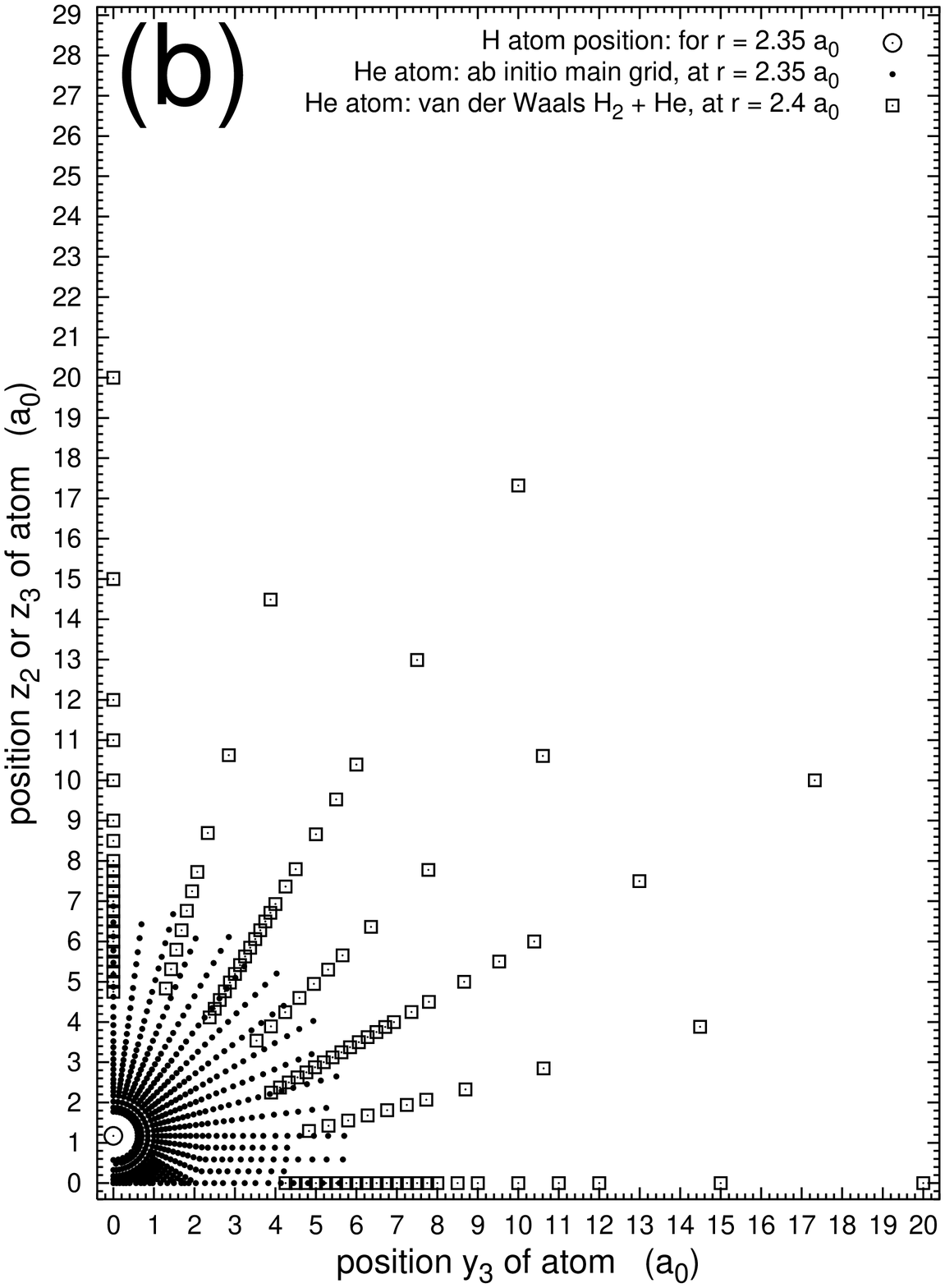}{4.3 true in}{0}{45}{45}{-255}{-15}{-25}
%
%
\caption[fig:closegrid]{The grid of fitted points for a given H$_2$-molecule
size~$r$, showing positions of the He~atom relative to a molecule lying on
the $z$-axis and centered at the origin.
(a)~The main grid of \abinitio\ points (solid dots), for $r = 2.35 \; \bohr$
(the H-atom position is shown by the open circle); crosses show positions of
``nearby'' random \abinitio\ points (in the same ``$r$-bin'', namely, with
$2.275 \; \bohr < r < 2.45 \; \bohr$).
(b)~The grid for the van der Waals H$_2 + {}$He points (open squares),
for $r = 2.4 \; \bohr$; the main grid of part~(a) is also visible (solid
dots).
}
\label{fig:closegrid}
\end{figure}

\begin{figure}   
%
%
\plotfiddle{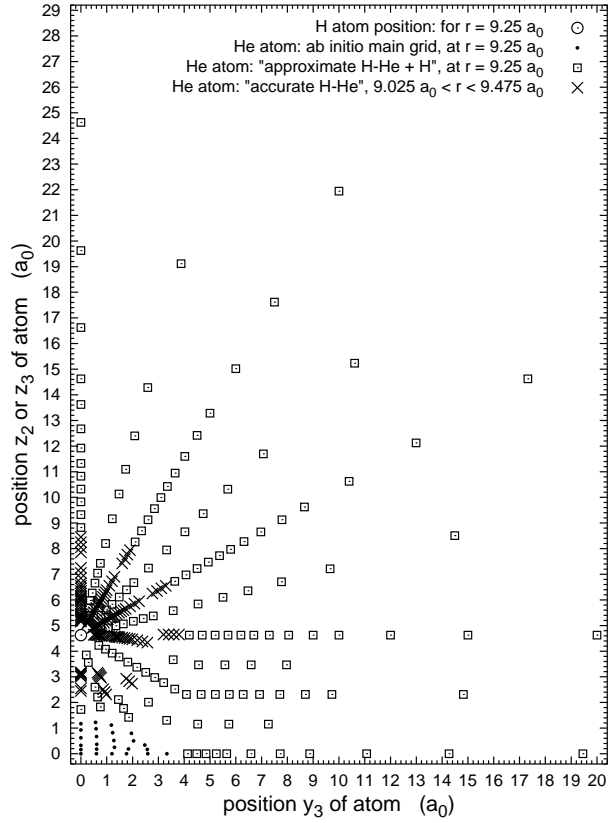}{4.3 true in}{0}{45}{45}{-150}{-15}
%
%
\caption[fig:fargrid]{The grid of fitted points for a large H$_2$ separation
$r = 9.25 \; \bohr$; unlike Fig.~\ref{fig:closegrid}, the \abinitio\ points
(solid dots) all lie relatively near the origin, between the H~atoms.  The
``approximate H-He${} + {}$H'' grid (open squares) at $r = 9.25 \; \bohr$
was designed to avoid other grids, including any ``nearby'' points
on the ``accurate H-He'' grid (crosses --- with
$9.025 \; \bohr < r < 9.475 \; \bohr$).
}
\label{fig:fargrid}
\end{figure}

The main set of \hehh\ \abinitio\ energies was computed for a grid of
$16\,703$ conformations,
defined in terms of the H$_2$ molecule size~$r$, the distance~$R_A$ of
the He~atom from one of the H~atoms, the angle~$\theta$ between
these two, and in some cases the distance~$z_3$ of the He~atom from the
midplane of the H$_2$ (see Fig.~\ref{fig:notation}).
The set of $R_A$~grid values
comprised $\{ 0.6$, 0.7, 0.85, 1.0, 1.1, 1.2, 1.3, 1.4, 1.525, 1.65,
1.775, 1.9, 2.05, 2.2, 2.35, 2.55, 2.75, 2.95, 3.2, 3.45, 3.7, 4.0, 4.3,
4.6, 4.95, 5.3, and~$5.7 \; \bohr \}$.  The H$_2$-size grid values~$r$
were taken from the same set as~$R_A$, supplemented by a set of extra values
at short distances $\{ 0.5$, 0.55, 0.65, 0.775, and~$0.925 \; \bohr \}$,
and a set at long distances $\{ 6.1$, 6.55, 7.0, 7.45, 7.9, 8.35, 8.8, 9.25,
9.7, 10.15, and~$10.6 \; \bohr \}$ --- Figure~\ref{fig:closegrid} shows a
typical grid example (a slice of the grid, at~$r = 2.35 \; \bohr$).
For large $\rm H - H$ separations (namely, $r \ge 6.1 \; \bohr$),
only conformations having $R_B \le 5.9 \; \bohr$
were computed (the He~atom lying more or less between the two
H~atoms) --- in other words, the shortest two of the three interatomic
distances were never allowed to exceed~$5.9 \; \bohr$.  An example of this
is shown in Figure~\ref{fig:fargrid} (a grid-slice at~$r = 9.25 \; \bohr$).

The angle~$\theta$ was in general taken at intervals
$\Delta\theta = 7.5^\circ$ (in $0^\circ \le \theta \le 180^\circ$);
for cases with $R_A = 5.7 \; \bohr$ or with $r = 0.5$, 0.55,
or~$0.65 \; \bohr$, intervals $\Delta\theta = 15^\circ$ were
used instead, and cases with $\theta > 90^\circ$ were treated specially.
Where the ray $\theta = 120^\circ$ reached the $y$-axis of
Figure~\ref{fig:notation} (i.e., the He~atom equidistant from the
two H~atoms), it was ``bent'' to follow the $y$-axis (since the grid needed
to cover only the first quadrant of the \hbox{$y$-$z$}~plane).  Similarly,
the rays $\theta = 102.5^\circ$, $105^\circ$, and~$97.5^\circ$ were ``bent''
to follow the lines $z_3 = r / 8$, $r / 4$, and~$3 r / 8$, respectively;
but in these cases, the rays were terminated when they reached the point
where the usual angular
separation $R_A \Delta\theta$ reached twice the separation between the
``bent'' rays (so as to avoid excessively closely spaced angles at
large~$R_A$).  For $\theta > 120^\circ$ (i.e., $\max\{ R_A, R_B \} < r$,
with the He~atom lying between the H~atoms), a tighter $\theta$-interval
was used if necessary to yield $R_A \Delta\theta \lesssim \Delta R_A$;
also, extra points were interspersed on the $y$-axis if necessary,
so that $\Delta y_3 \approx \Delta R_A$ on the $y$-axis.  An example of
this is shown in Figure~\ref{fig:closegrid}a.

\subsubsection{Set of ``random'' conformations}
\label{sssec:randomgrid}

Another set of \hehh\ \abinitio\ energies was computed for 3500~random
conformations.  The distances were chosen in the ranges
$0.6 \; \bohr < r < 9 \; \bohr$ and $0.9 \; \bohr < R_A < 5.7 \; \bohr$
using non-uniform probability distributions that had
broad peaks near~$1.5 \; \bohr$.
The angle~$\theta$ was chosen using a uniform distribution, subject to
the constraints that the the He~atom must lie in the first quadrant of the
\hbox{$y$-$z$}~plane and that no atom could lie further than~$5.7 \; \bohr$
from its nearest neighbor.  Also, ``near-duplicate'' conformations were not
permitted (i.e., where all cartesian coordinates agreed
within $0.01 \; \bohr$ with the coordinates of another conformation).
Figure~\ref{fig:closegrid}a shows those random
conformations that lie near the grid-slice $r = 2.35 \; \bohr$.

\subsubsection{Unfitted ``random'' conformations}
\label{sssec:randomtst}

The same probability distributions were used to obtain a second set of
3500 random \hehh\ conformations, for testing purposes.
The resulting \abinitio\ energies 
were used only to check the rms error at ``interstitial'' conformations;
they were {\it not\/} used in the surface fitting process.

\subsubsection{Accurate van der Waals He${} + {}$H$_2$ energies}
\label{sssec:generateheh2}

A set of 2145 accurate energies was generated for conformations in the
van der Waals well and outer repulsive wall, for H$_2$ molecules not
too far from equilibrium.  These energies were generated from
the accurate \abinitio\ energies and fitted surfaces of other
authors\cite{sk85,mr94,Tao94}, as described below.
These accurate van der Waals energies were given high weight in
the fit, with lower weight for smaller~$R$ or for $r$~further from
equilibrium (as discussed in \S~\ref{sssec:fitweights}).

Unlike the main grid, these conformations
were defined in terms of the distance and angle of the He~atom relative
to the H$_2$~{\it molecule}, i.e., in terms of $\{r,R,\gamma\}$
(rather than $\{r,R_A,\theta\}$).  The set of $r$~values comprised
$\{ 0.6$, 0.7, 0.8, 0.9, 1.1, 1.28, 1.449, 1.618, 1.8, 2.0, 2.4, 3.0,
and~$4.0 \; \bohr \}$; the set of $R$~values comprised $\{ 4.0$, 4.5, 5.0,
5.25, 5.5, 5.75, 6.0, 6.25, 6.5, 6.75, 7.0, 7.25, 7.5, 7.75, 8.0, 8.5,
9.0, 10.0, 11.0, 12.0, 15.0, and~$20.0 \; \bohr \}$.  The angle~$\gamma$
was generally taken at $15^\circ$ intervals
(in $0^\circ \le \gamma \le 90^\circ$); for $R = 20.0 \; \bohr$,
and for roughly half of the $R$~values at non-equilibrium~$r$,
intervals of $30^\circ$ were used instead.
Conformations generated at relatively
small~$R$ were discarded if they had an interaction energy of more than
a few~$\millih$ between the He~atom and the H$_2$ molecule (i.e., only
the van der Waals well and the very outermost part of the ``repulsive
wall'' were included).  An example of this grid is shown by the squares in
Figure~\ref{fig:closegrid}b (a grid-slice at~$r = 2.4 \; \bohr$).  Note that
there is an overlap with the main \abinitio\ grid.  In the outer part of this
overlap region, the interaction energy of He with the~H$_2$ molecule is
comparable to the uncertainty in the main-grid \abinitio\ energies,
so the fit depends largely on the generated van der Waals H$_2 + {}$He
energies.  In the inner part of the overlap region, the interaction energies
are larger and the generated van der Waals H$_2 + {}$He energies are given
fairly low weight (see \S~\ref{sssec:fitweights}), so both sets of energies
contribute to the fit.

Tao\cite{Tao94} used the complete fourth-order M{\o}ller-Plesset
approximation (MP4)\cite{mp34,pksb79} with a large basis set to
compute 69~accurate \abinitio\ \hehh\ energies
in the van der Waals well and repulsive wall (15~$R$ values in
$2 \; \bohr \le R \le 15 \; \bohr$), for three orientations
($\gamma = 0^\circ$, $45^\circ$, and~$90^\circ$); separations~$R$
over the full range were reported only for the equilibrium H$_2$ molecule
($r = 1.449 \; \bohr$), but a few ($R = 3.0$, 5.0, 6.5, and~$8.0 \; \bohr$)
were also reported at $r = 1.28$ and~$1.618 \; \bohr$.  
(Note that Tao reported a maximum van der Waals well depth of $47\;\microh$
for He plus equilibrium H$_2$, at $r = 1.449 \; \bohr$,
$R = 6.5\;\bohr$, $\gamma = 0^\circ$.)
This subset of
\abinitio\ energies was referred to as the ``Tao vdW'' points.

For $r = 1.28$, 1.449, and~$1.618 \; \bohr$, a relatively simple formula
was fitted (for $\gamma = 0^\circ$, $45^\circ$, and~$90^\circ$) to the
difference between the Tao\cite{Tao94} \abinitio\ energies and the
analytic surface of Schaefer \& K\"ohler\cite{sk85}, and used to
interpolate corrected energies at $R$ and~$\gamma$ values not computed
by Tao.  These generated energies were referred to as ``gen-Tao vdW''
points.

For $r \le 1.1 \; \bohr$ and $r \ge 1.8 \; \bohr$, energies were
obtained from a version of the Muchnick \& Russek\cite{mr94} surface
where the van der Waals term in their potential had been modified
slightly to better fit the newer, more accurate
Tao\cite{Tao94} \abinitio\ energies
(this surface is the one described in \S~\ref{sssec:modMuRu}).
These generated energies were referred to as ``gen-MR vdW'' points.

For $r = 0.9$, 1.1, 1.8, and~$2.0 \; \bohr$, energies from the
Schaefer \& K\"ohler\cite{sk85} surface were also used (since it
might give energies of comparable accuracy in this region).
These generated energies were referred to as ``gen-SK vdW'' points.

\subsubsection{Accurate ``H-He'' energies}
\label{sssec:generateheh}

A set of 1887 energies was generated for conformations consisting
of a relatively close-together $\rm H - He$ pair (of size~$R_A$)
with a very-distant H~atom (a distance $R'$ from the nearest
point on the line segment~$R_A$).  Since the contribution of
the isolated H~atom was completely negligible, this subset was
referred to as ``H-He'' points.

For $R_A \le 4 \; \bohr$, the H-He energies were generated
from 31 \abinitio\ $\rm He + H$ energies that we computed
at $R_A = \{0.5$, 0.55, 0.6, 0.65, 0.7, 0.75, 0.8, 0.9, 1.0, 1.1, 1.2, 1.3,
1.35, 1.4, 1.45, 1.5, 1.6, 1.7, 1.8, 1.9, 2.0, 2.2, 2.4, 2.6, 2.8, 3.0,
3.2, 3.4, 3.6, 3.8, and~$4.0 \; \bohr \}$ (note that we had also computed
9~other \abinitio\ energies at larger~$R_A$ values, which were not used).
For $R_A > 4 \; \bohr$, the H-He
energies were generated from the $\rm H - He$ pair-wise potential of
Dove \& Raynor\cite{dr78} (which used slightly-modified functions from
Gengenbach, Hahn, \& Toennies\cite{ght73}), with values chosen at
$R_A = \{4.25$, 4.5, 4.75, 5.0, 5.25, 5.5, 5.75, 6.0, 6.25, 6.5,
6.75, 7.0, 7.25, 7.5, 7.75, 8.0, 8.5, 9.0, 10.0, 11.0, 12.0, 15.0,
and~$20.0 \; \bohr \}$.  Distances to the isolated H~atom comprised
$R' = \{ 7.1$, 7.7, 8.3, 9.3, 10.3, 11.4, 13.1, 14.8, 16.5, 19.0,
22.0, and~$25.0 \; \bohr \}$.
Conformations were discarded
whenever pair-wise $\rm H - He$ or $\rm H - H$ energies from the isolated
H~atom would be non-negligible compared to the energy
at separation~$R_A$ of the relatively close-together $\rm H - He$ pair.
(Although they were negligibly small for all retained conformations,
pair-wise interaction energies of the isolated H~atom
with the other two atoms were nonetheless added
to the main H-He energy, for consistency with the ``H-He plus H'' energies
of \S~\ref{sssec:generatehehh} below).
For each pair $\{R_A,R'\}$, eight angular grid-points were chosen,
four with $r = R'$ at $\varphi = 90^\circ$, $120^\circ$, $150^\circ$,
and~$180^\circ$, and four with $R_B = R'$ at $\Psi = 90^\circ$,
$120^\circ$, $150^\circ$, and~$180^\circ$ 
(see Fig.~\ref{fig:notation}) --- a ``race-track'' shaped grid.  For
$R' > 2 \, R_A$, the two points at $\varphi = 90^\circ$ and $\Psi = 90^\circ$
were replaced by a single point at a distance~$R'$ from the midpoint of~$R_A$.
Higher weight was given to conformations with large $R_A$~values
(as discussed in \S~\ref{sssec:fitweights}).

\subsubsection{Approximate ``H-He + H'' energies}
\label{sssec:generatehehh}

A set of 830 approximate energies was generated to constrain the fit
in regions not covered by the above grids, where the two H~atoms were
fairly far apart ($5 \; \bohr \le r \le 19 \; \bohr$), and at least
one of the H~atoms was well separated from the He~atom so that
the interaction energy
of this isolated H~atom would be small compared to the uncertainty in our
\abinitio\ energies (but might not be small compared to the van der
Waals well depth).  Since the contribution of
the isolated H~atom was small but not completely negligible, this
subset was referred to as ``H-$\rm He + H$'' points.

The grid was similar to the main grid, but with
$r = \{ 5.0$, 6.55, 7.2, 7.9, 8.5, 9.25, 9.9, 10.6, 11.4, 14.8,
and~$19.0 \; \bohr \}$ and $R_A = \{ 0.8$, 1.1, 1.4, 1.7, 2.1, 2.5, 2.9,
3.3, 3.7, 4.2, 4.7, 5.2, 5.7, 6.2, 6.7, 7.3, 8.05, 9.0, 10.0, 12.0, 15.0,
and~$20.0 \; \bohr \}$; the minimum angular spacing was $15^\circ$
in~$\theta$.  Many parts of this grid were in fact more coarsely spaced
than this in $R_A$ and/or~$\theta$, and for $r = 7.2$, 8.5, 9.9, and~11.4
almost all gridpoints with $\theta < 105^\circ$ were omitted.
In addition, conformations relatively close to areas covered by the
other grids were omitted; this restriction eliminated cases where more than
one of the three pair-wise potentials would exceed~$\sim 0.1 \; \millih$.
Approximate energies were thus computed by
simply summing the three pair-wise potentials.
Cases where this computed energy would be expected to be quite
accurate (i.e., large~$R_A$) were given increased weight
(as discussed in \S~\ref{sssec:fitweights}).

\subsection{Ab initio computations and analysis}
\label{ssec:abinitio}

\subsubsection{Computational methods and CPU-time}
\label{sssec:compmethods}

The \abinitio\ computations and analysis of errors largely follow
the methods described in our previous
papers\cite{bo91,bo91paps,bkmp,bkmp2,bmkp02,bmkp02epaps}.
Energies were computed using the workstation version\cite{bf92}
of Buenker's \hbox{MRD-CI} program\cite{bu87}.  At the position of each
of the two hydrogen atoms, we used the $(9s3p2d)/[4s3p2d]$
Gaussian basis set from Boothroyd \etal\cite{bo91,bo91paps}
(with $d$-functions optimized for~H$_2$).  At the position of the
helium atom, we used a $(10s4p2d)/[5s4p2d]$ Gaussian basis set,
with $s$-functions taken from a (partially-decontracted) set from
Huzinga\cite{Huz65}, and the $p$- and $d$-functions chosen based on
hints from several papers\cite{Huz65,fpb84,pbtasm88}; the basis
set error for an isolated He~atom is only~$1.028 \; \millih$ with this
basis set.  These two basis sets are given in Table~\ref{tab:basis};
they comprise the largest combination of basis sets allowed
by array sizes in the \hbox{MRD-CI} program\cite{bf92}.
For most conformations, molecular
orbitals were obtained from closed shell~SCF;
if closed shell SCF iterations were
slow to converge, open shell~SCF or mixed (1~closed, 2~open) shell~SCF
was automatically used instead.  A configuration selection threshold of
$T = 2\;\microh$ was used for most points, with $T = 10\;\microh$ used
for 3006 high-energy cases at short distances (Buenker's MRD-CI program
automatically extrapolates to zero threshold, using truncated-CI
energies at~$T$ and at~$2T$ plus a parameterization of the estimated
effects of neglected configurations).  Multiple CI roots were computed,
to ensure that the ground state energy was always among the energies
reported by the program.  A fairly extensive set of reference
configurations was used for the CI calculation: the minimum size
of~$\Crsq$ (the sum of reference configuration $C^2$ values) was~0.969,
and the {\it average\/} size was~0.982.  (Note that additional reference
configurations were included automatically if the program estimated that
they would increase $\Crsq$ by more than 0.001 individually and by more
than 0.003 collectively).

The above relatively small basis set and relatively large selection
threshold were required by CPU-time considerations.  So long as the
\abinitio\ energies are of accuracy comparable to the desired ``chemical
accuracy,'' the most important factor in obtaining a good fitted
analytic surface is adequate coverage in the 3-dimensional conformation
space of~\hehh, which requires large numbers of \abinitio\ energies.
In total, over a year of CPU-time plus I/O-time was required for the
total of $23\,406$ \abinitio\ computations (this includes 3203 points
that were initially computed with only a single CI root, or with
$T = 10\;\microh$, and which were subsequently re-computed more
accurately).
These computations were distributed among about a dozen different
computers, mostly workstations of various types, but including
a couple of higher-performance machines.
For the faster computers the I/O-time tended to dominate (due to large
temporary files), but there was one exception to this: over two thirds of
the energies were computed on a single higher-end SGI with a fast disk, which
allowed computation of a typical \abinitio\ energy in about 10 minutes.
The other computers typically
required 30 to~60 minutes per point, with some older machines requiring
up to several hours per point.

\subsubsection{Small corrections applied to the MRD-CI energies}
\label{sssec:correctmrdci}

Extrapolation to zero
threshold (performed automatically by the \hbox{MRD-CI} program) and the
Davidson correction to full~CI are both standard procedures.
Although the \hbox{MRD-CI} program uses only the energy at thresholds
$T$ and~$2T$ to extrapolate to zero threshold, it also calculates the
energy at $3T$ and~$4T$; we therefore used these latter to make a very
small (rms size $0.05 \; \millih$) ``curvature
correction'' to the extrapolation (as
Boothroyd \etal\cite{bo91,bo91paps,bmkp02,bmkp02epaps} did for~H$_4$ --- but
unlike H$_4$, this was the {\it only\/} modification of the extrapolation).
The average size of the extrapolation was~$1.73 \; \millih$ for the
threshold $T = 2 \; \microh$, and about 3~times as large
for $T = 10 \; \microh$.
For the $17\,025$~cases where energies were obtained with
both $T = 10$ and~$2 \; \microh$, the rms difference between extrapolated
energies was~$0.17 \; \millih$.  This should be an overestimate of the
extrapolation error of either, suggesting that the rms error
in the extrapolation to zero threshold is about~$0.15 \; \millih$
for $T = 10 \; \microh$ (used for 3006 high-energy points) and
about~$0.1 \; \millih$ for $T = 2 \; \microh$ (used for the other
$17\,197$ points).  This error estimate is also supported by the fact
that, for 996~points where a single-root $T = 0.4 \; \microh$ case was also
computed, the rms difference was $0.12 \; \millih$ relative to the
5-root $T = 2 \; \microh$ cases.

As in Boothroyd \etal\cite{bo91,bo91paps,bmkp02,bmkp02epaps}, we made a
parameterized Davidson-type correction to full~CI, namely,
\begin{equation}
\label{eq:ddc}
 \Delta_{\rm full-CI} =
 \lambda_{DC}^{\!\scriptscriptstyle\rm(SCF)} \Delta E_{DC} =
 \lambda_{DC}^{\!\scriptscriptstyle\rm(SCF)}
 (1-\Crsq) \Delta E_{sd} / \Crsq \quad ,
\end{equation}
where $\Delta E_{sd}$ is the contribution of single and double
excitations to the energy, and $\Delta E_{DC}$ is the standard Davidson
correction.  For single-reference closed-shell
configurations with at least four electrons, if the effects of quadruple
excitations are assumed to be much larger than triple excitations,
then the above formula for the Davidson correction (with an expected
value of $\lambda_{DC} \sim 0.5$ for closed-shell \hehh) can be derived
theoretically\cite{bpb82,bur83,bbbp83,sbb87}.  We not only used multiple
references, we also used open and mixed shell SCF cases; we therefore
obtained values for the parameter~$\lambda_{DC}$ by minimizing the average
differences between MRD-CI energy values computed for the same point
with different reference sets (which thus had different sizes for the
Davidson correction).  We used
$\lambda_{DC}^{\!\scriptscriptstyle\rm(SCF)} = 0.52$, 0.16, and~0.32 for
closed, open, and mixed SCF cases, respectively; the average size of
$\Delta E_{DC}$ was
1.31, 1.90, and~$0.60\;\millih$, respectively,
yielding full-CI corrections averaging 0.68, 0.30, and~$0.19\;\millih$,
respectively.  For the $12\,077$~cases where the program
obtained improved reference sets, and thus two different-sized Davidson
corrections were available, the rms difference between the resulting energies
was~$0.16\;\millih$.  This may be a slight underestimate of the energy
uncertainty resulting from the Davidson correction, since on average the
initial reference set yielded a Davidson correction only 1.75~times as
large as the improved reference set, and we have no
information from cases where only one value of the Davidson correction
was available.
There are suggestions of a systematic uncertainty of order~0.1 in the
preferred value of~$\lambda_{DC}^{\!\scriptscriptstyle\rm(SCF)}$, leading
to a systematic uncertainty in the energy correction of
order~$0.13 \; \millih$.  We estimate the total uncertainty in
the Davidson-type correction to full-CI to be about~$0.3\;\millih$.

The final energy value~$E_{\rm final}$ was obtained from the estimated
full-CI energy using a London-type basis correction similar to that for
H$_4$ described in Boothroyd \etal\cite{bo91,bo91paps,bmkp02,bmkp02epaps}.
The H$_4$ London-type basis correction is obtained
by taking the difference between the H$_4$ London energy with the
accurate H$_2$ singlet curve and the H$_4$ London energy with the H$_2$
singlet curve computed using the finite basis set.  For \hehh, a London-type
basis correction was obtained by taking the H$_4$ basis correction, letting
one distance go to zero (to yield the He~atom), and replacing the
contribution from the zero distance by the isolated He~atom basis correction
of~$1.028 \; \millih$.  (After doing the algebra, this turns out to mean
taking the H$_2$ singlet basis correction for the H$_2$ molecule, adding
the isolated He~atom basis correction, and adding
half of the H$_2$ singlet basis correction for each of the $\rm H - He$
distances.)  This yields the correct
basis correction in the limit of a separated He~atom plus H$_2$ molecule.
The average size of this
correction was~$2.32 \; \millih$ (or $1.29 \; \millih$ if one excludes
the He~atom basis correction).  Boothroyd \etal\cite{bmkp02,bmkp02epaps}
concluded in the case of H$_4$ that the systematic error in the basis
correction might be as much as $1\;\millih$.  For \hehh,
the error should be somewhat smaller; one might estimate that the
error could be comparable to the non-H$_2$, non-He part of the basis
correction, i.e., about~$0.6\;\millih$.  Comparison of a number of energies
that were also computed using smaller basis sets suggests that this error
estimate is reasonable, as does comparison with energies computed
by Tao\cite{Tao94} using a larger basis set.
The error in the basis correction will be larger for geometries
where the He~atom is close to one (or both) of the H~atoms, and smaller
for geometries where the He~atom is relatively far from the H~atoms.

\subsubsection{Best MRD-CI energies and estimated errors}
\label{sssec:bestmrdci}

Where more than one energy value had been computed for a given
conformation, a weighted average of the ``best'' values
was obtained for use in the fitting
procedure.  (In general, multiple-root cases were preferred to
single-root cases, closed shell cases to open and mixed shell cases,
lower thresholds to higher ones, and smaller Davidson corrections to
larger ones.)  The extrapolation to zero threshold and the Davidson-type
correction to full~CI appear to yield largely ``random'' errors; the
systematic components of these, though not completely negligible,
are much smaller than the
systematic error in the basis correction.  Thus the total uncertainty in
the \abinitio\ energies probably comprises a ``random'' error of
about~$0.2\;\millih$ and a systematic error of order~$0.6\;\millih$.
The \abinitio\ energies are available from
EPAPS\cite{epaps} or from the authors\cite{bmp}.

\subsection{Functional Representation of the He-H$_2$  Surface}
\label{ssec:heh2surfeqns}

This section lays out the equations which underlie our fits to the
\hehh\ \abinitio\ data.  Since the Muchnick \& Russek\cite{mr94} surface
was the best of the previous surfaces, we based our analytic formulae
on theirs, but added terms to get more flexibility.  The surface is
defined in terms of the three interatomic distances
$\vec{\cal R} \equiv \{r,R_A,R_B\}$, where $r$~is the
$\rm H - H$ distance and $R_A$ and~$R_B$ are the two
$\rm He - H$ distances (see Fig.~\ref{fig:notation}).
As in Muchnick \& Russek\cite{mr94}, new (basically spheroidal)
variables are defined as
\begin{equation}
  {\bar R} \equiv { { R_A + R_B } \over 2 } \; , \qquad
  \eta \equiv { { R_A - R_B } \over r } \; .
      \label{eq:vardef}
\end{equation}
Note that ${\bar R} \rightarrow R$ and $\eta \rightarrow \cos\gamma$
for $R \gg r$, i.e., the usual Legendre function expansion terms
$P_L(\cos\gamma) / R^n$ in the van der Waals well can be replaced by
$P_L(\eta) / {\bar R}^n$.  However, the functional form of the fitted
surface does {\it not\/} explicitly contain either~$R$ or~$\gamma$.

\subsubsection{Functional form of our fitted surfaces}
\label{sssec:fittedeqns}

The full \hehh\ potential energy surface is given by
\begin{equation}
  V_{\rm He\,H_2}(\vec{\cal R}) = V_{\rm H_2}(r) + V_{int}(\vec{\cal R}) \; ,
      \label{eq:Vheh2}
\end{equation}
where
\begin{equation}
  V_{int}(\vec{\cal R}) = V_C(\vec{\cal R})
    + V_H(\vec{\cal R}) + V_{cb}(\vec{\cal R}) + V_d(\vec{\cal R})
    + V_D(\vec{\cal R}) \; .
      \label{eq:Vint}
\end{equation}
The $V_{\rm H_2}$~term is the isolated-H$_2$-molecule potential (which is
of course a function
only of the distance~$r$ between the two H~atoms); for this
we used the accurate H$_2$-molecule potential of Schwenke\cite{sh88}.
The interaction energy~$V_{int}$ between the H$_2$ ``molecule'' and the
He~atom was fitted by several terms, functions of all three interatomic
distances (all terms except~$V_D$ being generalizations of terms
used by Muchnick \& Russek\cite{mr94}).
The basic forms of these terms correspond to Coulomb,
Pauli, and dispersion terms, as discussed by Muchnick \& Russek\cite{mr94}.
However, the added flexibility (provided by the summations in the equations
below) largely blurs the
distinctions between these types of terms in our fitted surfaces.

The $V_C$~term has the form of a Coulomb term
(omitting the $\rm H - H$ contribution):
\begin{equation}
  V_C(\vec{\cal R}) = A_C(\vec{\cal R})
    \left[ 1 + \left( { 2 \over A_C(\vec{\cal R}) } - 1 \right)
      P_2(\eta) \right]
    \Big[ F_C(R_A) + F_C(R_B) \Big] \; ,
      \label{eq:Vc}
\end{equation}
where
\begin{equation}
  F_C(R_\nu) = { e^{-\lambda_C R_\nu} \over R_\nu }
    \left[ 1 + \sum_{j=1}^{J_C} { b_C(j) \over {R_\nu}^j } \right]
  \qquad \hbox{for} \quad \nu = A, B
      \label{eq:Fc}
\end{equation}
and
\begin{equation}
  A_C(\vec{\cal R}) = 2 - e^{-\beta_C {\displaystyle r}} \sum_{m=1}^{M_C} r^m
    \sum_{ { {L=0} \atop {L\;\rm even} } }^{L_C} P_L(\eta)
    \sum_{i=0}^{I_C} { a_C(m,L,i) \over {\bar R}^i } \; ;
      \label{eq:Ac}
\end{equation}
note that the lower limit $m = 1$ on the summation (rather than $m = 0$)
was chosen in order that $A_C(\vec{\cal R}) \rightarrow 2$ for
$r \rightarrow 0$ as well as for $r \rightarrow \infty$ (as in the
similar term of Muchnick \& Russek\cite{mr94}).
Parameters to be fitted include $\lambda_C$, $\beta_C$, and all of the
$a_C(m,L,i)$ and $b_C(j)$.  Our final adopted surface had
$\{ M_C, L_C, I_C, J_C \} = \{ 2, 2, 1, 5 \}$, but we tested fits with
ranges as high as $\{ 3, 4, 2, 5 \}$.
Note that Muchnick \& Russek\cite{mr94} used this $V_C$~term with
only the parameters $\lambda_C$, $\beta_C$, and $a_C(1,0,0)$.

The $V_H$~term looks like a Pauli-type interaction between the He~atom and
each of the H~atoms:
\begin{equation}
  V_H(\vec{\cal R}) = A_H \left[ 1 - e^{-\beta_H {\displaystyle r}}
    \sum_{m=0}^{M_H} r^m \sum_{ { {L=0} \atop {L\;\rm even} } }^{L_H}
    P_L(\eta) \sum_{i=0}^{I_H} { a_H(m,L,i) \over {\bar R}^i } \right]
    \Big[ F_H(R_A) + F_H(R_B) \Big] \; ,
      \label{eq:Vh}
\end{equation}
where $a_H(0,0,0) \equiv 1$ is a constant (in order that the lowest-order
form of $V_H$ vanish at $r = 0$, for consistency with the similar term
of Muchnick \& Russek\cite{mr94}), and
\begin{equation}
  F_H(R_\nu) = { e^{-\lambda_H R_\nu} }
    \left[ 1 + \sum_{j=1}^{J_H} { b_H(j) \over {R_\nu}^j } \right]
  \qquad \hbox{for} \quad \nu = A, B \; .
      \label{eq:Fh}
\end{equation}
Parameters to be fitted include $A_H$, $\lambda_H$, $\beta_H$, and all
of the $b_C(j)$ and $a_C(m,L,i)$ except for $a_H(0,0,0)$.
Our final adopted surface had
$\{ M_H, L_H, I_H, J_H \} = \{ 2, 4, 2, 5 \}$, but we tested fits with
ranges as high as $\{ 3, 6, 3, 5 \}$.
Note that Muchnick \& Russek\cite{mr94} used this $V_H$~term with
only the parameters $A_H$, $\lambda_H$, and~$\beta_H$.

The $V_{cb}$~term looks like a Pauli-type interaction between the He~atom and
that part of the electronic distribution drawn in to form the $\rm H - H$
covalent bond:
\begin{equation}
  V_{cb}(\vec{\cal R}) =
    e^{ -\beta_{cb} {\displaystyle r} - \Lambda_{cb}(r) {\bar R} }
    \sum_{m=0}^{M_{cb}} r^m
    \sum_{ { {L=0} \atop {L\;\rm even} } }^{L_{cb}} P_L(\eta)
    \sum_{i=I_{cb}}^{J_{cb}} { a_{cb}(m,L,i) {\bar R}^i } \; ,
      \label{eq:Vcb}
\end{equation}
where
\begin{equation}
  \Lambda_{cb}(r) = \varepsilon_{cb}
    + \zeta_{cb} e^{-\delta_{cb} {\displaystyle r}} \; .
      \label{eq:LAMBDAcb}
\end{equation}
Parameters to be fitted include $\beta_{cb}$, $\varepsilon_{cb}$,
$\zeta_{cb}$, $\delta_{cb}$, and all of the $a_{cb}(m,L,i)$.
Our final adopted surface had
$\{ M_{cb}, L_{cb}, I_{cb}, J_{cb} \} = \{ 2, 4, -1, 2 \}$, but we
tested fits with ranges as high as $\{ 3, 6, -2, 2 \}$.
Note that Muchnick \& Russek\cite{mr94} used this $V_{cb}$~term with
only the parameters $\beta_{cb}$, $\varepsilon_{cb}$,
$\zeta_{cb}$, $\delta_{cb}$, $a_{cb}(0,0,0)$, and $a_{cb}(0,2,0)$.

The $V_d$~term is a dispersion (long-range) interaction between the He~atom
and the H$_2$ molecule:
\begin{eqnarray}
  V_d(\vec{\cal R}) & = & - \, \fad \, A_d(r)
    \left\{ { 1 \over { {\bar R}^6 + {C_1}^6 } }
    \left[ 1 + f_{\alpha d} \, \alpha_d(r) P_2(\eta)
    + \!\!\! \sum_{ { {L=4} \atop {L\;\rm even} } }^{L_d}
    \!\! e^{\varepsilon(6,L) \, {\displaystyle r}}
    P_L(\eta) \sum_{m=1}^{M_d} a_d(6,L,m) \, r^m \right] \right. \nonumber \\
 & & \qquad \qquad
    + \, { { C_3 \left[ 1 + f_{\alpha d} \, \alpha_d(r) P_2(\eta) \right] }
    \over { {\bar R}^8 + {C_2}^8 } } \nonumber \\
 & & \qquad \qquad \left.
    + \sum_{ { {n=8} \atop {n\;\rm even} } }^{N_d}
    { 1 \over { {\bar R}^n + \left[ C_d(n) \right]^n } }
    \sum_{ { {L=0} \atop {L\;\rm even} } }^{L_d}
    e^{\varepsilon(n,L) \, {\displaystyle r}}
    P_L(\eta) \sum_{m=\min\{L,1\}}^{M_d} a_d(n,L,m) \, r^m
    \right\} \; ,
      \label{eq:Vd}
\end{eqnarray}
where
\begin{equation}
  A_d(r) = \left( 4.790521 \, r^2 - 2.494444 \, r + 3.008850 \right)
    e^{ -0.5891 \, {\displaystyle r} }
      \label{eq:Ad}
\end{equation}
and
\begin{equation}
  \alpha_d(r) = \left( 0.238248 \, r^2 - 0.126241 \, r \right)
    e^{ -0.8075 \, {\displaystyle r} }
      \label{eq:ALPHAd}
\end{equation}
(where $r$ is in~$\bohr$).
Note that the values of~$\varepsilon(n,L)$ were close to~$-1.0$ in all fits,
so that the term containing $C_2$ and~$C_3$ (retained for consistency with
the form of Muchnick \& Russek\cite{mr94}) has a significantly different
behavior from the $n = 8$ term in the last of the summations.  Note
also that only the $L = 0$ terms are allowed to have $m = 0$, since
the angular variation in~$V_d$ must vanish for $r \rightarrow 0$.  For
cases (such as the final fitted surface) with an upper limit $L_d < 4$,
the last term in the large square brackets is omitted (since $L = 4$ is the
lowest $L$ value allowed in that term).

\begin{figure}   
%
%
\plotfiddle{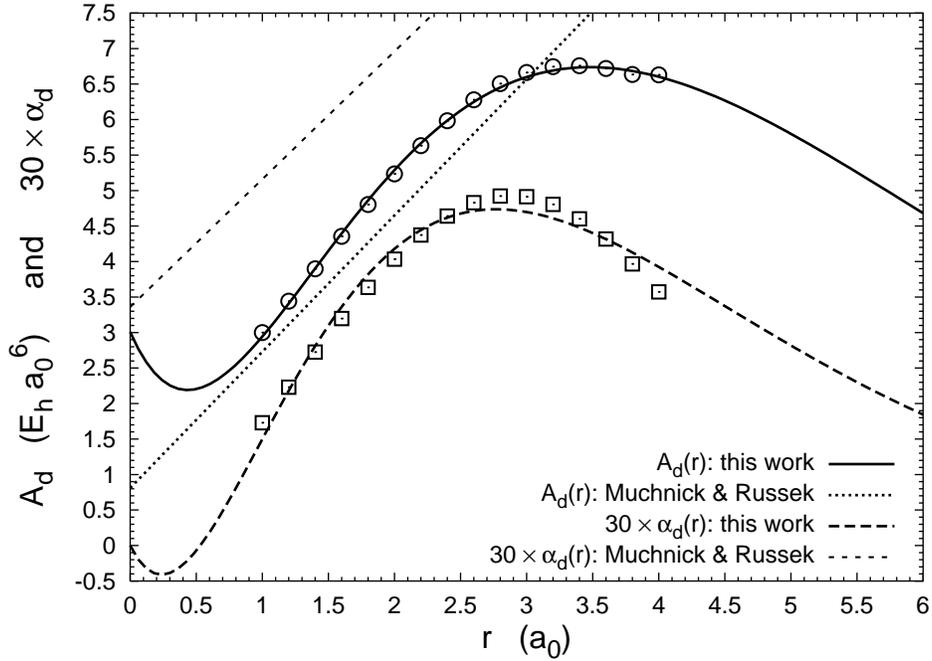}{3.6 true in}{-90}{50}{50}{-200}{295}
%
%
\caption[fig:thakkar]{Our fits $A_d(r)$ and $\alpha_d(r)$ (solid and dashed
lines, respectively) to the asymptotic parameters $A_d$
and $\alpha_d$ of Thakkar \etal\ (Ref.~\onlinecite{thccl92}: circles and
squares), as a function of the H$_2$~molecule size~$r$
(note that actual $\alpha_d$ values have been multiplied by a factor of 30
so as to be visible on the same plot as~$A_d$).
Values extrapolated to $r < 0.8 \; \bohr$ are of little practical
relevance, as H$_2$ is unbound at the corresponding energies.
The linear forms of $A_d(r)$
and $\alpha_d(r)$ (dotted and wide-dashed lines, respectively) that
were used by Muchnick \& Russek (Ref.~\onlinecite{mr94})
are also shown, although their $\alpha_d(r)$ is not strictly comparable,
since they used it in a term with $\eta^2$ rather than~$P_2(\eta)$.
}
\label{fig:thakkar}
\end{figure}

The 7 parameters quoted in the terms $A_d(r)$ and $\alpha_d(r)$ above
were obtained by fitting (with an accuracy of a few percent) to the
accurate values calculated by Thakkar \etal\cite{thccl92}
as a function of~$r$ for these asymptotic terms,
as shown in Figure~\ref{fig:thakkar}.
The remaining parameters to be fitted include $\fad$, $f_{\alpha d}$, $C_1$,
$C_2$, $C_3$, and all of the $C_d(n)$, $\varepsilon(n,L)$, and $a_d(n,L,m)$.
Since the parameters $\fad$ and~$f_{\alpha d}$ multiply a version of
the asymptotic long-range form (although ``softened'' at small~$R$
by the parameter~$C_1$), these
parameters might be expected to remain within a few percent of unity in
a fit; this did indeed prove to be the case, particularly for~$\fad$
(note that $\alpha_d$ is much smaller than~$A_d$, and thus $f_{\alpha d}$
is determined with less precision by the fitting process).

Our final adopted surface had $\{ N_d, L_d, M_d \} = \{ 8, 2, 1 \}$,
but we tested fits with ranges as high as $\{ 10, 4, 2 \}$.
Note that Muchnick \& Russek\cite{mr94} used this $V_d$~term with
only the parameters $C_1$, $C_2$, and~$C_3$; they also replaced
$P_2(\eta)$ with~$\eta^2$, and set $A_d(r) = 0.81 + 1.92 \, r$
and $\alpha_d(r) = 0.112 + 0.06 \, r$
(although they noted that these latter two functions really should have
used a decreasing exponential in~$r$).  Their versions of $A_d(r)$
and~$\alpha_d(r)$ are also illustrated in Figure~\ref{fig:thakkar},
although their $\alpha_d(r)$ is not strictly comparable, since it
multiplies $\eta^2$ rather than~$P_2(\eta)$.  (In spite of the fact
that their $A_d$ and~$\alpha_d$ increase without limit with
increasing~$r$, the $V_d$ term used by Muchnick \& Russek\cite{mr94}
remains bounded, since by definition ${\bar R} \ge r / 2$, and thus their
form of $V_d$ is proportional to~$r^{-4}$ for $r \gg C_1$.)

The $V_D$~term is a long-range interaction between the He~atom
and separated H~atoms:
\begin{equation}
  V_D(\vec{\cal R}) = S_D(r) \Big[ F_D(R_A) + F_D(R_B) \Big] \; ,
      \label{eq:VDheh}
\end{equation}
where
\begin{equation}
  F_D(R_\nu) = \sum_{ { {n=6} \atop {n\;\rm even} } }^{N_D}
    { a_D(n) \over { {R_\nu}^n + \left[ C_D(n) \right]^n } }
  \qquad \hbox{for} \quad \nu = A, B
      \label{eq:FDheh}
\end{equation}
and $S_D(r)$ is a switch function to turn $V_D$ on with increasing~$r$:
\begin{equation}
  S_D(r) = 1 - { B_D(r) \over
    { \left\{ 6^2 + \left[ B_D(r) \right]^2 \right\}^{1/2} } } \; ,
      \label{eq:SDheh}
\end{equation}
with
\begin{equation}
  B_D(r) = 4.790521 \left( r^2 + 6 \right)
    e^{ -0.5891 \, {\displaystyle r} }
      \label{eq:BDheh}
\end{equation}
(where $r$ is in~$\bohr$).
The softening parameter of~6 in the switch function~$S_D(r)$ above
(which regulates the switchover at relatively small~$r$)
may be considered to be a surface parameter, but the other constants
in the switch function were taken from $A_d(r)$ in Equation~(\ref{eq:Ad}),
so that the $V_D$ term would be turned on with increasing~$r$ as the
$V_d$ term was being turned off by~$A_d(r)$.
Other parameters to be fitted include all of the $C_D(n)$ and $a_D(n)$.
Our final adopted surface had $N_D = 8$,
but we also tested fits with $N_D = 10$.
Note that Muchnick \& Russek\cite{mr94} did {\it not\/} include
any such $V_D$~term.

\subsubsection{Other types of terms tested}
\label{sssec:otherterms}

Various index ranges were tried for the summations in the above equations.
In addition, it was tested whether a better fit could be obtained if the
terms $V_C$, $V_H$, and~$V_{cb}$ were multiplied by an exponential cutoff
in the overall size of the \hehh\ conformation, namely, by a factor
\begin{equation}
  e^{ -\beta \rho^\alpha } \; , \qquad \hbox{where} \quad
  \rho = \left( r^2 + {R_A}^2 + {R_B}^2 \right)^{1/2}
      \label{eq:cutoff}
\end{equation}
(with $\beta$ and~$\alpha$ being parameters that could be fitted).  It turned
out that such a cutoff did not improve the fit, and it was discarded.

It was tested whether adding a many body expansion term~$V_M$
to the \hehh\ PES would improve the fit:
\begin{equation}
  V_M(\vec{\cal R}) = \sum_{k=1}^{K_M} \; \sum_{j=0}^{ \min\{k,M_M-1-k\} }
    \left[ { p_A }^k { p_B }^j + { p_A }^j { p_B }^k \right]
    \sum_{i=1}^{ \min\{I_M,M_M-j-k\} } a_M(i,j,k) \, {p_r}^i \; ,
      \label{eq:Vmbe}
\end{equation}
where 
\begin{equation}
  p_r = r e^{ -\beta_M {\displaystyle r} } \; , \qquad \hbox{and} \qquad
  p_\nu = R_\nu e^{ -\beta_M R_\nu } \qquad \hbox{for} \quad \nu = A, B \; ,
      \label{eq:Pmbe}
\end{equation}
with parameters to be fitted including $\beta_M$ and all of the $a_M(i,j,k)$.
This~$V_M$ term could also be multiplied by an exponential cutoff as given
by Equation~\ref{eq:cutoff}.  Various orders~$M_M$ and individual index
limits~$K_M$ and~$I_M$ were tested, but $V_M$ always turned out to be of
very little help in improving the fit (compared to the other terms of
\S~\ref{sssec:fittedeqns}), so it was discarded.

\subsection{Our Approach to Fitting}
\label{ssec:overview}

This section outlines in general the steps followed in developing the
terms and optimizing the parameters in our analytical representation
of the \hehh\ surface.  The details of the equations are given in
\S~\ref{sssec:fittedeqns} above.

\subsubsection{Weights applied to fitted energies}
\label{sssec:fitweights}

In this section, the weights referred to are those that were used to
multiply the deviations between the fit and the data,
before squaring these weighted deviations to calculate the rms of the fit.
Note that, frequently, the ``weight'' in a fit is defined as that
applied to the square of the (unweighted) deviation, i.e., with this latter
definition the weight would be the {\it square\/} of the values reported
below.

Several different criteria were used to determine weight values.  If more
than one applied, the final combined weight used was the {\it product\/}
of the individual weight factors from the following separate criteria.

High-energy portions of the surface are less likely to be accessed
in collisions of a hydrogen molecule with a helium atom; also, higher
\abinitio\ energies have larger uncertainties than lower energies.  Thus
{\it all\/} points with high energy~$E$ were given reduced weight, namely,
\begin{equation}
  w_E(E) = \left\{
  \begin{array}{lll}
  1 & \quad , & E \le 0.2 \; \har \\
  ( 0.2 \; \har ) / E & \quad , & E > 0.2 \; \har
  \end{array} \right.
      \label{eq:wtE}
\end{equation}
i.e., a weight inversely proportional to the energy~$E$ for
cases with energies more than about~2.2 times the H$_2$ dissociation
energy above that of an equilibrium H$_2$ molecule plus a separated
He~atom.

For non-compact conformations where the He~atom was relatively far from
either H~atom, the (absolute) size of the uncertainties in
our \abinitio\ energies was expected to be smaller
than for more compact conformations (especially for the
error in the basis correction:
see \S~\ref{sssec:correctmrdci}).  Thus our \abinitio\ energies
(described in \S~\ref{sssec:maingrid} and \S~\ref{sssec:randomgrid})
were given an additional weight factor
\begin{equation}
  w_{nc}(R_m) = \left\{
  \begin{array}{lll}
    1 & \quad , & R_m \le 3 \; \bohr \\
    ( R_m - 2 ) & \quad , & 3 \; \bohr < R_m < 4 \; \bohr \\
    2 & \quad , & R_m \ge 4 \; \bohr
  \end{array}
  \right. \quad , \qquad \hbox{where} \quad R_m = \min\{R_A,R_B\}
      \label{eq:wtNC}
\end{equation}
(i.e., a weight increasing linearly from 1 to~2 as the shortest
$\rm He - H$ distance~$R_m$ is increased from 3 to~$4 \; \bohr$).

The van der Waals \hephh\ points described in \S~\ref{sssec:generateheh2}
are expected to have smaller (absolute) uncertainties as the distance~$R$
between the H$_2$~molecule and the He~atom increases; they should also
be most accurate for near-equilibrium H$_2$ molecule sizes (i.e.,
$r \sim 1.4 \; \bohr$).  For these
``vdW'' points, two weight factors $w_R$ and~$w_r$ were applied:
\begin{equation}
  w_R(R) = \left\{
  \begin{array}{lll}
    0.3 & \quad , & R \le 3 \; \bohr \\
    1 & \quad , & R = 4 \; \bohr \\
    10 & \quad , & R = 5 \; \bohr \\
    300 & \quad , & R = 6 \; \bohr \\
    1200 & \quad , & R = 6.5 \; \bohr \\
    3000 & \quad , & R \ge 10 \; \bohr
  \end{array}
  \right. \quad , \qquad
  w_r(r) = \left\{
  \begin{array}{lll}
    0.1 & \quad , & r \le 0.6 \; \bohr \\
    0.3 & \quad , & r = 0.9 \; \bohr \\
    1 & \quad , & r = 1.2 \; \bohr \\
    1 & \quad , & r = 1.62 \; \bohr \\
    0.3 & \quad , & r = 2.0 \; \bohr \\
    0.1 & \quad , & r = 2.4 \; \bohr \\
    0.01 & \quad , & r \ge 4.0 \; \bohr
  \end{array}
  \right.
      \label{eq:wtVDW}
\end{equation}
(where $w_R$ and~$w_r$ were interpolated linearly in between the distance
values specified above).
The ``Tao vdW'' points (\abinitio\ energies of Tao\cite{Tao94}) were given
an additional weight factor $w_T = 3$, and the ``gen-Tao vdW'' points
were given an additional weight factor $w_{gT} = 2$.
Note that quite a large number of fits were performed where $w_R$ was
either increased or decreased relative to Equation~(\ref{eq:wtVDW})
by about a factor of~3 for $R \gtrsim 5 \; \bohr$, but these
fits were discarded --- the weight $w_R$ of Equation~(\ref{eq:wtVDW})
appeared to work the best, neither over-emphasizing nor under-emphasizing
the van der Waals well energies relative to the \abinitio\ energies.

The ``H-He'' points described in \S~\ref{sssec:generateheh}
should be as accurate as any of our other
\abinitio\ energies for $R_A \lesssim 5 \; \bohr$ (where $R_A$ is the
{\it shorter\/} $\rm H - He$ distance for these points: $R_A \le R_B$);
they should be rather more accurate in the region of the $\rm H - He$
van der Waals well and further out.  Thus, a weight factor
$w_{HHe}(R_A) = \max\{ w_R(R_A) , 1.0 \}$ was applied to them, where
$w_R$ has the form given in Equation~(\ref{eq:wtVDW}).

The ``H-$\rm He + H$'' points described in \S~\ref{sssec:generatehehh}
are expected to be somewhat less accurate; a weight factor
$w_{HHe{+}H}(R_A) = \max\{ 0.01 \, w_R(R_A) , w_{nc}(R_A) \}$ was applied
to them, where $w_{nc}$ has the form given in Equation~(\ref{eq:wtNC}).

\subsubsection{Modified Muchnick \& Russek surface}
\label{sssec:modMuRu}

The first step was to get a very slightly modified version of the Muchnick \&
Russek\cite{mr94} surface, which we refer to as the ``modMR'' surface.
This differed from the surface of Muchnick \& Russek\cite{mr94} in that
we used the lowest-order version of our functional form for~$V_d$,
including use of our $A_d(r)$ and $\alpha_d(r)$ from Equations \ref{eq:Ad}
and~\ref{eq:ALPHAd}, and the fact that $\alpha_d(r)$ multiplied~$P_2(\eta)$
rather than~$\eta^2$.  We also added the lowest
order~$V_D$ term (i.e., $N_D = 6$) from Equation~(\ref{eq:VDheh}).  Since
we used our already-fitted versions (8~parameters)
of $A_d(r)$, $\alpha_d(r)$ and~$S_D(r)$ from Equations (\ref{eq:Ad}),
(\ref{eq:ALPHAd}), and~(\ref{eq:SDheh}), this initial ``modMR'' surface had
19~other basic parameters that could be fitted.  There were 9~short-range
non-linear parameters $\{ \lambda_C, \beta_C, a_C(1,0,0), \lambda_H,
 \beta_H, \beta_{cb}, \varepsilon_{cb}, \zeta_{cb}, \delta_{cb} \}$,
3~short-range linear parameters $\{ A_H, a_{cb}(0,0,0), a_{cb}(0,2,0) \}$,
3~long-range non-linear parameters $\{ C_1, C_2, C_D(6) \}$, and 4~long-range
linear parameters $\{ \fad, f_{\alpha d}, C_3, a_D(6) \}$.  For this
first surface, we obtained $a_D(6)$ from Gengenbach, Hahn, \&
Toennies\cite{ght73}, and set $C_D(6) = 6.7876662396 \; \bohr$ (actually,
$2^{1/6} \times 3.2 \; $\AA\ --- a reasonable softening value
to keep the $V_D$~term from contributing significantly at relatively
compact geometries).  We set the other parameters equal to the values used
by Muchnick \& Russek\cite{mr94}, and fitted only the parameters $\fad$
and~$f_{\alpha d}$, using only the ``Tao vdW'' and ``gen-Tao vdW'' points
described in \S~\ref{sssec:generateheh2}.  This procedure yielded values
of $\fad = 0.98234$ and~$f_{\alpha d} = 1.261$, reasonably close to unity
(i.e., at large $\rm H_2 - He$ separations~$R$, this surface still had a
form very close to that calculated by Thakkar \etal\cite{thccl92}).
The surface was almost identical to that of Muchnick \& Russek\cite{mr94},
but was a significantly better fit to the more recent (and more accurate)
van der Waals \abinitio\ energies of Tao\cite{Tao94} (unweighted rms error
of~$1 \; \microh$ for
these points, rather than~$3 \; \microh$).  This ``modMR'' surface was the
one used to generate the ``gen-MR vdW'' points described
in \S~\ref{sssec:generateheh2}.

\subsubsection{Optimization of fitted parameters}
\label{sssec:optpar}

Several hundred fits were subsequently performed to the full set of
$20\,203$ \abinitio\ energies and 4862~generated points, with various
ranges for the indices in the summations in the surface equations
of \S~\ref{sssec:fittedeqns}; the number of parameters actually
fitted varied from a couple of dozen to about~150.  Fitting was performed
using the NAG Fortran library non-linear fitting routine E04FDF to minimize
the weighted rms error with respect to the $20\,203$ \abinitio\ plus
4862 generated energies, using the weights of \S~\ref{sssec:fitweights}
(the covariance matrix was then produced by the NAG routine E04YCF) ---
a typical fit required a few hours on a higher-end
computer.  The first fit described in
the previous paragraph supplied initial values for the basic parameters
in the next fits, often with values shifted randomly; the added parameters
in the summations were either initialized to zero or to random values.
A reasonably good fit was frequently used as the basis for subsequent
larger fits (sometimes with random shifts applied to the parameters before
fitting).  There were some exceptions to this: the 3~basic long-range
non-linear parameters $\{ C_1, C_2, C_D(6) \}$ were not re-fitted, and other
long-range non-linear parameters were generally not fitted, but just given a
reasonable initial value.  Also, 3~of the basic long-range linear
parameters, namely, $\{ \fad, f_{\alpha d}, C_3 \}$, were not re-fitted
until near the end of the fitting process.  Note that in our final
adopted fit the values of $\fad \approx 0.9552$
and~$f_{\alpha d} \approx 1.149$ were near unity; since these parameters
multiply forms $A_d(r)$ and $\alpha_d(r)$ that had been fitted with an
accuracy of a few percent to the values of Thakkar \etal\cite{thccl92}
(as discussed in \S~\ref{sssec:fittedeqns}), this final adopted fit
yielded an asymptotic interaction between the H$_2$ molecule and the
He~atom very close to that calculated by Thakkar \etal\cite{thccl92}

\subsubsection{Selection of a good fitted PES}
\label{sssec:selectsurf}

For a quick comparison of the quality of all the different fits, we
considered the full weighted rms minimized by the fitting program
with respect to all $24\,804$ fitted points, and also the energy-weighted
rms for 9~subsets of these points, where only the weight~$w_E$ from
Equation~(\ref{eq:wtE}) had been applied.  These 9~subsets comprised:
\begin{enumerate}
\item all $20\,203$ \abinitio\ energies,
\item the $14\,585$ \abinitio\ energies with $E < 0.174 \; \har$ (i.e.,
below about twice the H$_2$ dissociation energy),
\item the 7177 \abinitio\ energies below
the H$_2$ dissociation energy ($E < 0$),
\item 931 \abinitio\ energies with near-equilibrium H$_2$
($1.2 \; \bohr \le r \le 1.62 \; \bohr$) and a not-too-distant He
($2 \; \bohr < R \le 4 \; \bohr$),
\item 468 fairly-close ``H-He'' points (with
$2 \; \bohr < R_A \le 4 \; \bohr$),
\item 143 van der Waals ``H-He'' points (with
$6 \; \bohr \le R_A < 10 \; \bohr$),
\item 231 ``Tao vdw'' and ``gen-Tao vdw'' $\rm H_2 + He$ van der Waals
points with near-equilibrium H$_2$ (i.e.,
$1.28 \; \bohr \le r \le 1.62 \; \bohr$,
and $6 \; \bohr \le R < 10 \; \bohr$),
\item 434 ``gen-MR vdW'' and ``gen-SK vdW'' van der Waals points
with small-H$_2$
($r < 1.2 \; \bohr$ and $6 \; \bohr \le R < 10 \; \bohr$),
\item 714 ``gen-MR vdW,'' ``gen-SK vdW,'' and ``H-$\rm He + H$''
van der Waals points with large-H$_2$
($r > 1.62 \; \bohr$ and $6 \; \bohr \le R < 10 \; \bohr$).
\end{enumerate}
For a few cases where the above rms values suggested that the surface was
among the best of the fits, a collection of weighted, energy-weighted,
and unweighted rms values was considered for a more comprehensive list of
about 100 subsets.

The fitting program produced estimated uncertainties of the fitted
parameters (from the NAG covariance routine E04YCF).  When an enlarged
fit did not significantly reduce the rms error and yielded new parameters
whose values were not significant (i.e., comparable to or smaller than
the estimated errors in these parameters), this suggested that the point of
diminishing returns had been reached --- larger fits would presumably
yield spurious ``wiggles'' rather than an improved fit.  This criterion was
used to obtain a near-optimum set of index ranges for the summations in the
surface equations of \S~\ref{sssec:fittedeqns} (although a number
of larger fits were performed, with different initial
parameter values, to make certain that no significant improvement
was indeed possible).  Finally, for a few of the ``best'' surfaces
with near-optimum index ranges (and also a few very large fits), scatterplots
were made of the error vs.\ the \abinitio\ energy, and a couple of dozen
contour plots of each surface were examined, as well as over a hundred
plots of energy vs.\ one of the \hehh\ distances or angles (in these latter
plots, the surfaces could be directly compared to the \abinitio\ energies
along some cut of the PES).  Our adopted \hehh\ PES was chosen in this
manner from among the ``best'' surfaces with near-optimum index ranges.
A Fortran program to compute this analytic PES
(including analytic first derivatives)
is available from EPAPS\cite{epaps} or from the authors\cite{bmp}.

We have used a relatively large number of linear parameters in order to
fit our \hehh\ data.  Our final BMP surface uses 112~parameters, with a
few of the non-linear parameters being given reasonable constant
values as discussed in \S~\ref{sssec:fittedeqns}, and a few others being
set by early (small) fits and held constant thereafter (such as the
parameters in $A_d$ and~$\alpha_d$); in the
final fitted surface, the 9~non-linear
parameters that were refitted at that point were all significant (at
better than the 10-$\sigma$ level), and only 2~of the
87~linear parameters had values smaller than their estimated errors, with
only 3~others having less than a 3-$\sigma$ significance level.
These 112~parameters of our BMP surface were used to fit a total
of $25\,065$ points ($20\,203$ of which are \abinitio\ energies), a 224:1
ratio of points to parameters (180:1 if one considers \abinitio\ points only).
For comparison, Truhlar and Horowitz\cite{th78} fitted 287 \abinitio\ H$_3$
points with about 23 parameters, a 12:1 ratio.  Our BKMP2 H$_3$
surface\cite{bkmp2} fitted 8559 H$_3$ points
(7591 \abinitio\ points) with about 120 parameters, a 71:1 ratio
(63:1 for \abinitio).  Aguado \etal\cite{ag94} fitted 6101
\abinitio\ H$_4$ energies with 865 linear parameters, a 7:1 ratio.
Our BMKP H$_4$ surface\cite{bmkp02} uses 400 parameters to fit 61547
points (48180 \abinitio\ points), giving a ratio of 154:1
(120:1 for \abinitio).  Thus the final fitted BMP \hehh\ surface of
the present work actually uses relatively few fitting parameters for
the number of points fitted, minimizing the risk of ``overfitting,''
i.e., of spurious wiggles {\it between\/} fitted points.

Added confidence in our error estimates is provided by the recent
extremely accurate H$_3$ surface of Wu \etal\cite{wu99}  Their
\abinitio\ energies and fitted surface are an order of magnitude more
accurate than our earlier BKMP2 H$_3$ surface\cite{bkmp2}; they found that
our H$_3$ error estimates were essentially correct, both for our
H$_3$ \abinitio\ energies and our fitted H$_3$ surface.  They report
no spurious wiggles in our BKMP2 H$_3$ surface with sizes larger than
its quoted accuracy.  This suggests that our error estimates should
be reasonable for the similarly-computed \hehh\ \abinitio\ energies
of the present work, and that our fitted BMP \hehh\ surface should
likewise be free of large spurious wiggles between fitted points.
Finally, random ``interstitial'' unfitted conformations were also tested
directly, as described in \S~\ref{sssec:interstitial}.

\section{Discussion}
\label{sec:discussion}

\subsection{Accuracy of analytic He-H$_2$ surfaces}
\label{ssec:surfacc}

\begin{figure}   
%
%
\plotfiddle{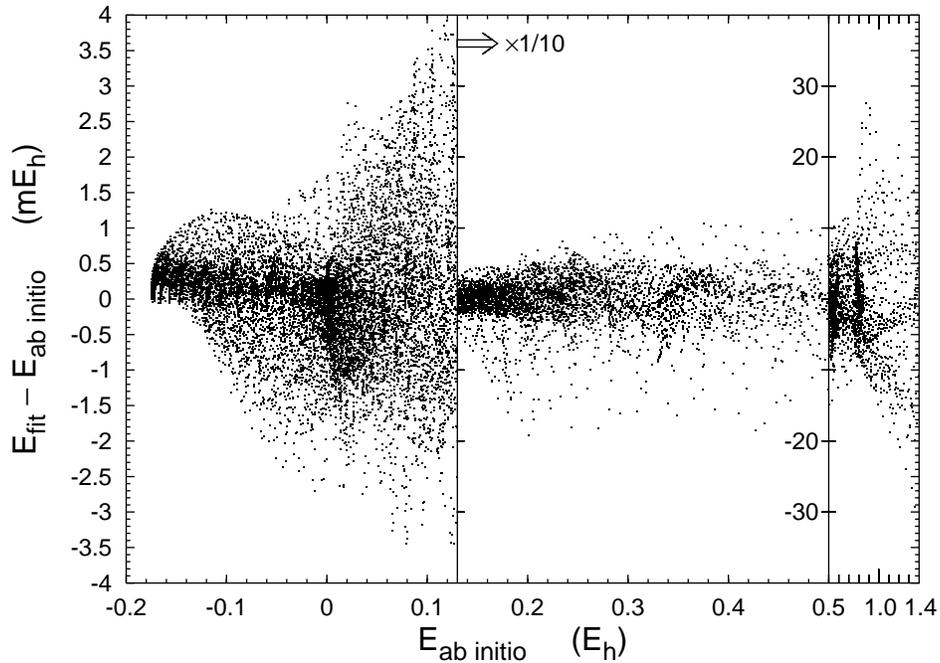}{3.6 true in}{-90}{50}{50}{-200}{295}
%
%
\caption[fig:scatterplot]{Scatterplot of the deviations of the adopted ``BMP''
\hehh\ PES relative to the $20\,203$ \abinitio\ energies, as a function of
energy~$E$.  Note that for $E > 0.13 \; \har$ the vertical scale is compressed
by a factor of~10; for $E > 0.5 \; \har$, the horizontal scale is compressed
likewise.  (There are $13\,397$, 4592, and 1991 points in these three parts
of the plot, respectively; 223~points lie offscale to the right.)
Denser bands of points (usually nearly vertical) result from the discrete
$r$~values of the main \abinitio\ grid.
Note that most of the extreme outliers for $E \gtrsim 0.2 \; \har$ arise
from the conical intersection of the ground
state with the first excited state.
}
\label{fig:scatterplot}
\end{figure}

Our adopted surface (the ``BMP'' \hehh\ PES) contains a total of
112~parameters; even much larger surfaces (up to 175 parameters)
yielded no significant improvement (having rms values only a few
percent smaller).  Figure~\ref{fig:scatterplot} shows a scatterplot
of the deviations relative to our \abinitio\ energies.  The surface
fits best at low energy ($E \lesssim 0.15 \; \har$, i.e., energies
below about twice the H$_2$-molecule dissociation energy, relative
to the energy of an equilibrium H$_2$~molecule plus a distant He~atom).
However,
even up to quite high energies, the surface lies within a few~$\millih$
of most of the \abinitio\ points; it turns out that most of the outliers
are due to the fact that the surface can only produce an approximate
fit to the conical intersection with the first excited state (this is
discussed in more detail below, in \S~\ref{sssec:cusps}).

Table~\ref{tab:rms} compares the rms errors of four earlier analytic
\hehh\ surfaces with rms errors of the adopted (BMP) surface of this
work, for various subsets of \abinitio\ and generated energies; the
modified Muchnick \& Russek (modMR) surface of \S~\ref{sssec:modMuRu}
(which was used to generate some of the constraining van der Waals points
for non-equilibrium H$_2$) is also shown.
``Energy-weighted'' deviations were used to obtain the rms values
of Table~\ref{tab:rms},
i.e., energies above about two H$_2$-dissociation energies were given
reduced weight~$w_E$ according to the formula of Equation~(\ref{eq:wtE}).
The only exception to this is the last line of Table~\ref{tab:rms}
(subset~44), where the {\it fully weighted\/} rms values are reported
for {\it all points\/} used in our \hehh\ fit (for our BMP surface,
this is the rms value that the fitting program
tried to minimize, as described in \S~\ref{sssec:fitweights} and
\S~\ref{sssec:optpar}).

It should be noted that the earlier analytic \hehh\ surfaces were fitted
to \abinitio\ points in relatively restricted regions, compared to the
region covered by the \abinitio\ and generated points of the present work.
The rms values given in {\it italics\/} in Table~\ref{tab:rms} indicate
subsets of points that cover regions where the corresponding surface
was not fitted, i.e., where energies predicted by these previous surfaces
are being tested by our new \abinitio\ energies.  (The only unconstrained
region of our fitted BMP surface is at very small interatomic
separations, i.e., very high energies; this is discussed
in \S~\ref{sssec:extrap}.)

The Wilson, Kapral, \& Burns\cite{wkb74} (WKB) surface was fitted to
a set of
\abinitio\ energies in the region $0.75\;\bohr \le r \le 5.0\;\bohr$,
$0.0\;\bohr \le R \le 5.0\;\bohr$, plus the three pair-wise potentials.
The Dove \& Raynor\cite{dr78} (DR) surface was essentially the same,
with improved versions of the pair-wise potentials.  Since the interaction
part of these surfaces was fitted using only four free parameters, they
lack sufficient flexibility for accuracy even in their fitting region.

The Schaefer \& K\"ohler\cite{sk85} (SK) surface was fitted to a set
of \abinitio\ energies in the region $1.28\;\bohr \le r \le 1.618\;\bohr$,
$3.0\;\bohr \le R \le 15.0\;\bohr$ plus a somewhat less accurate set in
the region $0.9\;\bohr \le r \le 2.0\;\bohr$,
$1.5\;\bohr \le R \le 8.0\;\bohr$; it is quite accurate in these regions.
However, it is designed to be interpolated on
a grid in the region $0.9\;\bohr \le r \le 2.0\;\bohr$, $R \ge 1.6\;\bohr$,
and therefore can be {\it highly\/} inaccurate when extrapolated outside
this region (just how inaccurate depends on the extrapolation formula used
--- we just used an spline in~$r$, even for extrapolation, but fitted
exponentials to extrapolate the $R$~spline to small~$R$).

The Muchnick \& Russek\cite{mr94} (MR) surface was fitted to a set of
energies in the region $1.2\;\bohr \le r \le 1.6\;\bohr$,
$0.0\;\bohr \le R \le 15.0\;\bohr$;
the very similar modMR surface of \S~\ref{sssec:modMuRu} differs from the MR
surface only in the parameters describing the $\rm H_2 - He$
and $\rm H - He$ van der Waals wells.  These surfaces are accurate in the
part of their fitting regions that lies at $R \gtrsim 3 \; \bohr$.
However, although the MR
surface was fitted to \abinitio\ energies down to $R = 0$, it is a relatively
poor fit at small~$R$ (subset~9 in Table~\ref{tab:rms}).  This is due both to
the fact that Muchnick \& Russek\cite{mr94} had only a few \abinitio\ points
in this region and to the fact that a good fit in this region requires a
good deal of flexibility in the fitted surface (we found in our first,
smallest fits that significant improvements at small~$R$ could be obtained
only at the expense of worsening the fit in the van der Waals well, unless
more parameters were added).  Even at intermediate~$R$
($2 \; \bohr < R \le 3 \; \bohr$: subsets 6, 10, and~14 in
Table~\ref{tab:rms}), the SK surface does somewhat better than the MR
surface, provided that $r$ is not too large, i.e., $r \le 2 \; \bohr$
(our BMP surface is of course significantly better than either, in these
regions of small-to-intermediate~$R$).

Table~\ref{tab:rms} demonstrates that, overall, the MR surface\cite{mr94} is
the best of the previous surfaces considered here.  Nonetheless,
even the MR surface does quite
poorly overall, with an rms error of~$28.6 \; \millih$ relative to our
\abinitio\ energies (subset~1 in Table~\ref{tab:rms}).  The
present BMP surface is an order of magnitude improvement, with an
rms error of~$1.42 \; \millih$ relative to the \abinitio\ energies --- of
the same order as the estimated \abinitio\ error of~$\sim 0.6 \; \millih$.
The advantage of the BMP surface is largest in the region that can be
sampled by dissociative collisions, but even below the dissociation energy
(i.e., $E < 0.0 \; \har$: subset~3 in Table~\ref{tab:rms}) the
present BMP surface (rms of~$0.48 \; \millih$) remains a significant
improvement over the MR surface (rms of~$4.48 \; \millih$).

For the restricted region of not-too-large H$_2$-molecules
($r \le 2 \; \bohr$) plus a not-too-close He~atom ($R \gtrsim 3 \; \bohr$),
subsets 7, 11, 15, and~31 in Table~\ref{tab:rms} show that
both the SK and MR surfaces\cite{sk85,mr94} do nearly as well as the present
BMP surface, but the earlier WKB and DR surfaces\cite{wkb74,dr78} are an order
of magnitude worse even in this region (and have no van der Waals well
at all).
As one might expect, both the SK and MR surfaces do better
when extrapolated to small~$r$ (subsets 4 to~7 in Table~\ref{tab:rms}) than
when extrapolated to large~$r$ (subsets 16 to~19), due presumably
to the fact that reducing~$r$ tends to reduce the anisotropy with
respect to~$\gamma$ at reasonable $R$
values, while for large~$r$ the anisotropy can become very large
(in fact, for $R \sim r / 2$,
the He~atom approaches one of the H~atoms for $\gamma \rightarrow 0$).

We performed a number of further tests to assess in more detail the quality
of our BMP surface (and of the previous surfaces),
as discussed in the following subsections.

\subsubsection{Test for ``interstitial'' wiggles in the BMP surface}
\label{sssec:interstitial}

The rms errors of the 3500
unfitted ``random'' conformations of \S~\ref{sssec:randomtst}
(subsets 23 to~25 in Table~\ref{tab:rms})
were compared to the rms errors of the 3500
fitted ``random'' conformations of \S~\ref{sssec:randomgrid}
(subsets 20 to~22), which had
been generated with the same probability distribution.  The similarity of
these two sets of rms values is evidence that our BMP surface does not
have large spurious wiggles between fitted \abinitio\ points.  The fact
that the rms values for these sets of ``random'' conformations are also
similar to the overall \abinitio\ rms values (subsets 1 to~3, dominated by
the main grid of \S~\ref{sssec:maingrid}) yields added reassurance.

These ``random'' points of these subsets (20 to~25 in Table~\ref{tab:rms})
avoid duplicate conformations
by a relatively small ``avoidance margin'' of~$\sim 0.01\;\bohr$
(see \S~\ref{sssec:randomgrid}), in order to keep the fraction of
``avoided'' conformation space small even in regions where the main
grid of conformations was dense.  We also considered rms errors of
our BMP surface for
``maximally interstitial'' cases with larger avoidance margins: e.g.,
for a case where points from fitted subset~20 were
discarded if they lay within a 3\% avoidance margin,
the remaining 2949 points had an
(energy-weighted) rms error of~$1.356\;\millih$ ($0.528\;\millih$ for the
1140 points with $E < 0$), while the corresponding 2742
points from unfitted subset~23 had an rms error
of~$1.374\;\millih$ ($0.536\;\millih$ for the 1063 points with $E < 0$).
(Recall that
the typical main-grid spacing was $\sim 7$\% in $r$ and~$R_A$,
$\sim 7$\% in $\theta$ for $\theta > 120^\circ$, and $\sim 13$\% in
$\theta$ for $\theta < 120^\circ$: see \S~\ref{sssec:maingrid}.)
As expected, there was good agreement between rms values of fitted and
unfitted ``random'' subsets independent of the size of the avoidance margin.

It would of course have been very surprising if any interstitial wiggles
had showed up; as discussed at the end of \S~\ref{sssec:selectsurf},
the density of fitted \abinitio\ points is high enough that there is
no room for interstitial wiggles (unless we had used a great many more
parameters in the fit).  However, there are nonetheless certain regions
where the BMP surface {\it does\/} have errors significantly larger
than its typical ones, as discussed further below.

\subsubsection{Contour plots of the BMP surface}
\label{sssec:bmpcontour}

\begin{figure}   
%
%
\plotfidtwo{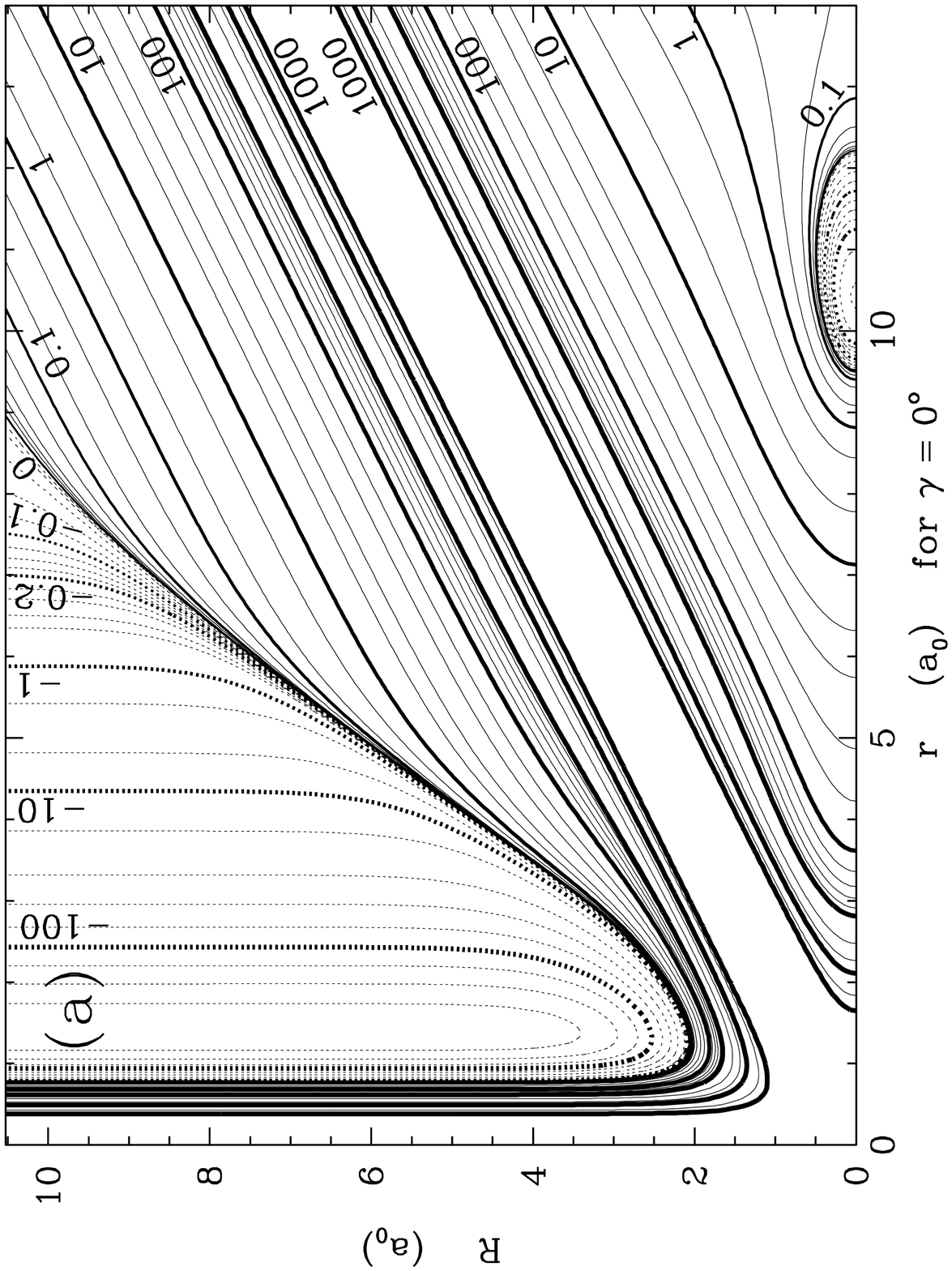}{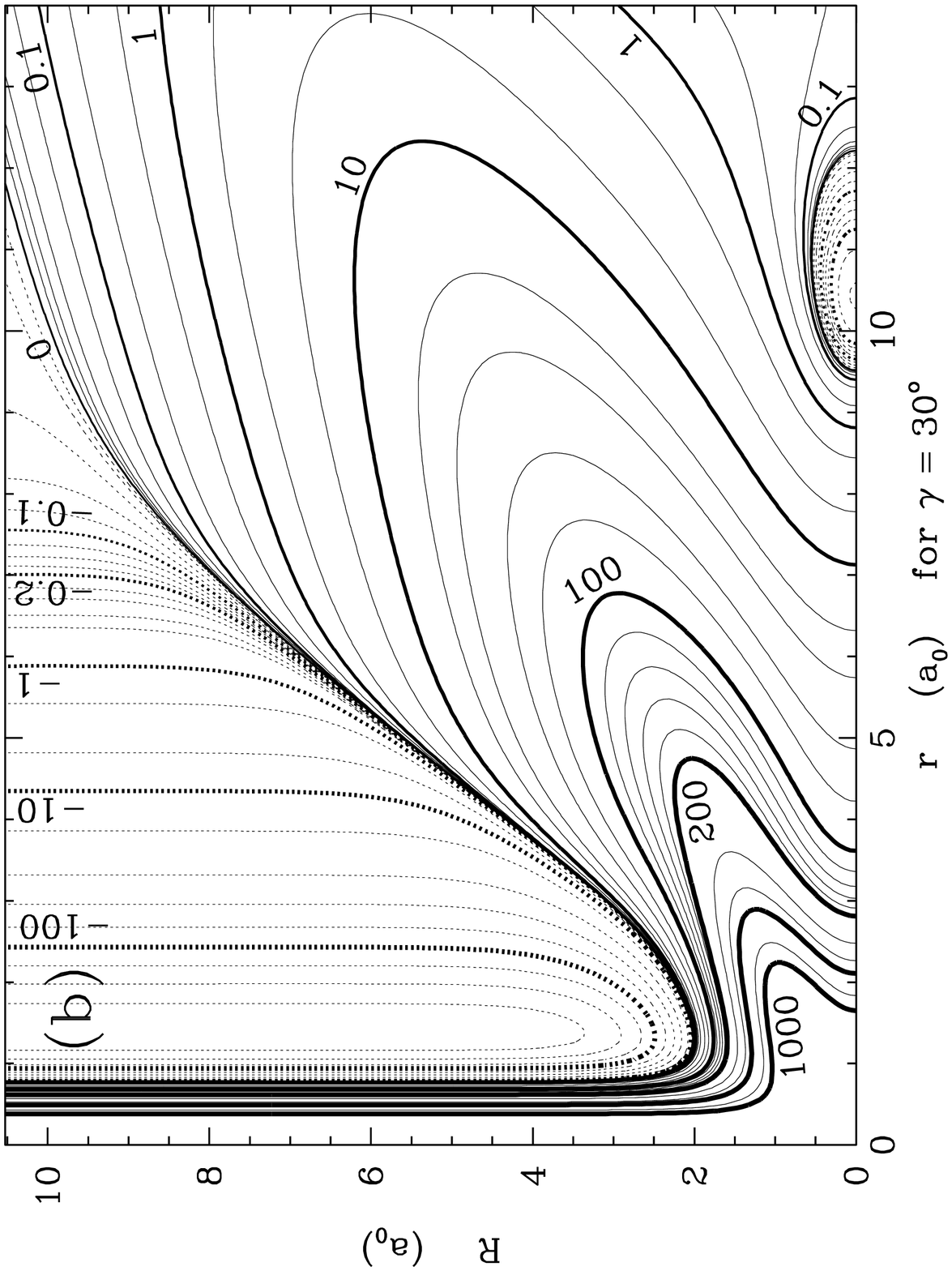}{2.4 true in}{-90}{33}{33}{-250}{195}{-10}
\plotfidtwo{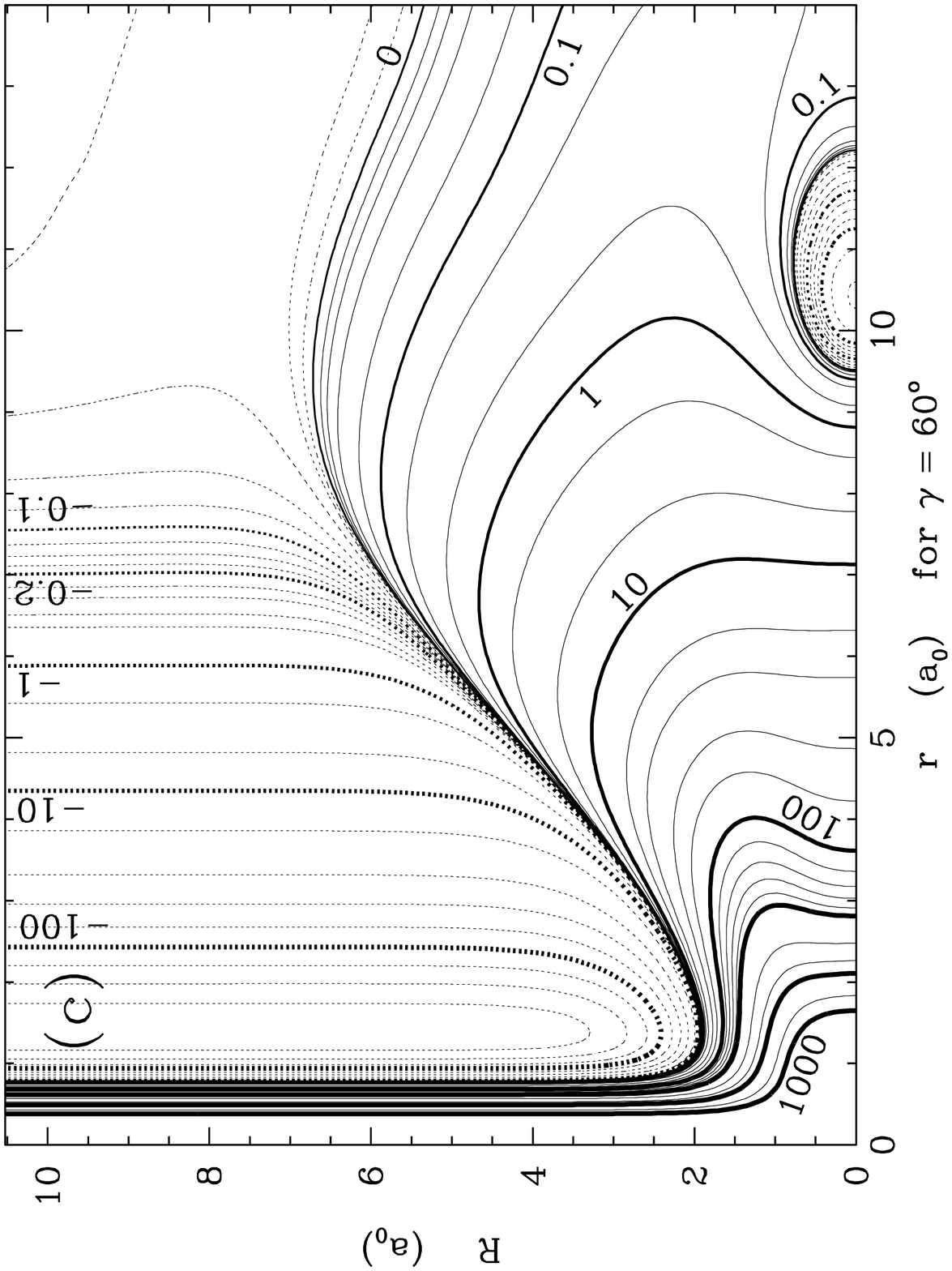}{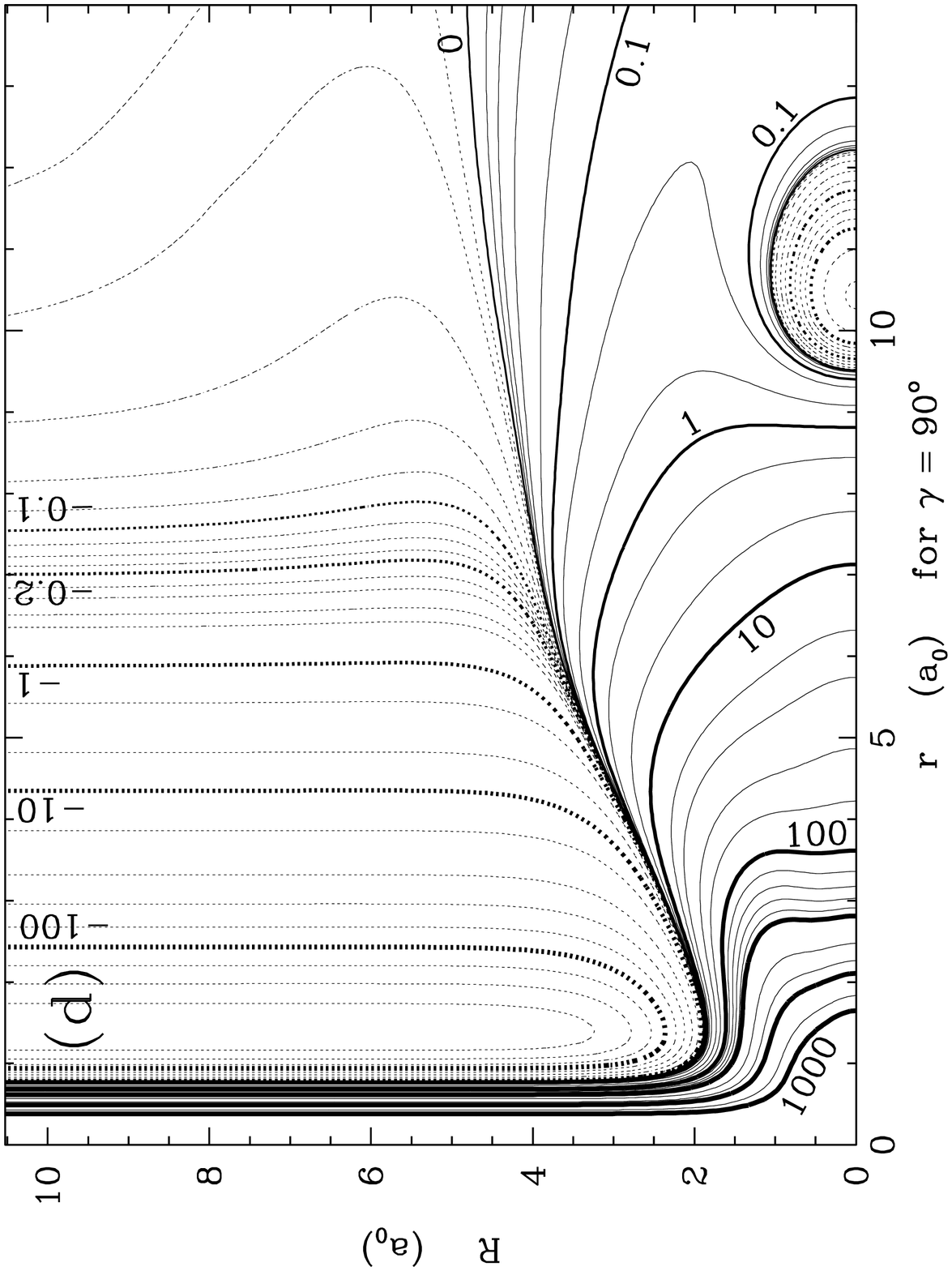}{2.4 true in}{-90}{33}{33}{-250}{195}{-10}
%
%
\caption[fig:contourplot]{Contour plots of the fitted ``BMP'' \hehh\ surface
of this paper as a function of $\rm H - H$ separation~$r$ and $\rm H_2 - He$
separation~$R$, at four angles~$\gamma$ (see Fig.~\ref{fig:notation}).
Positive contour levels (solid
lines) lie at energies $E = \{ {}${\it 0}, 0.005, 0.01, 0.02, 0.05,
{\it 0.1}, 0.2, 0.5, {\it 1}, 2, 5, {\it 10}, 20, 30, 50, 70, {\it 100},
120, 140, 160, 180, {\it 200}, 300, 400, {\it 500}, 750,
and~{\it 1000\/}$ \; \millih \}$ relative to
separated $\rm H + H + He$ (numbers in
{\it italics\/} correspond to heavy contour lines --- most of these are also
labelled in the plots).  Negative contour levels (dotted lines) lie at
energies $E = \{ -0.005$, $-0.01$, $-0.025$, $-0.05$, $-0.075$,
${\it -0.1}$, $-0.125$, $-0.15$, $-0.175$, ${\it -0.2}$, $-0.25$, $-0.3$,
$-0.4$, $-0.5$, ${\it -1}$, $-2$, $-5$, ${\it -10}$, $-20$,
$-40$, $-60$, $-80$, ${\it -100}$, $-120$, $-140$, and~$-160 \; \millih \}$.
(a)~$\gamma = 0^\circ$, (b)~$\gamma = 30^\circ$, (c)~$\gamma = 60^\circ$,
(d)~$\gamma = 90^\circ$.
Note that the small ($E \sim -0.3 \; \millih$) ``basin'' near
$(r = 10.5 \; \bohr, R = 0)$ is spurious: see text.
}
\label{fig:contourplot}
\end{figure}

\begin{figure}   
%
%
\plotfiddle{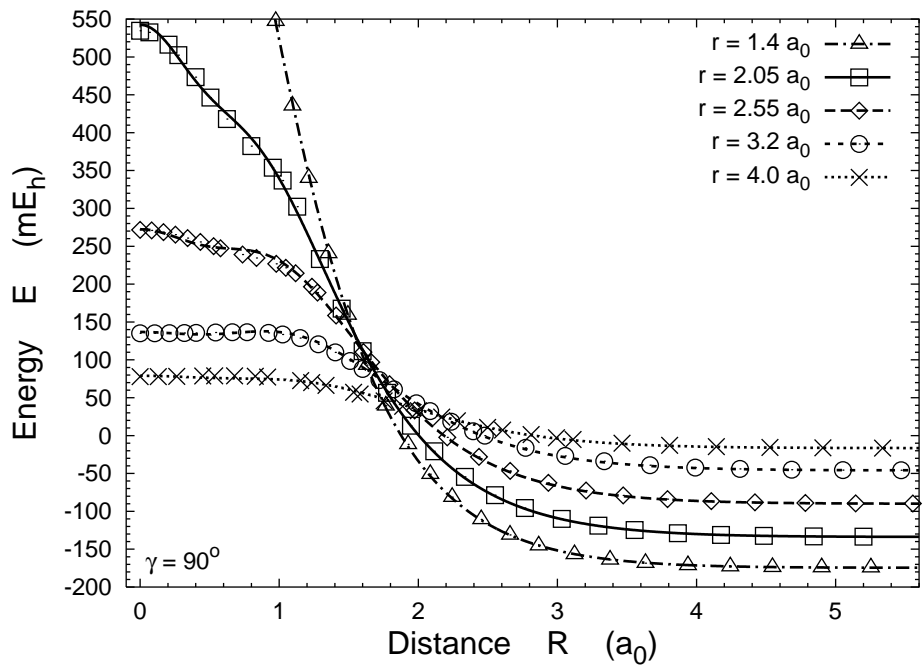}{3.6 true in}{-90}{50}{50}{-200}{295}
%
%
\caption[fig:plateau]{Comparison of the BMP fit (lines)
with \abinitio\ energies (symbols)
as a function of the $\rm H_2 - He$ distance~$R$ at $\gamma = 90^\circ$
(T-shaped orientation), for five $\rm H - H$ distances $r = 1.4$, 2.05,
2.55, 3.2, and~$4.0 \; \bohr$.
}
\label{fig:plateau}
\end{figure}

The contour plots of Figure~\ref{fig:contourplot} show the overall shape
of our fitted BMP surface for four orientations~$\gamma$ of the
$\rm H_2 - He$ separation~$R$ relative to the $\rm H - H$ separation~$r$.
Note that the total potential~$V_{\rm He\,H_2}$ is represented
in Figures~\ref{fig:contourplot} and~\ref{fig:plateau} (not the
interaction energy~$V_{int}$), and thus its gradient gives the total force
on the H and He atoms.

In Figure~\ref{fig:contourplot}a, with $\gamma = 0^\circ$, the He~atom
lies very close to one of the H~atoms for $R \approx 0.5 \, r$, yielding
a double-sided ``wall'' in the contour plot.
In Figure~\ref{fig:contourplot}b, this ``wall'' is replaced
by a ridge, since for $\gamma = 30^\circ$ the He~atom is closest to
the H~atom when $R = 0.5 \, r$.  The ridge-like structure at lower left in
Figure~\ref{fig:contourplot}c ($\gamma = 60^\circ$) is probably
due to the attempt by the surface to fit the nearby conical intersection
with the first excited state,
which is discussed in more detail in \S~\ref{sssec:cusps} below.

From the slopes of the contour lines in Figure~\ref{fig:contourplot}d
(a T-shaped orientation, i.e., $\gamma = 90^\circ$), one can see that,
as $R$ decreases, the He~atom does not begin to ``push apart'' the
H~atoms until it is quite close to the H$_2$~molecule: the partial
derivative $(\partial V / \partial r)_{R,\gamma}$ remains positive
for $r \gtrsim 1.5 \; \bohr$ until $R$ reaches $\sim 1.5 \; \bohr$,
between the 100 and $200 \; \millih$ contours.
(There is a large repulsive force between the He~atom and the
H$_2$~molecule that begins much further out, as may be seen from the
significant reduction in depth of the H$_2$ potential well as $R$ is
reduced below $\sim 4 \; \bohr$).  The cuts at constant~$r$ in
Figure~\ref{fig:plateau} also illustrate this effect --- the
curves for larger~$r$ values lie above those for $r = 1.4 \; \bohr$,
except at small~$R$.
This illustrates an effective barrier to dissociation of H$_2$ by~He:
at least in the T-shaped geometry, the He~atom must have a high
enough energy to sample the part of the surface where
$(\partial V / \partial r) < 0$, i.e., considerably more than just the
the H$_2$ dissociation energy.

For an equilibrium H$_2$ ($r = 1.4 \; \bohr$) and $\gamma = 90^\circ$,
Muchnick \& Russek\cite{mr94} found that, with their surface, the presence
of the He~atom actually caused an inward force on the H$_2$~molecule
for $2.36 \; \bohr \le R \lesssim 6.5 \; \bohr$,
i.e., $(\partial V_{int} / \partial r )_{R,\gamma} > 0$ in this region;
for the BMP surface of the present work, the corresponding region is
$2.03 \; \bohr \le R \le 6.29 \; \bohr$.
As Muchnick \& Russek\cite{mr94} pointed out, this ``force reversal'' is
a real effect, even visible in the \abinitio\ energies.
Nonetheless, this ``force reversal'' is only in the interaction energy, and
is a minor effect compared to the stability provided by
H$_2$-molecule potential; as may be seen
from Figure~\ref{fig:contourplot}, the equilibrium H$_2$ size is almost
independent of~$R$, except at small~$R$.

Figure~\ref{fig:plateau} also illustrates that the nearly-vertical parts of
the contours at the lower left in Figure~\ref{fig:contourplot}d
($\gamma = 90^\circ$) are a real
reflection of the \abinitio\ data (and
not an artifact in the fit arising from fitting the smaller-$\gamma$
region where the He~atom passes close to one of the H~atoms).
For $\gamma = 90^\circ$ and $r \sim 3 \; \bohr$, there really is a
(well-fit)
plateau in the energy as a function of~$R$ for $R \lesssim 1.5 \; \bohr$.
For the $r \le 2.55 \; \bohr$ cases plotted in Figure~\ref{fig:plateau},
this plateau is more tilted and lies at higher energy; the ``wiggles''
of up to $\sim 10 \; \millih$ visible in the fitted BMP surface there are
probably due partly to the reduced weight given to high-energy points,
but are probably also connected to the relatively nearby conical intersection
with the first excited state (see \S~\ref{sssec:cusps} below).

\begin{figure}   
%
%
\plotfiddle{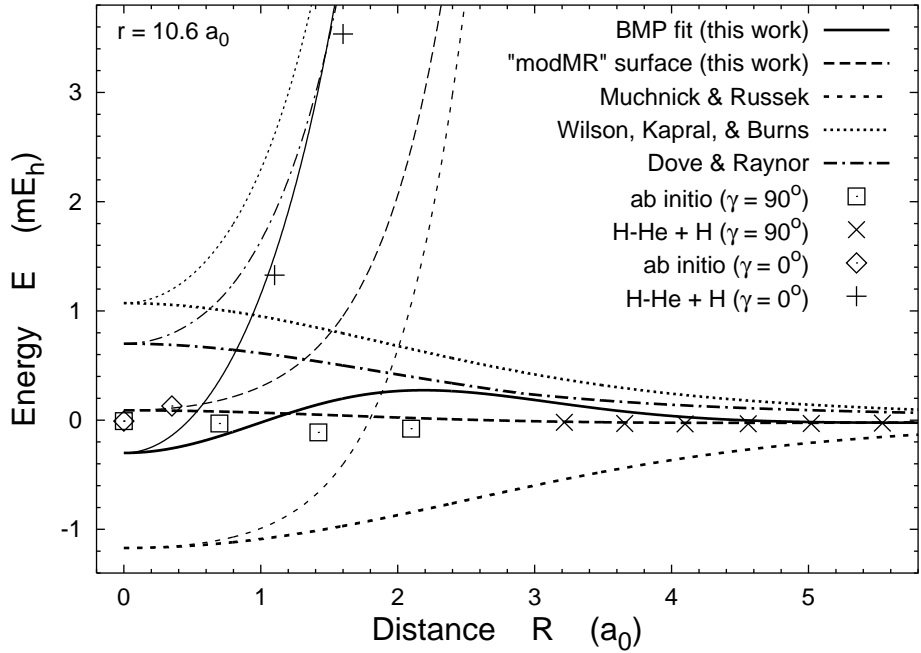}{3.6 true in}{-90}{50}{50}{-200}{295}
%
%
\caption[fig:basin]{Potential for the He~atom ``between'' widely-separated
H~atoms ($r = 10.6 \; \bohr$), for displacements~$R$ of the He~atom
at two angles~$\gamma$ (from the center point).  For $\gamma = 90^\circ$
(T-shaped: heavy lines), as $R$ increases, the He~atom is moving out from
between the H~atoms.  For $\gamma = 0^\circ$ (linear: light lines),
as $R$ increases,
the He~atom approaches one of the H~atoms and encounters the base of
the repulsive ``wall.''  Note that the
``modMR'' surface is identical to the Muchnick \& Russek
(Ref.~\onlinecite{mr94}) surface,
except for the long-range (dispersion) terms. 
}
\label{fig:basin}
\end{figure}

Note that in Figure~\ref{fig:contourplot} the $\sim 0.3 \; \millih$ ``basin''
near $\{ r = 10.5 \; \bohr , \; R = 0 \}$, and also the $\sim 0.3 \; \millih$
``ridge'' in Figures~\ref{fig:contourplot}c and~\ref{fig:contourplot}d near
$\{ r = 12 \; \bohr, \; R = 2 \; \bohr \}$, are probably
spurious features, where our fitted BMP surface overcompensates slightly
due to having to fit the nearby repulsive walls (the spurious features
being comparable in size
to the expected errors in the \abinitio\ energies in this region).
Figure~\ref{fig:basin} illustrates that there is no real evidence for
this ``basin'' and ``ridge'' in the \abinitio\ energies or generated points.
These (presumably) spurious features in our BMP fit are smaller
even than the rms error with which the surface fits the low-energy points,
and furthermore lie in a region
of the surface not likely to be sampled, i.e., by the time the H~atoms are
that far apart in a dissociative collision, the He~atom is not likely
to lie half-way between them.  Consequently, there seemed to be little
point in attempting to smooth out the ``ridge'' and ``basin'' by
increasing the weight on the points in this region (especially since
this might lead to a poorer fit elsewhere).

Figure~\ref{fig:basin} also shows that the previous \hehh\ surfaces
have a larger spurious ``bulge'' or ``basin'' than our BMP surface in this
region.  Through the use of $A_d(r)$ from Equation~(\ref{eq:Ad}), which
has an exponential cutoff at large~$r$, the modMR surface
of \S~\ref{sssec:modMuRu} largely eliminates the spurious ``basin,''
but, like the
MR surface itself, is a poor fit
to the base of the repulsive ``wall'' in the $\gamma = 0^\circ$ direction.
(In fact, the limit $r \rightarrow \infty$ at constant~$R_A$ was not
constrained at all in the Muchnick \& Russek\cite{mr94} fit, and thus
it is not surprising that their resulting MR surface
and the similar modMR surface yield a very poor fit to the
pair-wise H-He potential.)  The WKB and DR surfaces did
include a pair-wise H-He potential, yielding a reasonable fit to the
base of the repulsive H-He ``wall'' in the $\gamma = 0^\circ$ direction
(although their H-He potentials were less accurate at higher energies).
However, their three-body terms did
not have a van der Waals well, but rather a relatively long positive
``tail'' at large separations, resulting in the ``bulge'' (in
the $\gamma = 90^\circ$ direction) shown by these potentials at
small~$R$ in Figure~\ref{fig:basin} (where $r = 10.6\;\bohr$ and
$R_A = R_B \ge 5.3\;\bohr$).

In the following subsections, we consider the quality of our BMP fit in
various other regions of \hehh\ conformation space.

\subsubsection{Conical intersection with the first excited state}
\label{sssec:cusps}

\begin{figure}   
%
%
\plotfidtwo{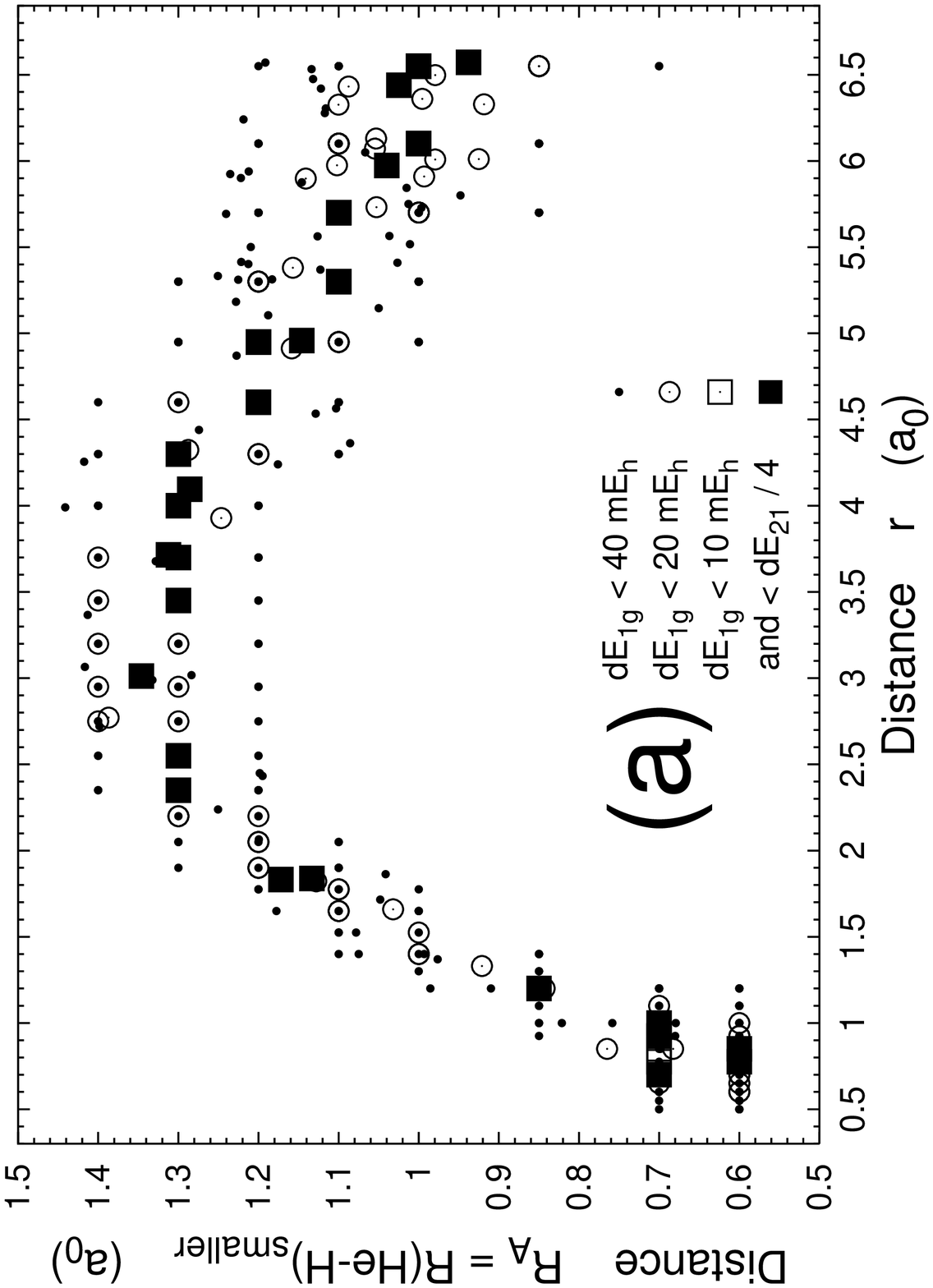}{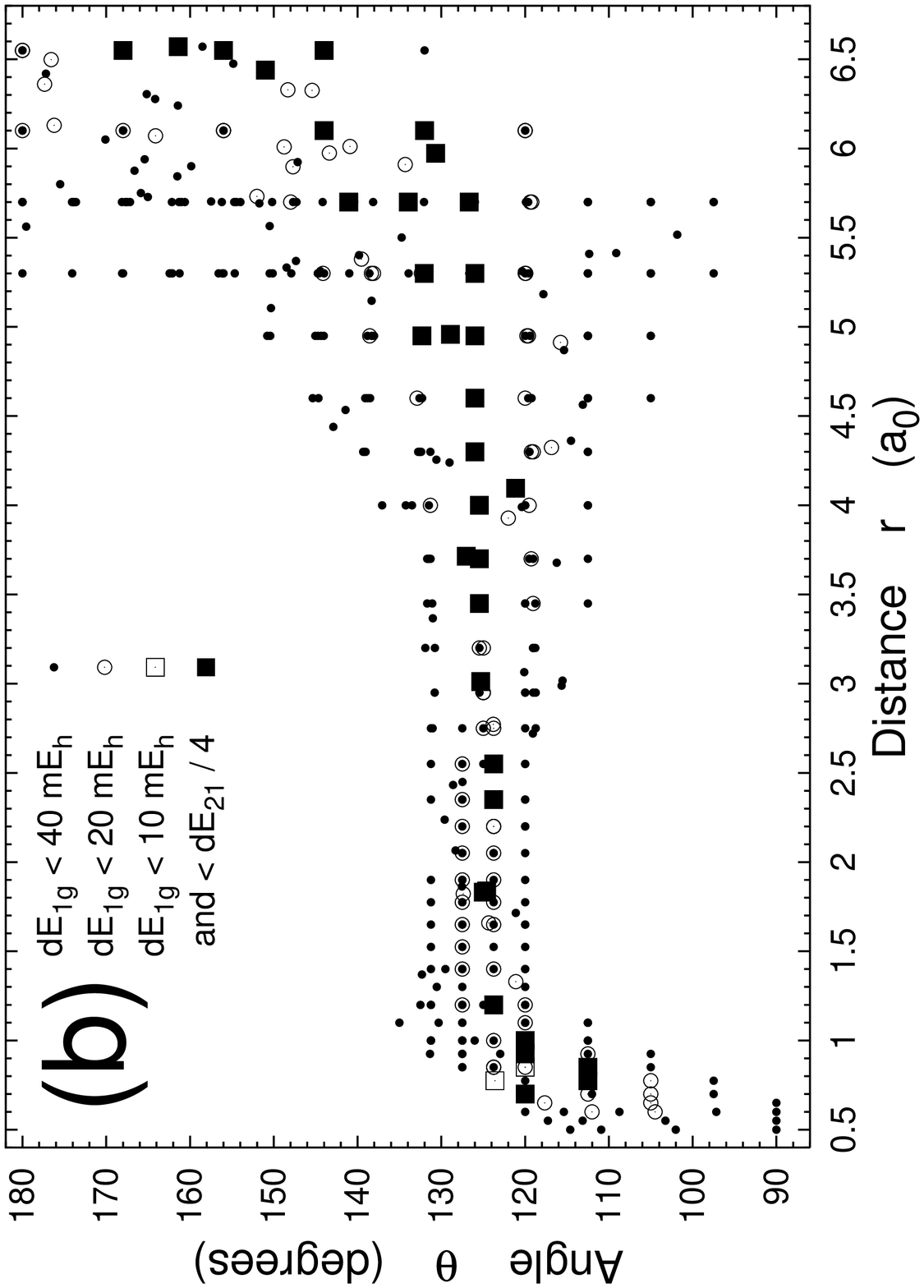}{2.1 true in}{-90}{33}{33}{-250}{200}{-15}
\plotfidtwo{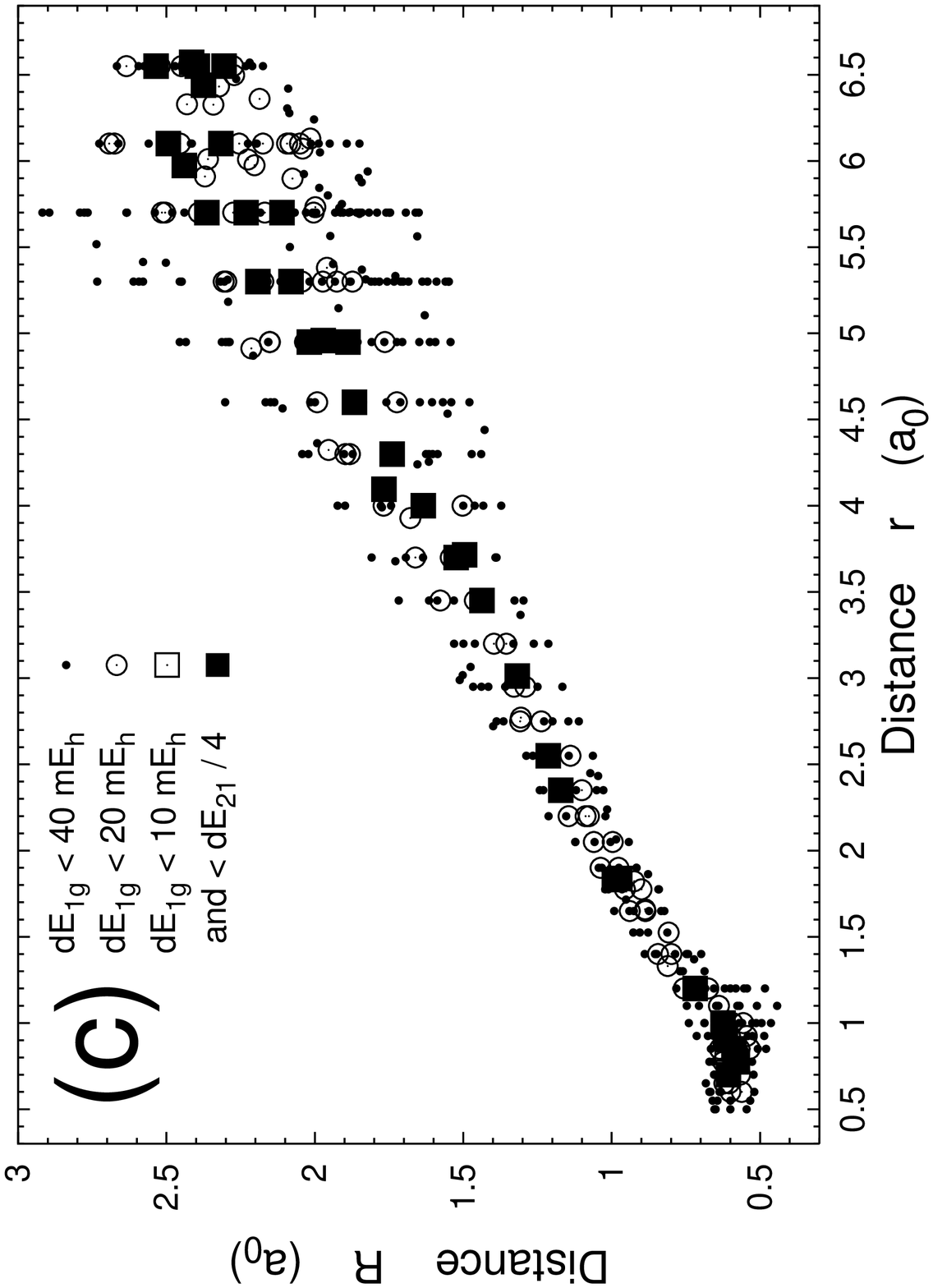}{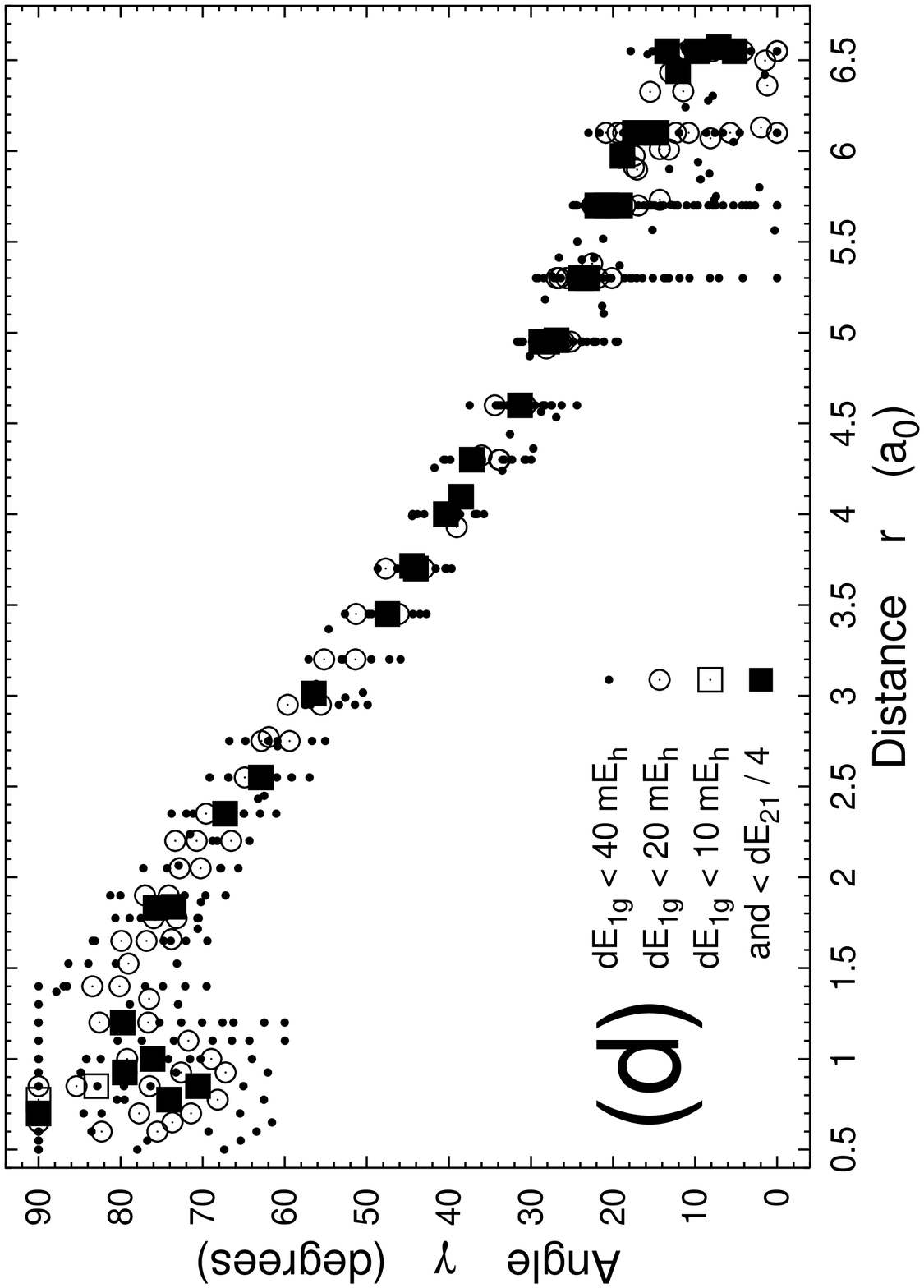}{2.1 true in}{-90}{33}{33}{-250}{200}{-15}
\plotfidtwo{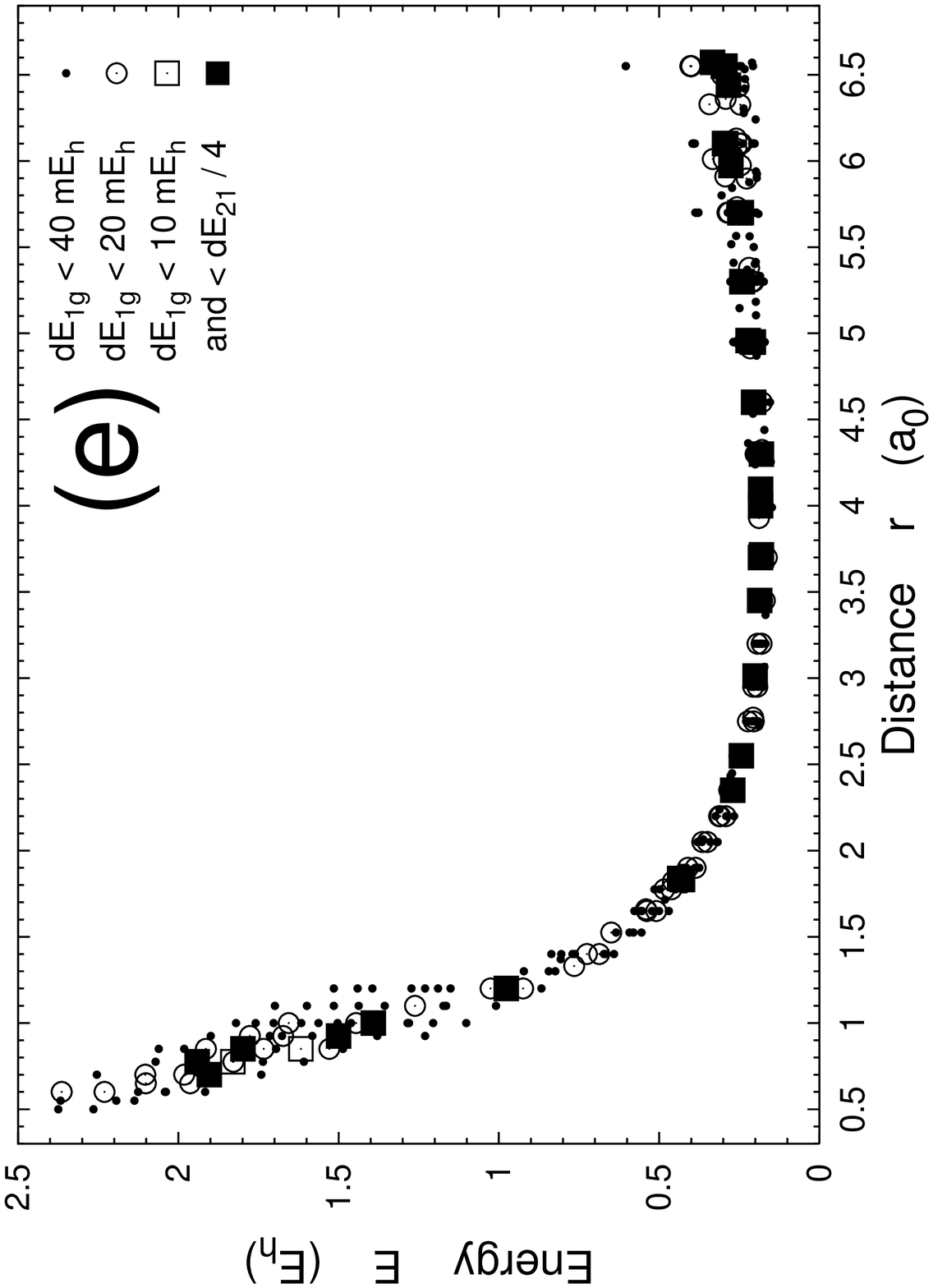}{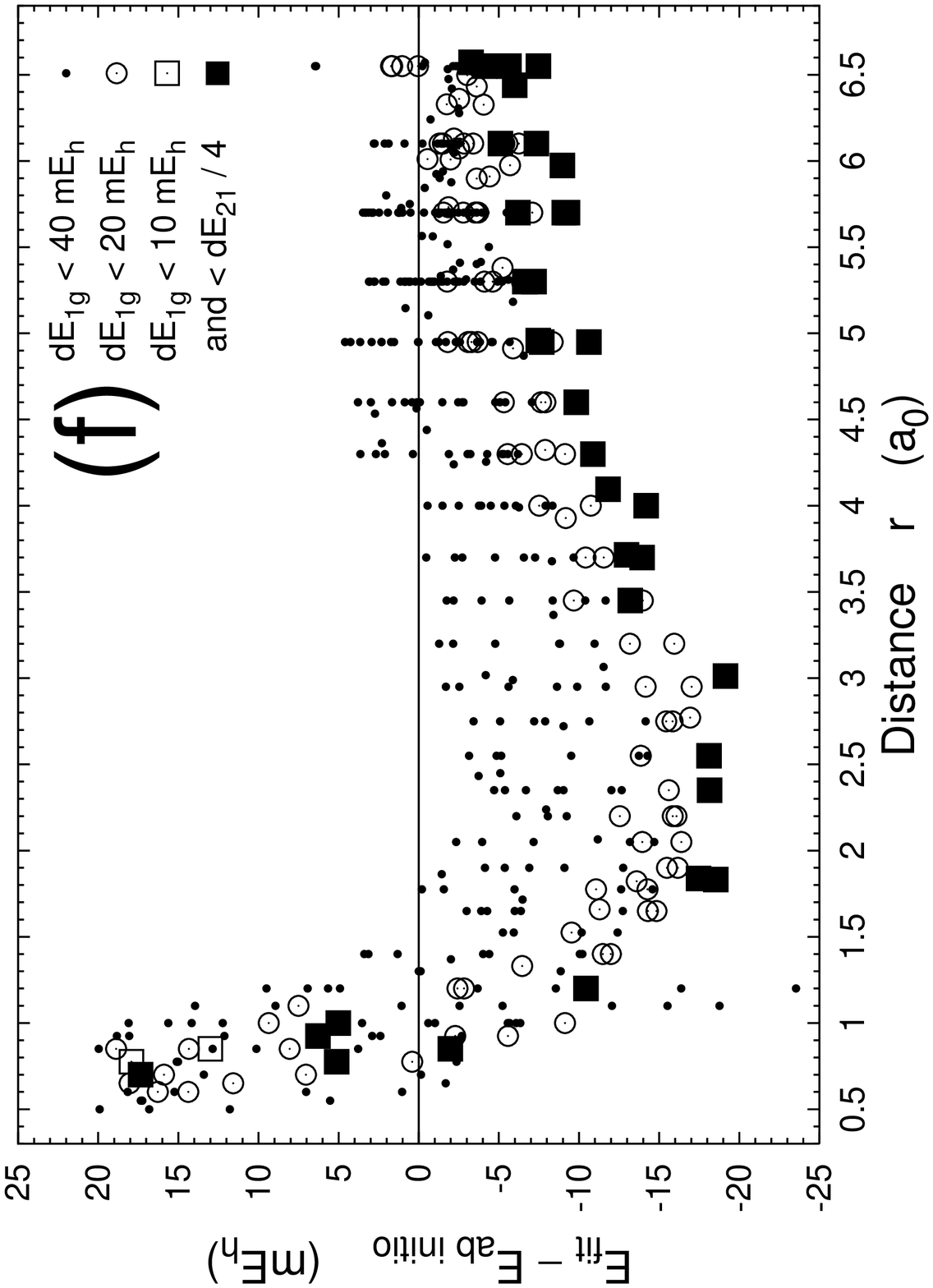}{2.1 true in}{-90}{33}{33}{-250}{200}{-15}
%
%
\caption[fig:conintpos]{The approximate position of the conical intersection
between the ground state and the first excited state, 
as a function of the $\rm H - H$
separation~$r$.  Plotted points lie close to the conical intersection, as
they have small differences $dE_{1g}$ between
ground state and first excited state energies ($20 - 40 \; \millih$ for
dots, $10 - 20 \; \millih$ for open circles, $0 - 10 \; \millih$ for open
and filled squares --- filled squares also have $dE_{1g} < 0.25 \, dE_{21}$,
where $dE_{21}$ is the energy difference between the first and second
excited states).
(a)~The smaller $\rm He - H$ separation~$R_A$,
(b)~the angle~$\theta$ of the He~atom relative to the closer H~atom,
(c)~the distance~$R$ to the center of the H$_2$ molecule,
(d)~the angle~$\gamma$ relative to the H$_2$ molecule,
(e)~the energy~$E$ relative to separated $\rm H + H + He$, and
(f)~the error in the fitted BMP surface near the conical intersection.
}
\label{fig:conintpos}
\end{figure}

The conical intersection of the ground state with the first
(electronic) excited state forms a curved (1-dimensional)
line in the 3-dimensional conformation space of~\hehh.  Since we used the
\hbox{MRD-CI} program to calculate the first few excited states as well
as the ground state, we were able to estimate the position of this conical
intersection by considering the energy difference between the ground
state and first excited state in the \abinitio\ energies.
Figure~\ref{fig:conintpos} shows this estimated position: the squares,
with ground state and first excited state lying at nearly the same
energy, should lie closest to the conical intersection.  The position
of the conical intersection is thus determined to an accuracy of
$\sim 0.03 \; \bohr$ in~$R_A$ and $\sim 3^\circ$ in~$\theta$,
for $1 \; \bohr \lesssim r \lesssim 6 \; \bohr$.  Figure~\ref{fig:conintpos}d
shows that, in contrast to the H$_3$ surface, the conical intersection
for \hehh\ does not in general occur at high-symmetry conformations
(i.e., conformations having $\gamma = 90^\circ$).

Somewhat better estimates of the position of the conical intersection
(as well as estimates of its opening angle in the coordinate directions
orthogonal to the line of the conical intersection, as a function of
position along that line) could be obtained
by plotting the energies of
the ground and lowest excited state(s) as a function of position for
gridpoints near the estimated conical intersection.  Even better
accuracy could be obtained by
the computation of new \abinitio\ energies to home in on the position
of the conical intersection (e.g., as was done for H$_4$ in a limited
region by Boothroyd \etal\cite{bmkp02}).  Procedures such as these
would be required if one
wished to include the conical intersection explicitly in a fitted
surface; however, they are beyond the scope of the present work.

\begin{figure}   
%
%
\plotfidtwo{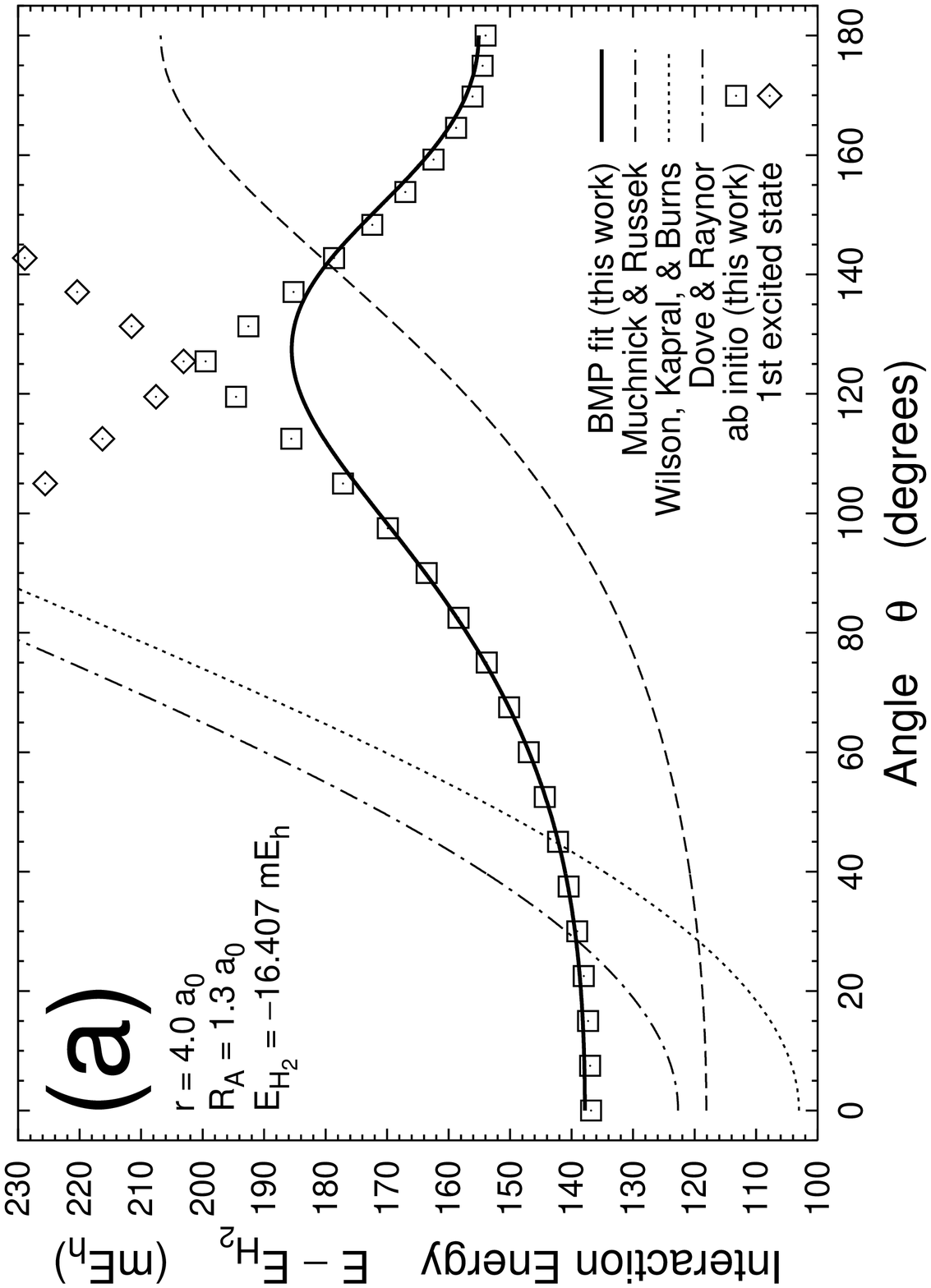}{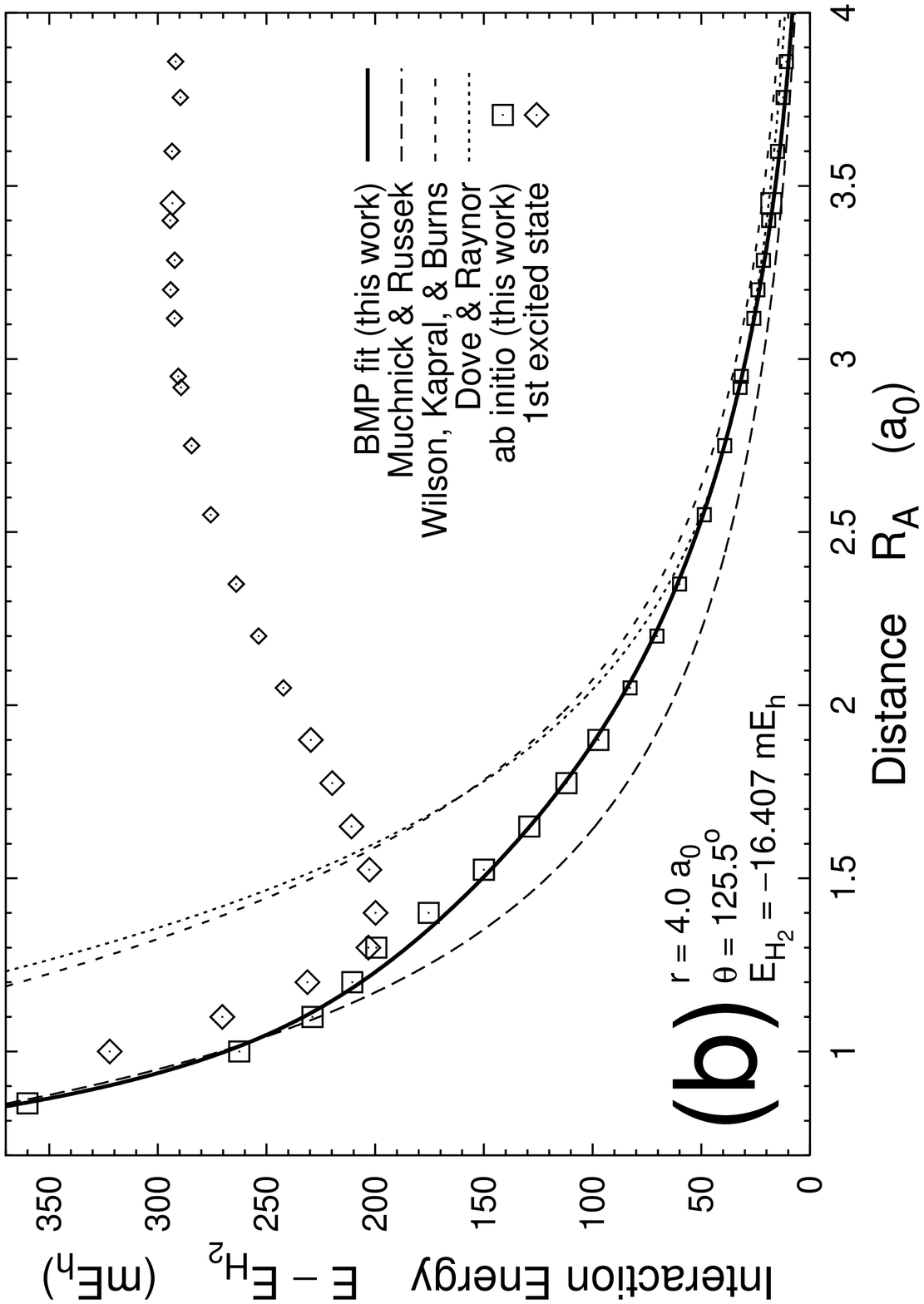}{2.3 true in}{-90}{35}{35}{-260}{200}{-20}
\plotfidtwo{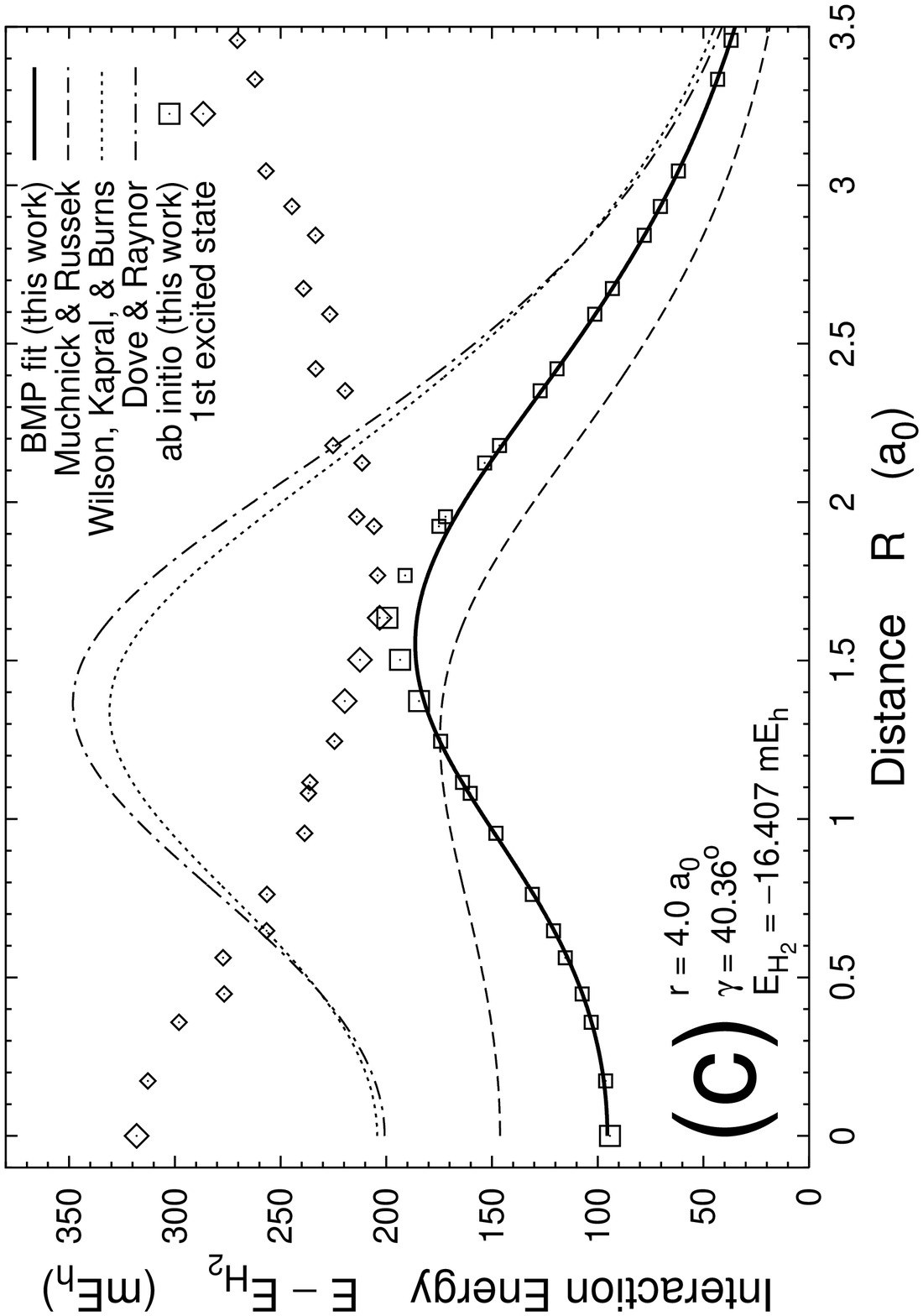}{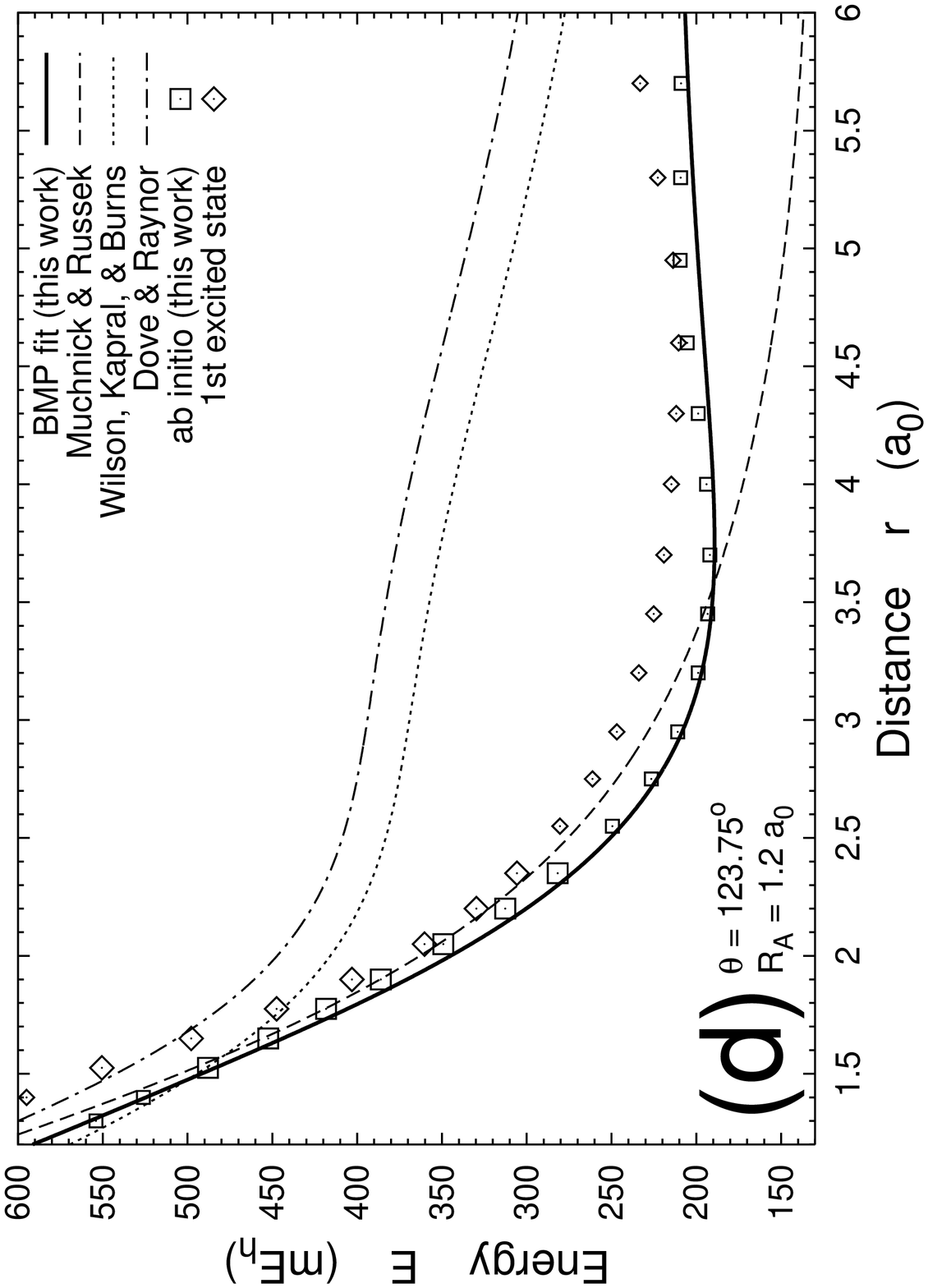}{2.3 true in}{-90}{35}{35}{-260}{200}{-20}
%
%
\caption[fig:conintcut]{Four different cuts through (or very near) conical
intersection points of the \hehh\ surface, showing \abinitio\ ground state
energies (squares) and first excited state energies (diamonds).  Smaller
symbols indicate points near but not on the cut; their energies have been
shifted for plotting purposes by the difference in the BMP surface energy
corresponding to moving these points onto the cut (as expected, this
works well for the ground state, but less well for the first excited state).
Also shown are the fitted BMP surface of this work (solid line) and the
previous surfaces of Muchnick \& Russek (Ref.~\onlinecite{mr94}) (dashed line),
Wilson, Kapral, \& Burns (Ref.~\onlinecite{wkb74}) (dotted line),
and Dove \& Raynor (Ref.~\onlinecite{dr78}) (dot-dashed line).
The Schaefer \& K\"ohler surface (Ref.~\onlinecite{sk85}) is not plotted,
as this would require extrapolating it far beyond its range of validity.
}
\label{fig:conintcut}
\end{figure}

As may be seen from Figure~\ref{fig:conintpos}e, the conical intersection
occurs at energies $E \gtrsim 0.2 \; \har$, i.e., above twice the
H$_2$ dissociation energy relative to $\rm H_2 + He$.
Figure~\ref{fig:conintcut} shows various cuts through points on the
conical intersection (note that Figs.~\ref{fig:conintcut}a,
\ref{fig:conintcut}b, and~\ref{fig:conintcut}c show
three different cuts through the same conical intersection point, while
Fig.~\ref{fig:conintcut}d shows a fourth cut through two different
conical intersection points).  The BMP surface can be seen
to ``round off'' the conical intersection, missing by as much
as~$20 \; \millih$ (although it still does much better
than the previous surfaces in this region of the \hehh\ surface).
Figure~\ref{fig:conintpos}f, showing errors of the BMP surface relative
to the \abinitio\ energies at points near the conical intersection, also
shows the surface lying quite far below the cusp
of the conical intersection (black squares) for $r \gtrsim 1 \; \bohr$,
and (at least in some cases)
overcompensating at points slightly further away (black dots).  As a
result, the conical intersection is the source of many, perhaps most,
of the outliers at $E \gtrsim 0.2 \; \har$ in the
scatterplot of Figure~\ref{fig:scatterplot}.

\subsubsection{The H$_2$ + He van der Waals well}
\label{sssec:vdw}

\begin{figure}   
%
%
\plotfiddle{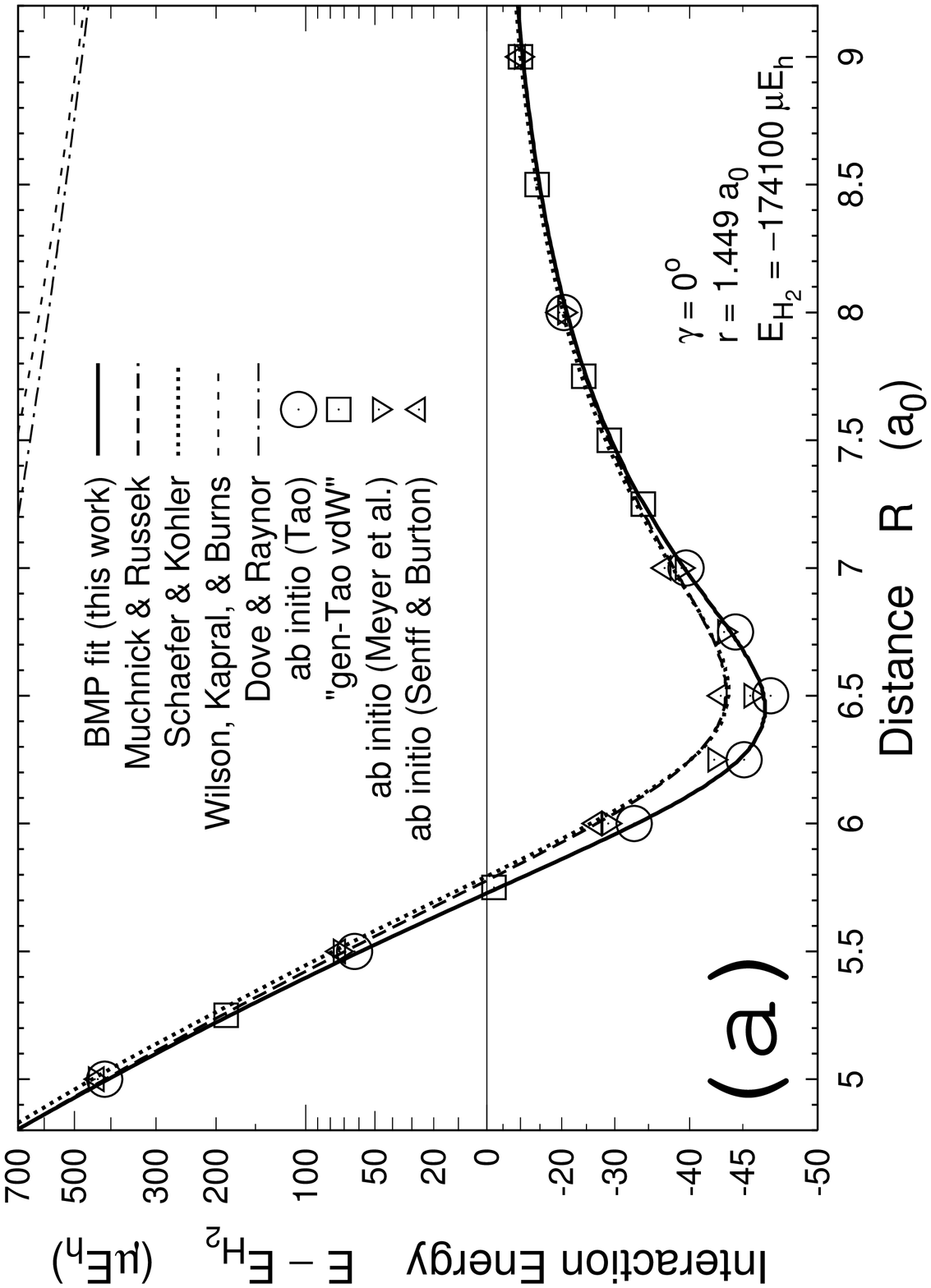}{2.3 true in}{-90}{36}{36}{-160}{205}
\plotfiddle{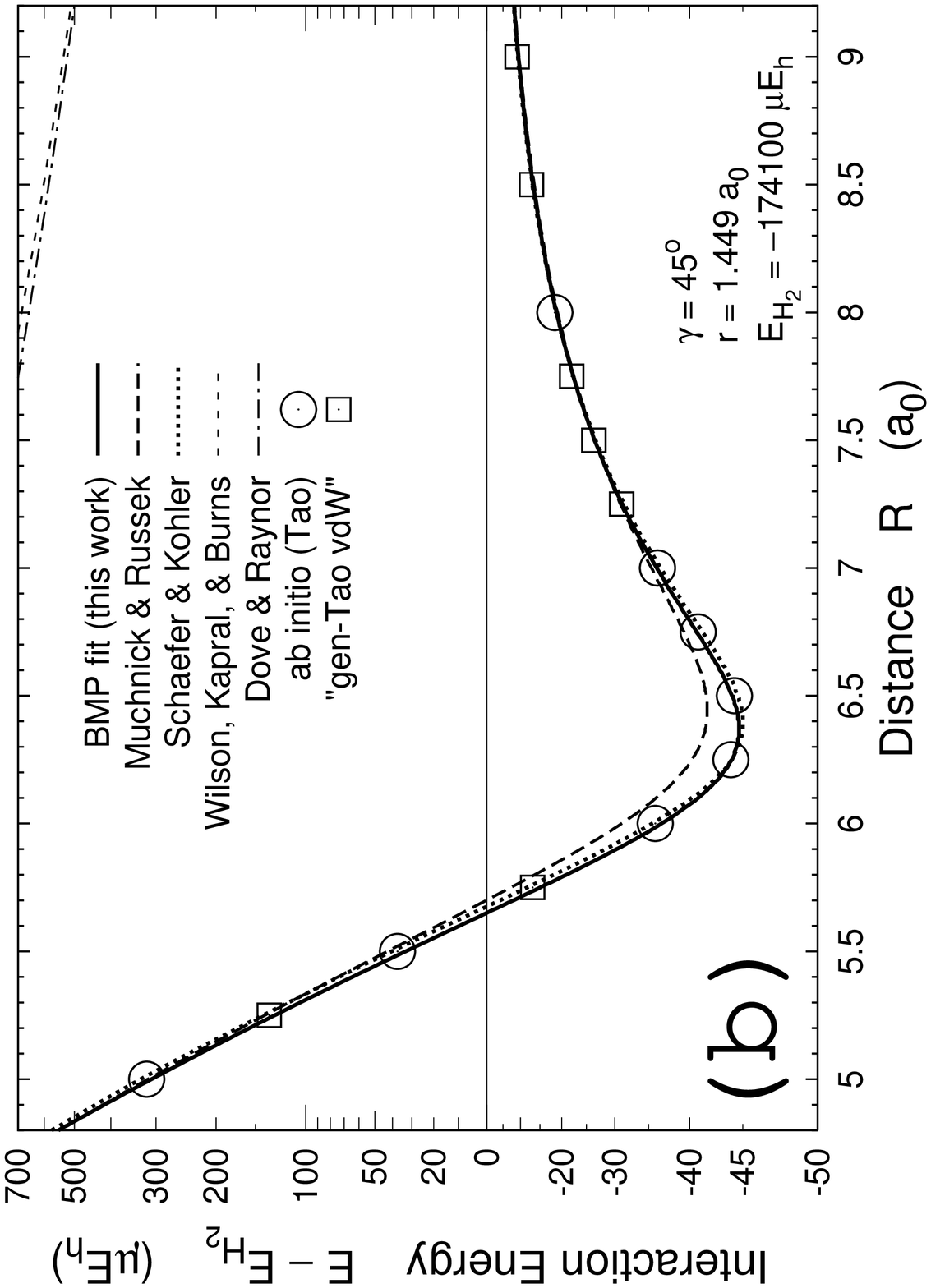}{2.3 true in}{-90}{36}{36}{-160}{205}
\plotfiddle{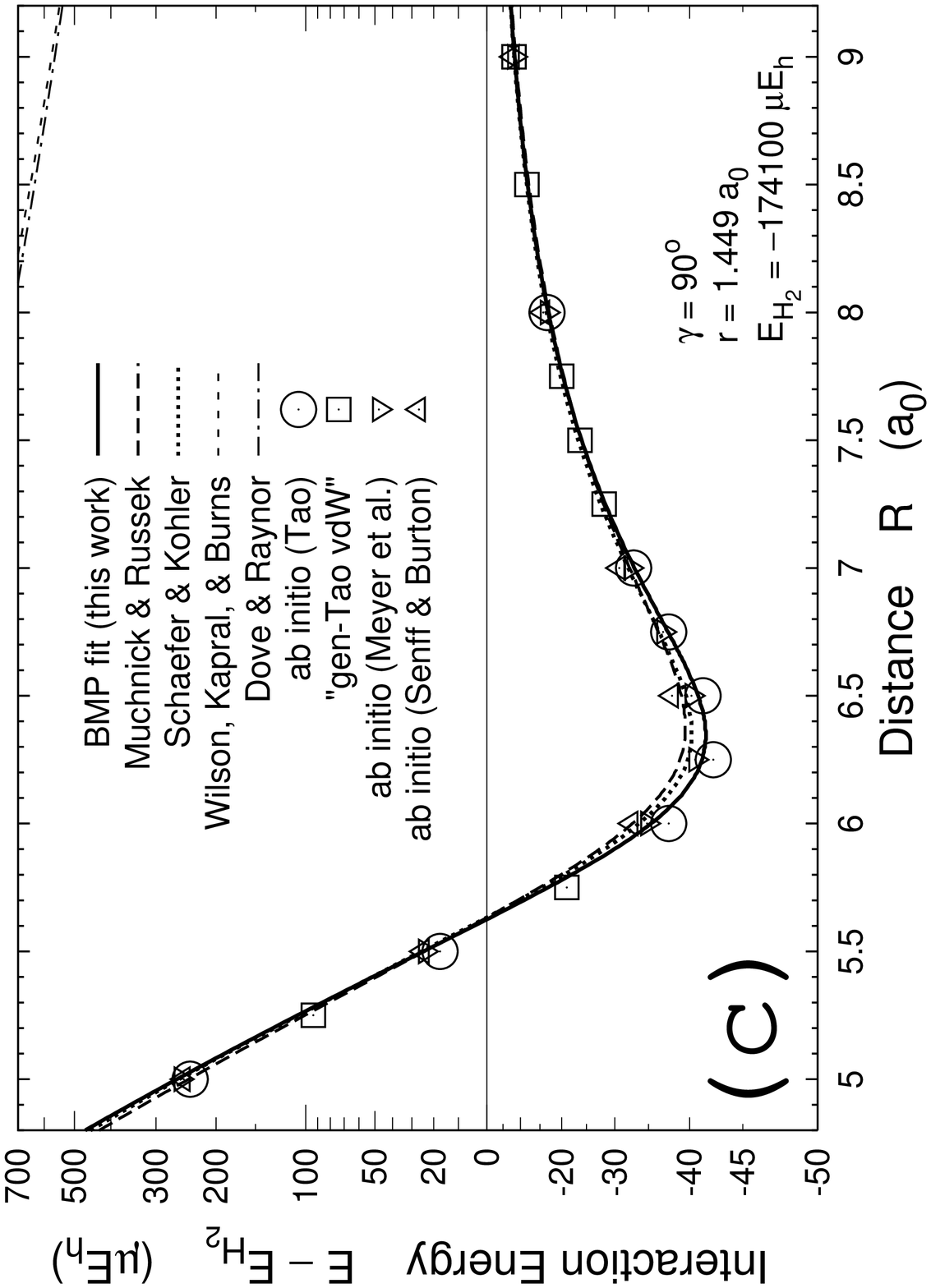}{2.3 true in}{-90}{36}{36}{-160}{205}
%
%
\caption[fig:vdw]{Comparison of the fitted BMP surface (solid lines)
in the van der Waals well at three different orientations with the
accurate \abinitio\ energies of Tao (Ref.~\onlinecite{Tao94}) (circles)
and the accurate ``gen-Tao vdW'' energies generated
from them (squares); some less-accurate earlier \abinitio\ energies
(triangles) are also shown, as well as some earlier analytic
surfaces (dotted/dashed lines: as in Fig.~\ref{fig:conintcut}).
Note the ``shifted-logarithmic'' energy scale, used to emphasize
effects at the bottom of the van der Waals well.
(a)~$\gamma = 0^\circ$, (b)~$\gamma = 45^\circ$, (c)~$\gamma = 90^\circ$.
}
\label{fig:vdw}
\end{figure}

Figure~\ref{fig:vdw} compares fitted \hehh\ surfaces to \abinitio\ van
der Waals energies --- note that the unusual ``shifted-logarithmic'' energy
scale emphasizes effects at the bottom of the van der Waals well, while
also showing a significant portion of the ``repulsive wall'' at smaller~$R$.
One can see that our BMP \hehh\ surface fits the
depth of the van der Waals well to within about~$1 \; \microh$, when
compared to the accurate \abinitio\ energies of Tao\cite{Tao94}
(which have $1.28 \; \bohr \le r \le 1.618 \; \bohr$); from subset~31
in Table~\ref{tab:rms}, the rms difference is $0.8 \; \microh$
over the range $6 \; \bohr \le R < 10 \; \bohr$.
The previous SK and MR surfaces\cite{sk85,mr94} agree nearly as well
(i.e., within a
few~$\microh$; rms values of 1.8 and~$2.6 \; \microh$, respectively),
since they were fitted to the fairly accurate \abinitio\ energies of
Meyer \etal\cite{mhk80}
However, the earlier WKB and DR surfaces\cite{wkb74,dr78}
do not contain a van der Waals well at
all --- they are visible only at the upper right of Figure~\ref{fig:vdw}.

Note that the ``torque reversal'' at $R \approx 6.1 \; \bohr$ noted by
Muchnick \& Russek\cite{mr94} is also visible in Figure~\ref{fig:vdw}.
For small angles~$\gamma$, the van der Waals well is slightly deeper
but the ``repulsive wall'' is slightly further out than
for $\gamma$~near~$90^\circ$.  Thus the $\gamma = 0^\circ$ (linear)
orientation is energetically favored at $R > 6.1 \; \bohr$, while the
$\gamma = 90^\circ$ (T-shaped) orientation is favored at smaller
distances~$R$.

\begin{figure}   
%
%
\plotfiddle{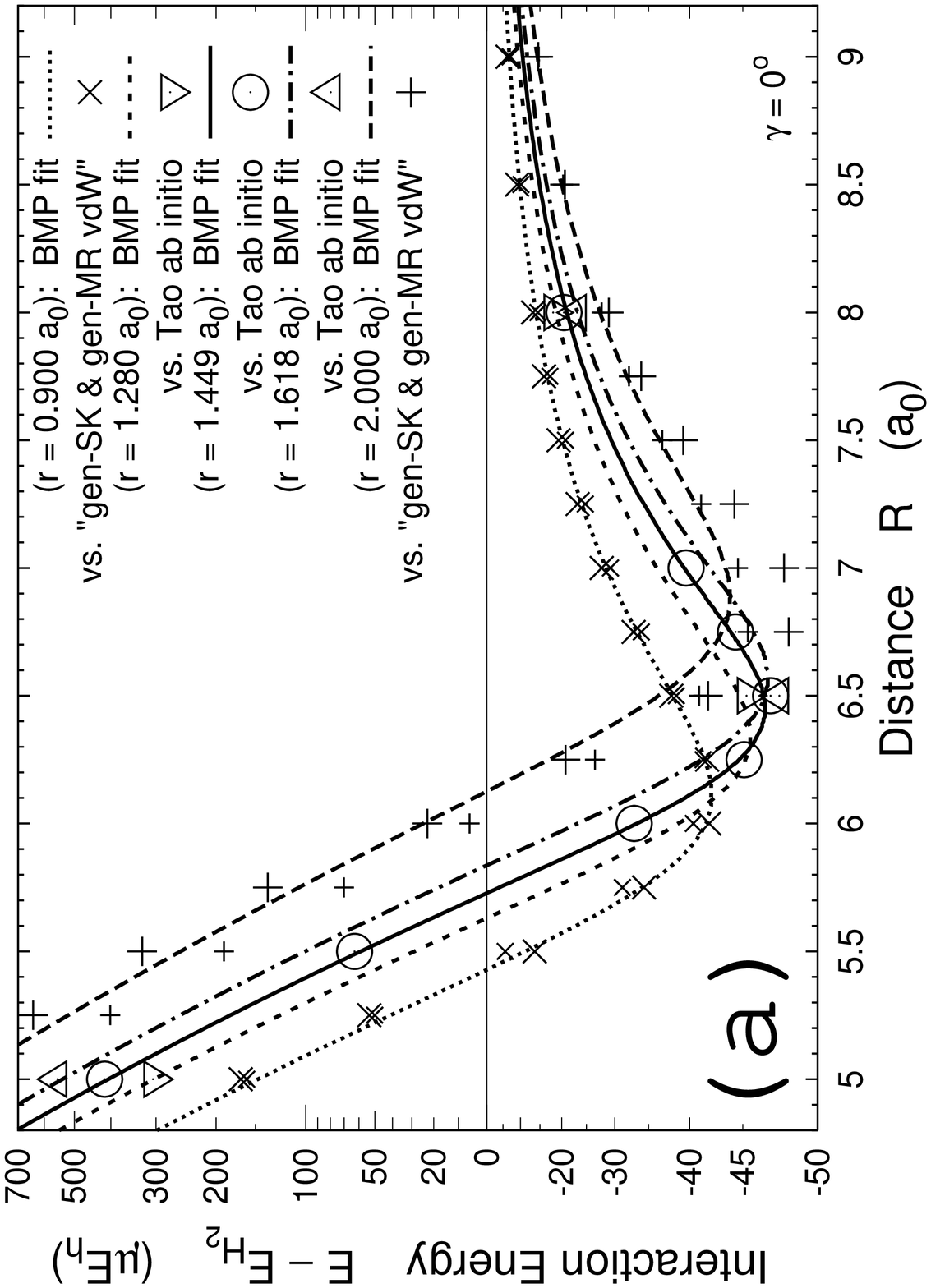}{2.3 true in}{-90}{36}{36}{-160}{205}
\plotfiddle{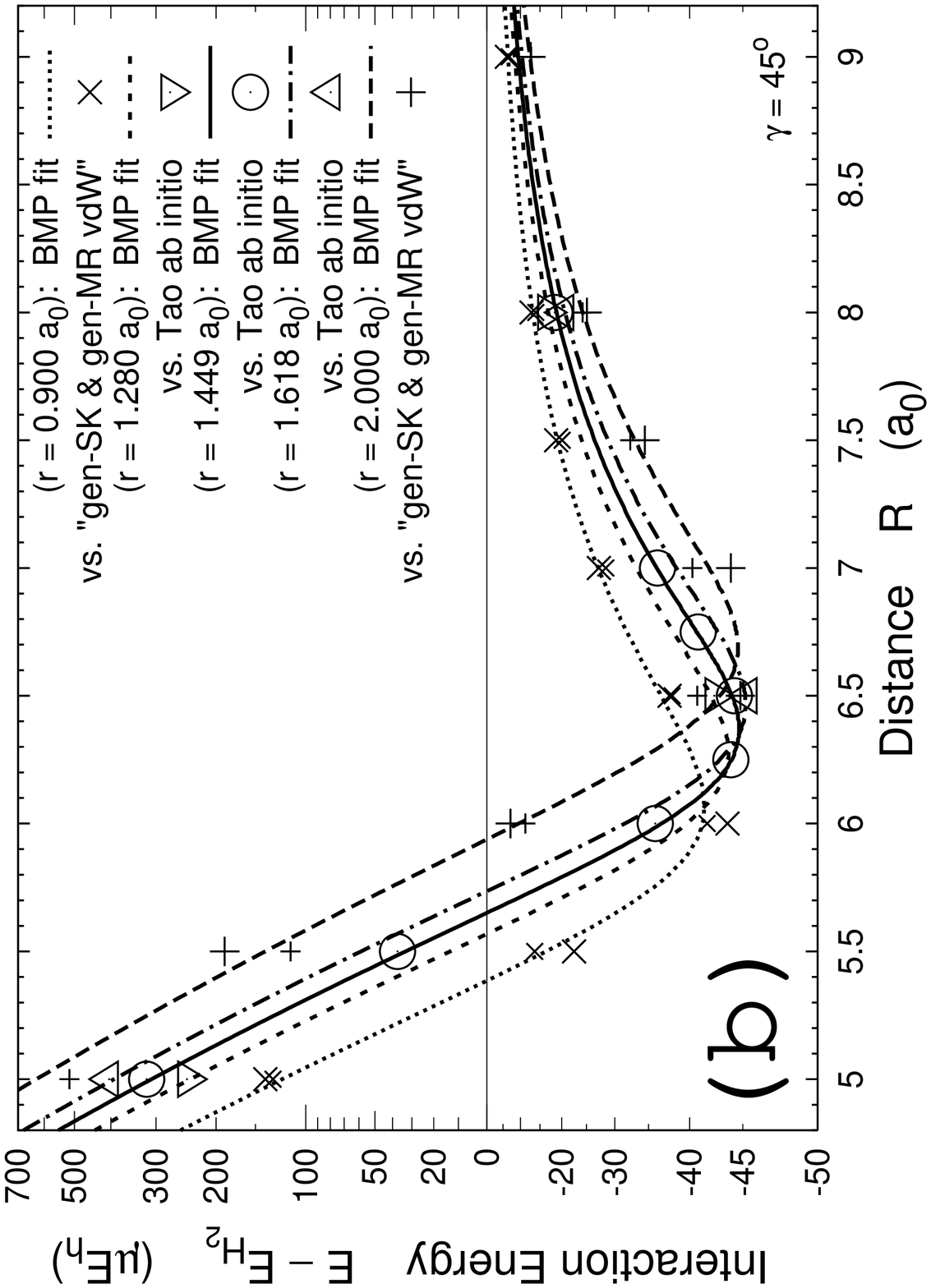}{2.3 true in}{-90}{36}{36}{-160}{205}
\plotfiddle{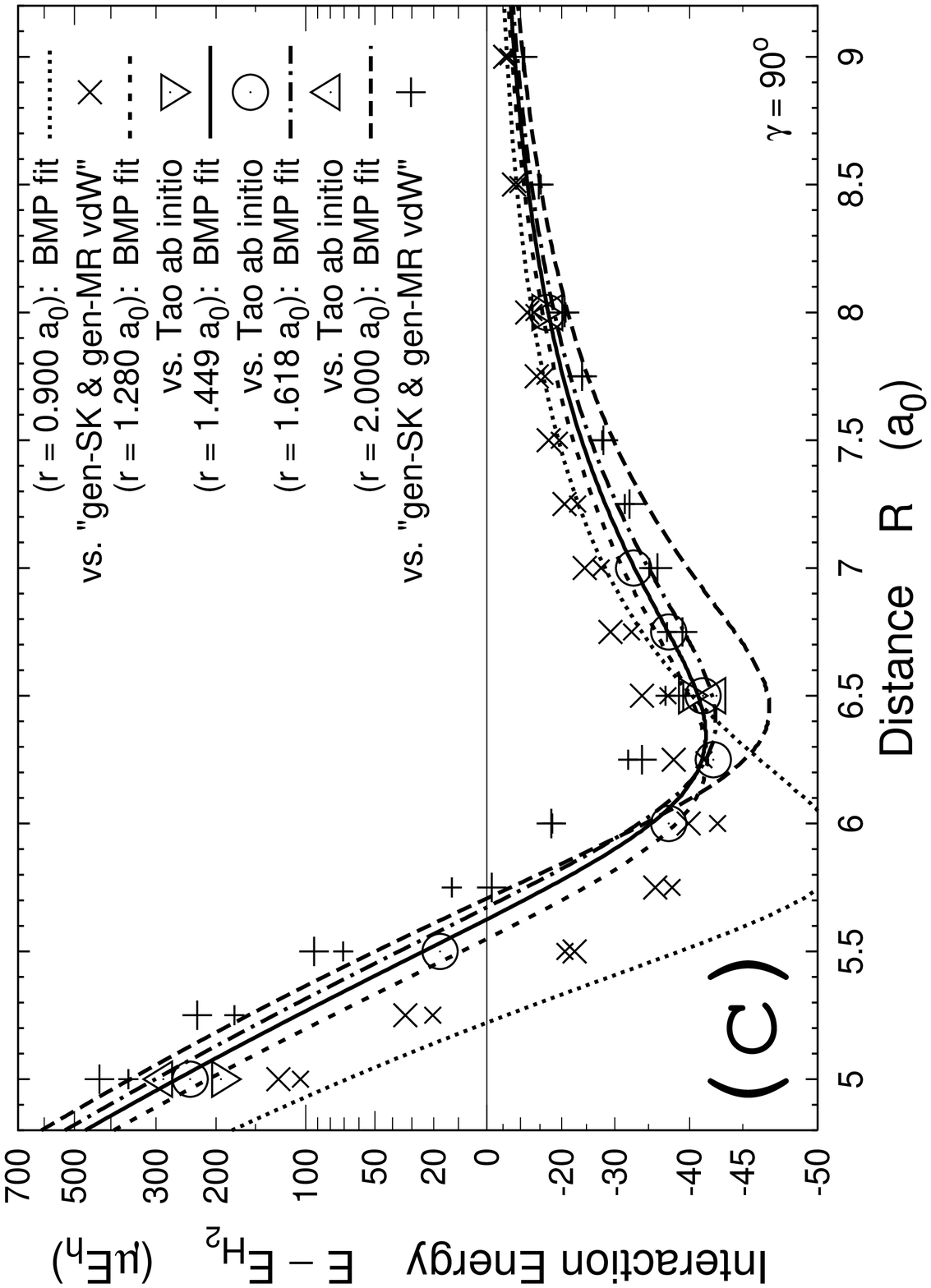}{2.3 true in}{-90}{36}{36}{-160}{205}
%
%
\caption[fig:rvdw]{Effect of H$_2$-molecule size~$r$ on the van der
Waals well, at three different orientations (note
that $E_{\rm H_2} = \{ -83571.4$, $-171346.4$, $-174100.0$, $-167643.3$,
and~$-138203.3 \; \microh \}$ for $r = \{ 0.9$, 1.28, 1.449, 1.618,
and~$2.0 \; \bohr \}$, respectively);
the energy scale is the same as in Fig.~\ref{fig:vdw}.
The fitted BMP surface (lines) is compared to the
near-H$_2$-equilibrium \abinitio\ energies of Tao (Ref.~\onlinecite{Tao94})
(circles and triangles)
and, at more extreme $r$~values, the much-less-accurate generated
energies described in \S~\ref{sssec:generateheh2}
(``gen-SK vdW'': larger crosses and plusses,
``gen-MR vdW'': smaller ones).
For the sake of clarity, the ``gen-Tao vdW'' points are omitted.
(a)~$\gamma = 0^\circ$, (b)~$\gamma = 45^\circ$, (c)~$\gamma = 90^\circ$.
}
\label{fig:rvdw}
\end{figure}

Figure~\ref{fig:rvdw} considers the variation of the van der Waals well
with the size~$r$ of the H$_2$ molecule (using the same energy scale as
Fig.~\ref{fig:vdw}).  For $r = 1.28$, 1.449,
and~$1.618 \; \bohr$, accurate \abinitio\ energies from Tao\cite{Tao94}
are available, which our BMP surface fits very well (rms of $0.8 \; \microh$:
see subset~31 in Table~\ref{tab:rms}).  For the more
extreme distances $r = 0.9$, 1.1, 1.8, and~$2.0 \; \bohr$, only the
much-less-accurate ``gen-SK vdw'' and ``gen-MR vdW'' generated points
of \S~\ref{sssec:generateheh2} were available.  The former were obtained
from the SK surface\cite{sk85}, which at those $r$~values
was fitted to the lower-accuracy extreme-position \abinitio\ energies of
Meyer \etal\cite{mhk80}; the latter were obtained from the modMR surface
of \S~\ref{sssec:modMuRu}, fitted to the accurate near-H$_2$-equilibrium
\abinitio\ energies from Tao\cite{Tao94} with the $r$-dependence taken
from Thakkar \etal\cite{thccl92}  These two sets of generated points
agree with each other to
within about $5 \; \microh$ in the van der Waals well, and are
fitted to roughly this accuracy by our BMP surface (rms
of $\sim 3 \; \microh$ over the range $6 \; \bohr \le R < 10 \; \bohr$,
from subsets 30 and~32 in Table~\ref{tab:rms};
largest difference $\sim 10 \; \microh$).
As the size~$r$ of the H$_2$ molecule is increased, the van der Waals well
moves slightly further out; the depth also changes slightly, but this is
less well constrained by the \abinitio\ energies.

At $r > 2 \; \bohr$,
there are no accurate \abinitio\ energies available in the region of
the van der Waals well (some of our \abinitio\ energies may lie nearby,
but these have estimated errors larger than the typical depth of the
van der Waals well).  The generated ``vdW'' energies in this region
are thus expected to be quite inaccurate --- the SK and modMR
surfaces differ from each other by~$\sim 0.1 \; \millih$ in this region.
These generated points were therefore
given quite low weight, only enough to keep the fitted surface
from developing large spurious features in this region.  This is the
cause of the relatively large rms error of~$0.1233 \; \millih$ of the
BMP surface shown by Table~\ref{tab:rms} (subset~33) for the ``vdW'' at
$r = 2.4, 3.0, 4.0 \; \bohr$ and $6 \; \bohr \le R < 10 \; \bohr$.

\subsubsection{The interaction region}
\label{sssec:interact}

\begin{figure}   
%
%
\plotfidtwo{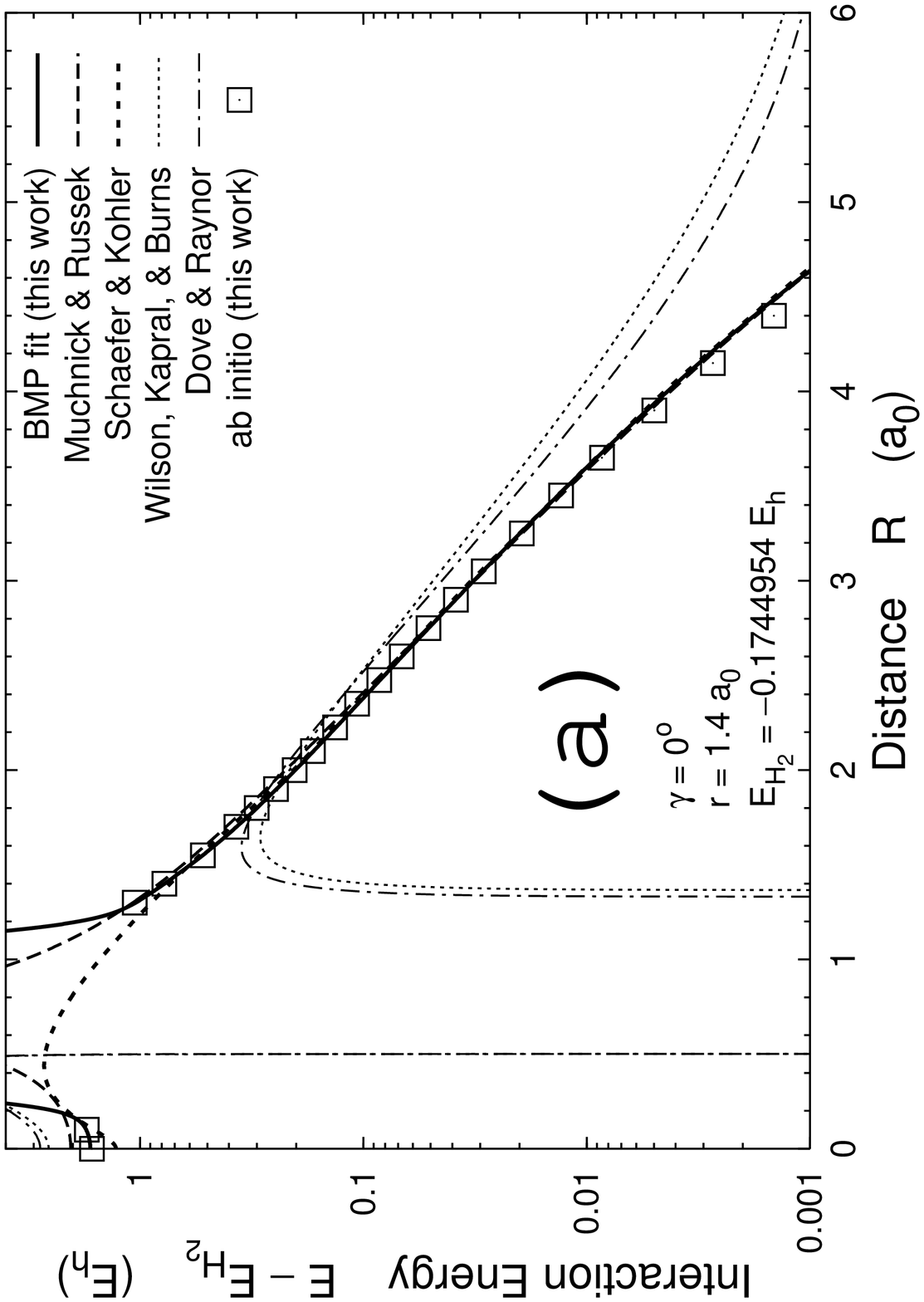}{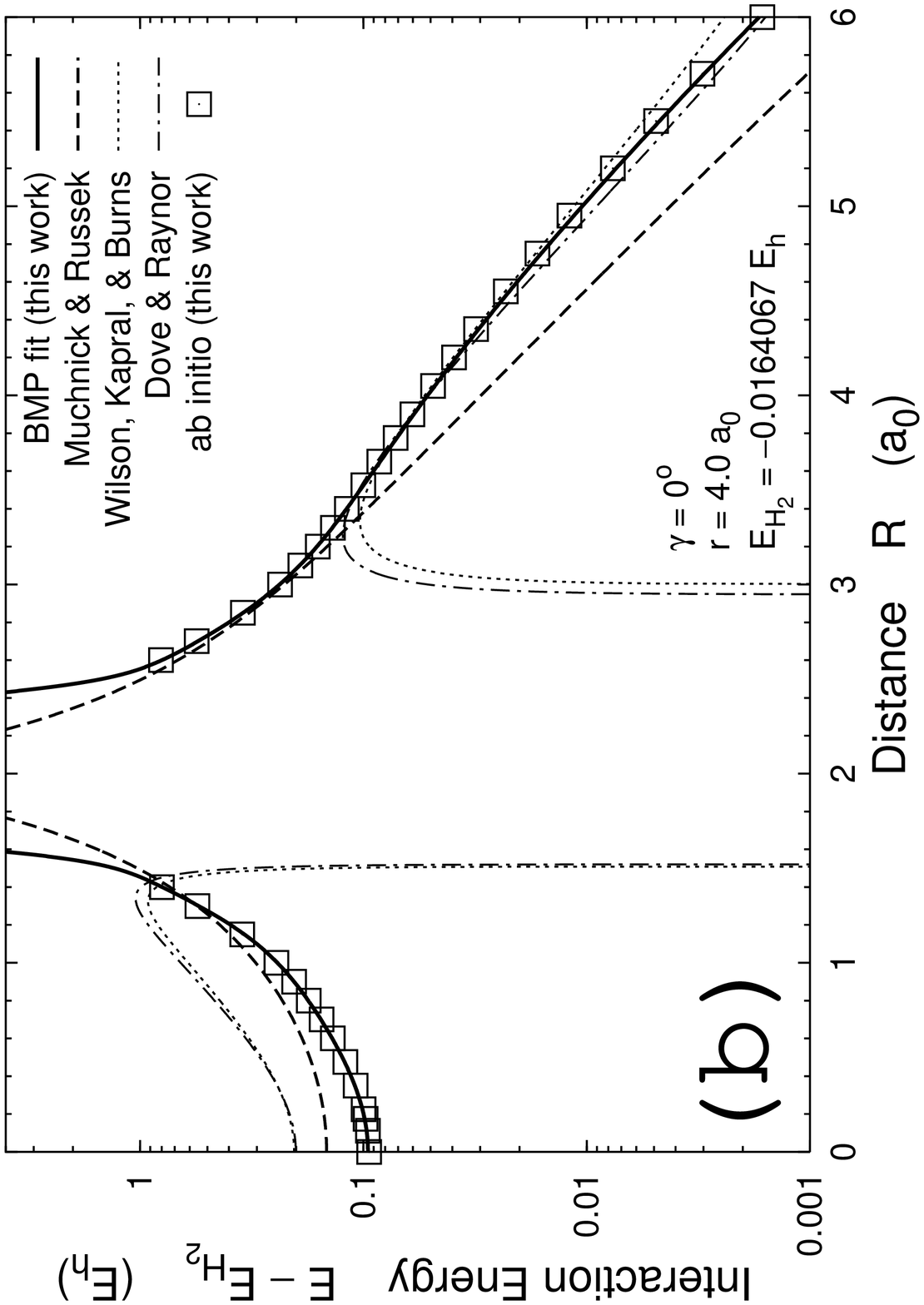}{2.3 true in}{-90}{35}{35}{-260}{200}{-20}
\plotfidtwo{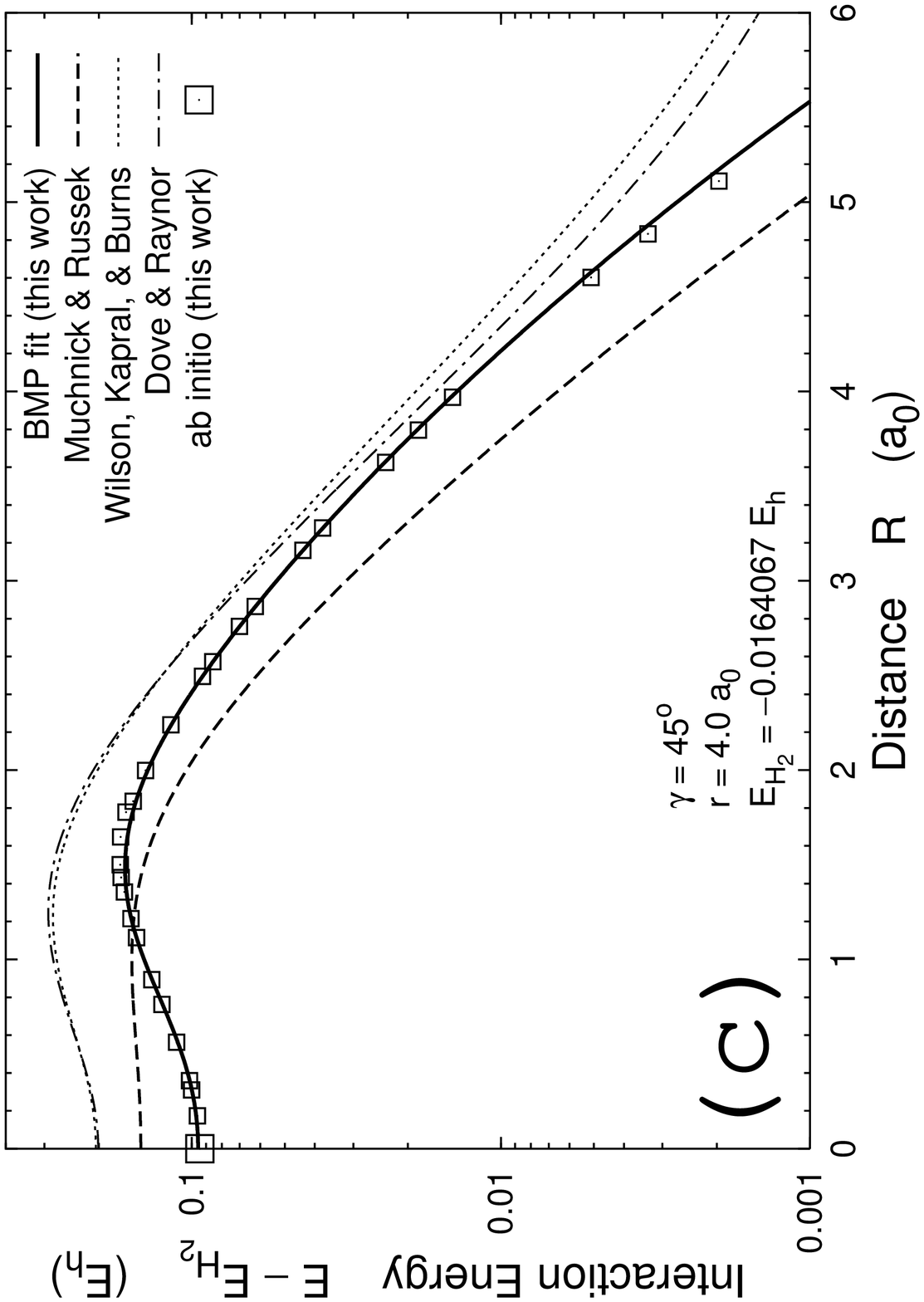}{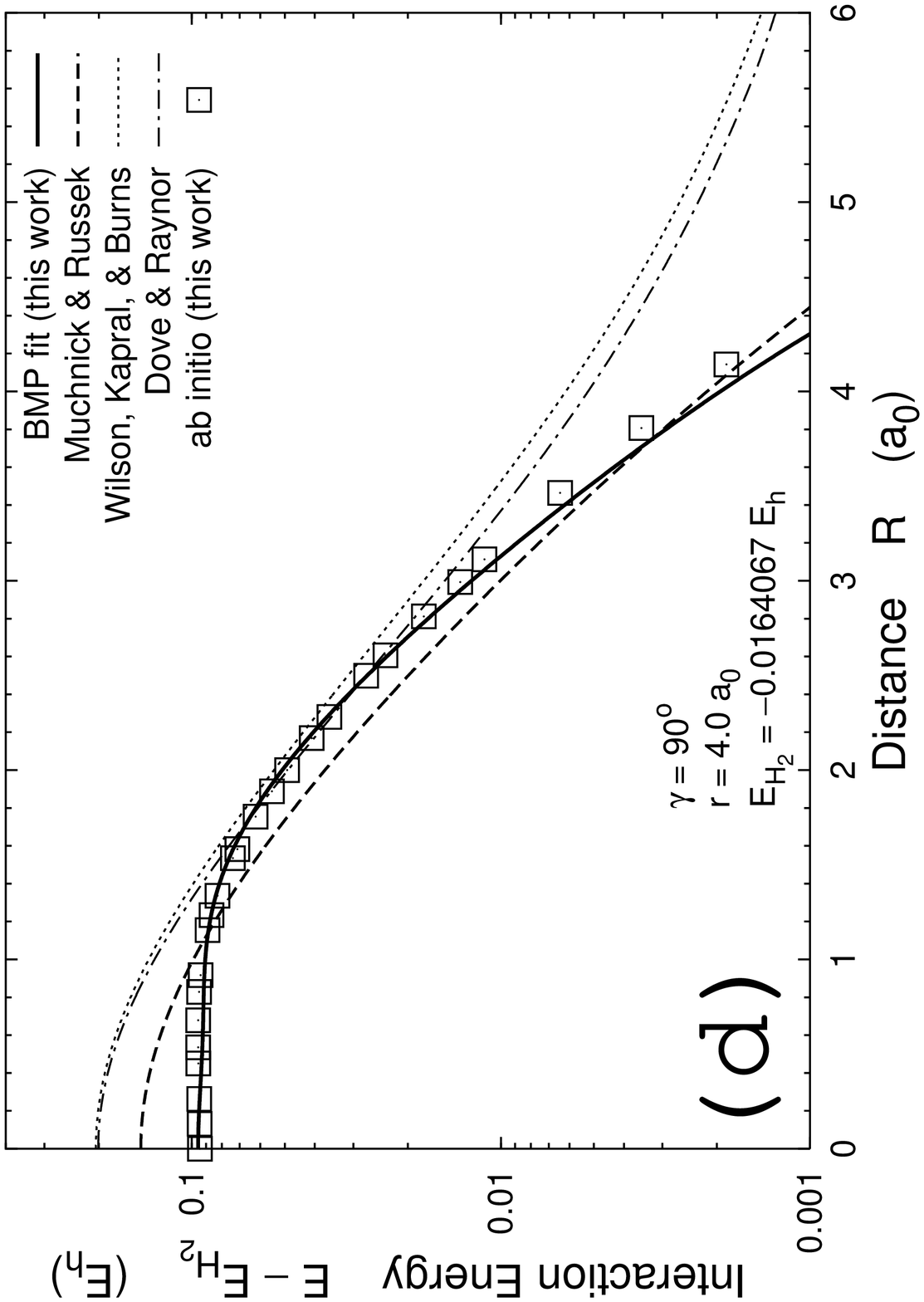}{2.3 true in}{-90}{35}{35}{-260}{200}{-20}
%
%
\caption[fig:wall]{Comparison of the fitted BMP surface of this work
(and also previous analytic surfaces) to the \abinitio\ energies, for two
H$_2$-molecule sizes~$r$.  Only for the lowest-energy points does the
(largely systematic) uncertainty of~$\sim 0.6 \; \millih$ in
the \abinitio\ energies approach or exceed the size of the symbols.
(a)~$r = 1.4 \; \bohr$, $\gamma = 0^\circ$: near equilibrium H$_2$;
the ``hole'' is visible in the Wilson, Kapral, \& Burns and Dove \& Raynor
surfaces (Refs.~\onlinecite{wkb74,dr78}), but the other surfaces are
quite accurate.
(b)~$r = 4.0 \; \bohr$, $\gamma = 0^\circ$: important for dissociation and
high-excitation H$_2$; the ``hole'' in the Wilson,
Kapral, \& Burns and Dove \& Raynor surfaces
is at even lower energy, and even the Muchnick \& Russek
surface (Ref.~\onlinecite{mr94}) has relatively large fractional errors.
The Schaefer \& K\"ohler surface (Ref.~\onlinecite{sk85}) is not
plotted, as this is too far to extrapolate beyond its range of validity.
(c)~$r = 4.0 \; \bohr$, $\gamma = 45^\circ$: similar to~(b), but note
expanded vertical scale.  Smaller symbols correspond
to points near but not quite on the $\gamma = 45^\circ$ cut, shifted as
in Fig.~\ref{fig:conintcut}.
(d)~$r = 4.0 \; \bohr$, $\gamma = 90^\circ$: similar to~(c).
}
\label{fig:wall}
\end{figure}

As mentioned above, Table~\ref{tab:rms} shows that the earlier
SK and MR surfaces\cite{sk85,mr94} do
relatively well for not-too-large H$_2$-molecules ($r < 2 \; \bohr$), even
up to relatively large interaction energies.  An example of this,
with $r = 1.4\;\bohr$, is
illustrated in Figure~\ref{fig:wall}a (note the logarithmic energy scale).
Even at $R \sim 1.3$ to~$2 \; \bohr$, these two surfaces have errors of only a
few percent for near-equilibrium H$_2$ --- the BMP surface of the present
work is an order of magnitude more accurate, with errors of a fraction
of a percent in this region, but nonetheless all three of these
surfaces are almost indistinguishable in Figure~\ref{fig:wall}a.  Even
the much-earlier WKB and DR surfaces\cite{wkb74,dr78} are not so very much
worse in the outer part of this region, but, as may be seen
in Figure~\ref{fig:wall}a, they have an unphysical ``hole'' at
$R \lesssim 1.5 \; \bohr$ for small angles~$\gamma$ (i.e., the ``hole''
is at small $\rm H - He$ separations~$R_A$).

Table~\ref{tab:rms} shows that, for all the fitted surfaces considered there,
errors at small~$r$ (i.e., $r < 1.2\;\bohr$: subsets 4 to~7) are very
similar to the errors at $r \sim 1.4\;\bohr$ (subsets 8 to~11) that are
illustrated in Figure~\ref{fig:wall}a.
In contrast, at larger $\rm H - H$ separations~$r$ (such as would arise
from dissociative collisions and highly-excited H$_2$ molecules), the
SK surface\cite{sk85} is undefined, and even the
MR surface\cite{mr94} does quite poorly, with errors
of order~50\% visible in Figure~\ref{fig:wall}b, c, and~d
(for $r = 4 \; \bohr$).  At these larger $r$~values, the
WKB and DR surfaces\cite{wkb74,dr78}
have errors of order a factor of~2, and Figure~\ref{fig:wall}b shows that
the ``hole'' in these surfaces appears at lower interaction energy
(and, in fact, at a lower total energy, though this is less obvious from
the figure).  A systematic numerical search found that
the lowest part of the lip of this ``hole'' lies at energies
$E = 53$ and~$87 \; \millih$ for the WKB and DR surfaces, respectively ---
less than two H$_2$ dissociation energies above equilibrium
$\rm H_2 + He$.
In this region, the BMP surface of the present work is much more accurate
than any previous surface, with typical errors only slightly larger than
for equilibrium H$_2$ molecules.  This is also illustrated by the other
cuts through this region shown in the previous Figure~\ref{fig:conintcut}.

\subsubsection{Extrapolation to very short distances}
\label{sssec:extrap}

The SK surface\cite{sk85} contains no repulsive
Coulomb terms, and thus does not have the correct form at small
$\rm H - He$ separations --- an example can be seen in
Figure~\ref{fig:wall}a.  The WKB and DR surfaces\cite{wkb74,dr78}
do even worse: they turn over and become negative at small
$\rm H - He$ separations ($R_A$ or $R_B \lesssim 1 \; \bohr$: the ``hole''
in these surfaces), as may be seen in Figure~\ref{fig:wall}a and~b.
The MR surface\cite{mr94} behaves roughly as an
exponentially damped Coulomb repulsion for small~$R_A$,
yielding reasonable behavior at short distances.

No constraints were placed on our fitted BMP surface at $\rm H - He$
distances less than those of the most compact \abinitio\ geometries
(namely~$R_A = 0.6 \; \bohr$), which lie at energies $E \gtrsim 1 \; \har$.
Some of our earlier fitted surfaces turned
over into an unphysical ``hole'' at shorter distances than this, but
our final BMP \hehh\ surface becomes strongly repulsive there instead.
An example of this behavior may be seen in Figure~\ref{fig:wall}a and~b:
above $\sim 1 \; \har$ the BMP curve (solid line) rises very steeply.
Extrapolating from the \abinitio\ energies, the short-distance behavior
at $\rm H - He$ distances $R_A < 0.6 \; \bohr$ ought to be intermediate
between the BMP curve and that of the
MR surface\cite{mr94} (long-dashed line), but closer
to the latter for $R_A \lesssim 0.55 \; \bohr$.
For cases where an extremely hard repulsive
core at $R_A < 0.6 \; \bohr$ is undesirable, the Fortran program
of our BMP analytic surface\cite{epaps,bmp} contains an option allowing
the user to switch over at small~$R_A$ or~$R_B$ (gradually, with
continuous first derivatives) to alternate forms, which behave more
like an exponentially damped Coulomb repulsion there.

{\it All\/} of the analytic surfaces include the H$_2$-molecule
potential $V_{H_2}$ as a separate term in the surface formulae.
The WKB and MR papers\cite{wkb74,mr94} left the choice of $V_{H_2}$
formula unspecified, while the Fortran program for the DR surface\cite{dr78}
used the spline fit of Kolos \& Wolniewicz\cite{kw65} for~$V_{H_2}$
(which is almost as good as Schwenke's\cite{sh88} $V_{H_2}$ formula).
For consistency, we used the accurate H$_2$-molecule potential of
Schwenke\cite{sh88} for {\it all\/} the surfaces.  This formula
behaves well at most $\rm H - H$ separations~$r$, but begins to
behave poorly at $r \lesssim 0.1 \; \bohr$.  Our Fortran program therefore
switches over at $r \approx 0.12 \; \bohr$ to a polynomial in $1/r$,
which in turn becomes a pure $1/r$
repulsive potential for $r \lesssim 0.045 \; \bohr$.
(The first derivative with respect to~$r$ is continuous at these
switchover points $r \approx 0.12 \; \bohr$ and $r \approx 0.045 \; \bohr$.)

\subsection{Prospects for further improvement}
\label{ssec:prospects}

Significant improvements over the BMP surface of this paper are
possible, but would require a major effort.

Improvements in the
van der Waals well for strongly non-equilibrium H$_2$ molecules
($r < 1.2 \; \bohr$ and $r > 1.6 \; \bohr$) would require very
accurate \abinitio\ calculations in this region (such as those
performed for equilibrium H$_2$ molecule sizes by Tao\cite{Tao94}).
Once these \abinitio\ energies were available, increased flexibility
(i.e., larger index ranges)
in the $V_d$ and~$V_D$ terms should allow the van der Waals well to
be well represented over a wide range of H$_2$ molecule sizes.

Improvements in the interaction region would require, in the analytic
functional form, explicit inclusion of the conical intersection
with the first excited state.  To do this, one would first have to
pin down its location and its opening angle in the perpendicular
directions (it is possible that these could be determined sufficiently
well from the existing \abinitio\ energies of the present work; the
alternative would be to use an explicit search via
new \abinitio\ computations, as was done for H$_4$ in a limited
region by Boothroyd \etal\cite{bmkp02}).
The position and opening angle would then have to be fitted to an
analytic form, which could then be added to the fitted surface --- the
magnitude of this new term (and perhaps its extent in the directions
perpendicular to the locus of the
conical intersection) would need to be parameterized and fitted as a
function of position along the locus of the conical intersection.
The other terms in the fitted surface would then not have to try to fit the
sharp peak very near the conical intersection, allowing a considerable
improvement in accuracy there.

\section{Conclusions}
\label{sec:conclusions}

A new set of $23\,703$ \abinitio\ energies was computed for
\hehh\ geometries where the interaction energy is expected to
be non-negligible, using Buenker's multiple reference (single and) double
excitation configuration interaction (\hbox{MRD-CI}) program\cite{bu87,bf92};
the lowest excited states were computed as well as the ground state energy.
These new \abinitio\ energies have an estimated rms ``random'' error
of~$\sim 0.2 \; \millih$
and a systematic error
of~$\sim 0.6 \; \millih$ (0.4~kcal/mol).
The position (in the 3-dimensional conformation space of \hehh)
of the conical intersection between the ground state and
the first excited state has been roughly mapped out; unlike H$_3$,
this conical intersection for \hehh\ does {\it not\/} lie at
high-symmetry conformations, but rather along a curved line
in conformation space.

These new \abinitio\ energies were used to test previous analytic
\hehh\ surfaces.  Even the best of the previous surfaces, that of
Muchnick \& Russek\cite{mr94}, does quite poorly for very large
H$_2$-molecule sizes (such as would be encountered in dissociative
collisions or highly excited H$_2$ molecules), although both this surface
and that of Schaefer \& K\"ohler\cite{sk85} are fairly accurate
for not-too-large H$_2$ molecules ($r \lesssim 2 \; \bohr$) with a
not-too-close He~atom ($R \gtrsim 3 \; \bohr$).

A new analytic \hehh\ surface (the BMP surface) was fitted to
$20\,203$ of the new \abinitio\ energies
(and to an additional 4862 points generated at large separations) ---
the other 3500 new \abinitio\ energies were used only to test
``interstitial'' conformations.
This BMP surface fits the interaction region at the ``chemical accuracy''
level ($\sim 1 \; \millih$) required for reaction dynamics; the overall
energy-weighted rms error of~$1.42 \; \millih$ (0.89~kcal/mole)
is comparable to the accuracy of the \abinitio\ energies
(note that this energy-weighting specifies lower weight for
high-energy points, i.e., $E > 0.2 \; \har$, but a weight of unity
everywhere else: see \S~\ref{sssec:fitweights}).
For the $14\,585$ \abinitio\ energies that lie below twice the H$_2$
dissociation energy, the new BMP \hehh\ surface has an rms error
of~$0.95 \; \millih$ (0.60~kcal/mole).
This surface is an order of magnitude better
than previous surfaces in the interaction region, and also yields a
slight improvement in the fit to the recent van der Waals energies
of Tao\cite{Tao94}.

For relatively compact conformations (i.e., with energies above twice the
H$_2$ dissociation energy), the conical intersection between the ground
state and the first excited state is the largest source of error in the
analytic surface.  The BMP surface ``rounds off'' the conical intersection,
yielding errors of up to $\sim 20 \; \millih$ there.
The position of this conical intersection forms a curved 1-dimensional
locus in the 3-dimensional conformation space of~\hehh; its
approximate position has been mapped out, but trying to include the conical
intersection explicitly in an analytic surface would require a more accurate
map of its position, and is beyond the scope of the present paper.

The \abinitio\ energies and a Fortran program for the analytic BMP
\hehh\ surface of the present work
are available from EPAPS\cite{epaps} or from the authors\cite{bmp}.

\acknowledgements

This work was supported by the Natural Sciences and Engineering Research
Council of Canada.

\clearpage

\begin{table}
\caption{
Gaussian basis sets
}
\label{tab:basis}

\bigskip

\begin{tabular}{lllll}
\noalign{\smallskip}
Basis set & Type & Exponent ($\bohr^{-2}$) & Coefficient \\
\noalign{\smallskip}
\tableline
\noalign{\smallskip}
H: $(9s3p2d)/[4s3p2d]$   & $s$ & \phn887.22     & 0.000112 \\
                         &     & \phn123.524    & 0.000895 \\
                         &     & \phnn27.7042   & 0.004737 \\
                         &     & \phnnn7.82599  & 0.019518 \\
                         &     & \phnnn2.56504  & 0.065862 \\
                         &     & \phnnn0.938258 & 0.178008 \\
                         & $s$ & \phnnn0.372145 & 1.0 \\
                         & $s$ & \phnnn0.155838 & 1.0 \\
                         & $s$ & \phnnn0.066180 & 1.0 \\
                         & $p$ & \phnnn2.1175   & 1.0 \\
                         & $p$ & \phnnn0.77     & 1.0 \\
                         & $p$ & \phnnn0.28     & 1.0 \\
                         & $d$ & \phnnn1.76     & 1.0 \\
                         & $d$ & \phnnn0.62     & 1.0 \\
\noalign{\smallskip}
\tableline
\noalign{\smallskip}
He: $(10s4p2d)/[5s4p2d]$ & $s$ &    3293.694    & 0.00010 \\
                         &     & \phn488.8941   & 0.00076 \\
                         &     & \phn108.7723   & 0.00412 \\
                         &     & \phnn30.17990  & 0.01721 \\
                         &     & \phnnn9.789053 & 0.05709 \\
                         &     & \phnnn3.522261 & 0.14909 \\
                         & $s$ & \phnnn1.352436 & 1.0 \\
                         & $s$ & \phnnn0.552610 & 1.0 \\
                         & $s$ & \phnnn0.240920 & 1.0 \\
                         & $s$ & \phnnn0.107951 & 1.0 \\
                         & $p$ & \phnnn5.823125 & 1.0 \\
                         & $p$ & \phnnn2.1175   & 1.0 \\
                         & $p$ & \phnnn0.77     & 1.0 \\
                         & $p$ & \phnnn0.28     & 1.0 \\
                         & $d$ & \phnnn2.83871  & 1.0 \\
                         & $d$ & \phnnn1.0      & 1.0 \\
\end{tabular}
\end{table}

\clearpage

\begin{table}
\caption{Energy-weighted\protect\tablenotemark[1] rms errors of
analytic \hehh\ surfaces, in $\millih$
}
\label{tab:rms}

\bigskip

%
\begin{tabular}{lrr@{}lr@{}lr@{}lr@{}lr@{}lr@{}l}
\noalign{\smallskip}
Surface I.D.:\tablenotemark[2] & & \multicolumn{2}{c}{WKB} &
 \multicolumn{2}{c}{DR} & \multicolumn{2}{c}{SK} & \multicolumn{2}{c}{MR} &
 \multicolumn{2}{c}{modMR} & \multicolumn{2}{c}{BMP} \\
No.\ of parameters: & & \multicolumn{2}{c}{4} & \multicolumn{2}{c}{4} &
 \multicolumn{2}{c}{[750]\tablenotemark[3]} & \multicolumn{2}{c}{19} &
 \multicolumn{2}{c}{27} & \multicolumn{2}{c}{112} \\
\noalign{\smallskip}
Subset & $N_{pts}$ & \multicolumn{2}{c}{rms} & \multicolumn{2}{c}{rms} &
 \multicolumn{2}{c}{rms} & \multicolumn{2}{c}{rms} &
 \multicolumn{2}{c}{rms} & \multicolumn{2}{c}{rms} \\
\noalign{\smallskip}
\tableline
\noalign{\smallskip}
1: All fitted \abinitio &
 $20\,203$ & {\it 100.}&{\it 90} & {\it 100.}&{\it 90} &
   \multicolumn{2}{c}{{\it 5.1E+5}} &
   {\it 28.}&{\it 60} & {\it 28.}&{\it 35} & 1.&417 \\
2: ...\ $E < 0.174 \; \har$\tablenotemark[4] &
 $14\,585$ & {\it 37.}&{\it 80} & {\it 40.}&{\it 40} &
   \multicolumn{2}{c}{{\it 6.0E+5}} &
   {\it 20.}&{\it 24} & {\it 19.}&{\it 89} & 0.&955 \\
3: ...\ $E < 0.0 \; \har$\tablenotemark[5] &
 7177 & {\it 11.}&{\it 62} & {\it 10.}&{\it 58} &
   \multicolumn{2}{c}{{\it 1.1E+4}} &
   {\it 4.}&{\it 48} & {\it 4.}&{\it 44} & 0.&481 \\
4: ...\ $0.5 \; \bohr \le r < 1.2 \; \bohr$ &
 3744 & 26.&75 & 20.&95 & {\it 21.}&{\it 35} &
   {\it 7.}&{\it 80} & {\it 7.}&{\it 80} & 0.&875 \\
5: ...\ ...\ $R \le 2 \; \bohr$ &
 1646 & 37.&73 & 29.&57 & {\it 32.}&{\it 18} &
   {\it 10.}&{\it 87} & {\it 10.}&{\it 87} & 1.&205 \\
6: ...\ ...\ $2 \; \bohr < R \le 3 \; \bohr$ &
 857 & 17.&89 & 14.&18 & 1.&323 &
   {\it 6.}&{\it 20} & {\it 6.}&{\it 21} & 0.&624 \\
7: ...\ ...\ $R > 3 \; \bohr$ &
 1241 & 7.&11 & 4.&97 & 0.&449 &
   {\it 0.}&{\it 418} & {\it 0.}&{\it 421} & 0.&336 \\
8: ...\ $1.2 \; \bohr \le r \le 1.62 \; \bohr$ &
 2311 & 48.&62 & 47.&82 & {\it 22.}&{\it 14} & 15.&79 & 15.&79 & 1.&003 \\
9: ...\ ...\ $R \le 2 \; \bohr$ &
 907 & 75.&84 & 74.&80 & {\it 35.}&{\it 31} & 24.&69 & 24.&68 &  1.&463\\
10: ...\ ...\ $2 \; \bohr < R \le 3 \; \bohr$ &
 575 & 18.&72 & 18.&04 & 1.&497 & 6.&39 & 6.&40 & 0.&690 \\
11: ...\ ...\ $R > 3 \; \bohr$ &
 829 & 7.&39 & 5.&19 & 0.&433 & 0.&445 & 0.&450 & 0.&365 \\
12: ...\ $1.62 \; \bohr < r \le 2 \; \bohr$ &
 1918 & 74.&01 & 76.&16 & {\it 116.}&{\it 17} &
   {\it 25.}&{\it 34} & {\it 25.}&{\it 33} & 1.&351 \\
13: ...\ ...\ $R \le 2 \; \bohr$ &
 735 & 117.&85 & 121.&07 & {\it 187.}&{\it 64} &
   {\it 39.}&{\it 59} & {\it 39.}&{\it 58} & 2.&068 \\
14: ...\ ...\ $2 \; \bohr < R \le 3 \; \bohr$ &
 518 & 22.&31 & 25.&29 & 3.&42 &
   {\it 12.}&{\it 36} & {\it 12.}&{\it 34} & 0.&723 \\
15: ...\ ...\ $R > 3 \; \bohr$ &
 665 & 7.&71 & 5.&51 & 0.&458 &
   {\it 0.}&{\it 865} & {\it 0.}&{\it 889} & 0.&358 \\
16: ...\ $2 \; \bohr < r \le 4 \; \bohr$ &
 7136 & 130.&27 & 133.&13 & \multicolumn{2}{c}{{\it 1.1E+4}} &
   {\it 37.}&{\it 63} & {\it 37.}&{\it 64} & 1.&797 \\
17: ...\ ...\ $R > 3 \; \bohr$ &
 2806 & 10.&16 & 8.&30 & {\it 24.}&{\it 75} &
   {\it 7.}&{\it 64} & {\it 7.}&{\it 44} & 0.&456 \\
18: ...\ $4 \; \bohr < r \le 10.6 \; \bohr$ &
 5094 & {\it 113.}&{\it 78} & {\it 109.}&{\it 56} &
   \multicolumn{2}{c}{{\it 1.0E+6}} &
   {\it 29.}&{\it 32} & {\it 28.}&{\it 34} & 1.&318 \\
19: ...\ ...\ $R > 4 \; \bohr$ &
 1244 & {\it 8.}&{\it 44} & {\it 6.}&{\it 40} & {\it 84.}&{\it 99} &
   {\it 18.}&{\it 47} & {\it 17.}&{\it 87} & 0.&588 \\
\noalign{\smallskip}
\tableline
\noalign{\smallskip}
%
20: Fitted ``random'' \abinitio\tablenotemark[6] &
 3500 & {\it 57.}&{\it 82} & {\it 64.}&{\it 25} &
   \multicolumn{2}{c}{{\it 2.9E+5}} &
   {\it 27.}&{\it 36} & {\it 26.}&{\it 95} & 1.&411 \\
21: ...\ $E < 0.174 \; \har$\tablenotemark[4] &
 2962 & {\it 36.}&{\it 35} & {\it 39.}&{\it 61} &
   \multicolumn{2}{c}{{\it 3.1E+5}} &
   {\it 20.}&{\it 81} & {\it 20.}&{\it 32} & 0.&994 \\
22: ...\ $E < 0.0 \; \har$\tablenotemark[5] &
 1367 & {\it 13.}&{\it 92} & {\it 13.}&{\it 03} &
   {\it 5.}&{\it 77}                 &
   {\it 5.}&{\it 07} & {\it 5.}&{\it 04} & 0.&527 \\
\noalign{\smallskip}
\tableline
\noalign{\smallskip}
%
%
23: Unfitted ``random'' \abinitio\tablenotemark[7] &
 3500 & {\it 57.}&{\it 07} & {\it 63.}&{\it 35} &
   \multicolumn{2}{c}{{\it 2.6E+5}} &
   {\it 27.}&{\it 74} & {\it 27.}&{\it 31} & 1.&400 \\
24: ...\ $E < 0.174 \; \har$\tablenotemark[4] &
 2943 & {\it 35.}&{\it 58} & {\it 38.}&{\it 41} &
   \multicolumn{2}{c}{{\it 2.8E+5}} &
   {\it 21.}&{\it 05} & {\it 20.}&{\it 52} & 0.&964 \\
25: ...\ $E < 0.0 \; \har$\tablenotemark[5] &
 1349 & {\it 14.}&{\it 10} & {\it 13.}&{\it 12} &
   {\it 14.}&{\it 50}                  &
   {\it 4.}&{\it 93} & {\it 4.}&{\it 89} & 0.&538 \\
%
%
\noalign{\smallskip}
\tableline
\noalign{\smallskip}
%
26: All ``vdW'' &
 2145 & {\it 2.}&{\it 037} & {\it 1.}&{\it 698} & {\it 0.}&{\it 134} &
   0.&609 & 0.&609 & 0.&129 \\
27: ...\ $R < 6 \; \bohr$ &
 587 & {\it 3.}&{\it 67} & {\it 3.}&{\it 03} & {\it 0.}&{\it 251} &
   1.&164 & 1.&164 & 0.&237 \\
28: ...\ $6 \; \bohr \le R < 10 \; \bohr$ &
 1098 & {\it 0.}&{\it 930} & {\it 0.}&{\it 830} & {\it 0.}&{\it 0379} &
   [0.&0114] & [0.&0013] & 0.&0507 \\
29: ...\ ...\ $r = 0.6, 0.7, 0.8 \; \bohr$\tablenotemark[8] &
 186 & {\it 0.}&{\it 491} & {\it 0.}&{\it 404} & {\it 0.}&{\it 0023} &
   [0.&0020] & [0.&0000] & 0.&0031 \\
30: ...\ ...\ $r = 0.9, 1.1 \; \bohr$\tablenotemark[9] &
 248 & {\it 0.}&{\it 731} & {\it 0.}&{\it 640} & [0.&0011] &
   [0.&0014] & [0.&0011] & 0.&0017 \\
31: ...\ ...\ $r = 1.28, 1.449, 1.618 \; \bohr$\tablenotemark[10] &
 231 & {\it 1.}&{\it 004} & {\it 0.}&{\it 906} & 0.&0018 &
   0.&0026 & 0.&0011 & 0.&0008 \\
32: ...\ ...\ $r = 1.8, 2.0 \; \bohr$\tablenotemark[9] &
 248 & {\it 1.}&{\it 139} & {\it 1.}&{\it 038} & [0.&0024] &
   [0.&0042] & [0.&0024] & 0.&0037 \\
33: ...\ ...\ $r = 2.4, 3.0, 4.0 \; \bohr$\tablenotemark[8] &
 185 & {\it 1.}&{\it 084} & {\it 0.}&{\it 952} & {\it 0.}&{\it 0922} &
   [0.&0271] & [0.&0000] & 0.&1233 \\
34: ...\ $R \ge 10 \; \bohr$ &
 460 & {\it 0.}&{\it 273} & {\it 0.}&{\it 270} & 0.&0004 &
   0.&0005 & 0.&0001 & 0.&0001 \\
\noalign{\smallskip}
\tableline
\noalign{\smallskip}
%
35: All ``H-He'' &
 1887 & {\it 114.}&{\it 69} & {\it 106.}&{\it 25} &
   \multicolumn{2}{c}{{\it 4.9E+4}} &
   {\it 44.}&{\it 54} & {\it 44.}&{\it 00} & 1.&122 \\
36: ...\ $R_A \le 2 \; \bohr$ &
 1008 & {\it 156.}&{\it 90} & {\it 145.}&{\it 35} &
   \multicolumn{2}{c}{{\it 1.2E+3}} &
   {\it 59.}&{\it 88} & {\it 59.}&{\it 18} & 1.&455 \\
37: ...\ $2 \; \bohr < R_A < 4 \; \bohr$ &
 468 & 3.&65 & 2.&75 &
   \multicolumn{2}{c}{{\it 4.5E+3}} &
   {\it 16.}&{\it 59} & {\it 16.}&{\it 19} & 0.&714 \\
38: ...\ $4 \; \bohr < R_A < 6 \; \bohr$ &
 228 & 0.&416 & 0.&0959 &
   \multicolumn{2}{c}{{\it 3.8E+3}} &
   {\it 0.}&{\it 680} & {\it 0.}&{\it 639} & 0.&0978 \\
39: ...\ $R_A \ge 6 \; \bohr$ &
 183 & 0.&0431 & 0.&0008 &
   \multicolumn{2}{c}{{\it 1.6E+5}} &
   {\it 0.}&{\it 0068} & 0.&0038 & 0.&0014 \\
\noalign{\smallskip}
\tableline
\noalign{\smallskip}
%
40: All ``H-$\rm He + H$'' &
 830 & {\it 26.}&{\it 03} & {\it 22.}&{\it 19} &
   \multicolumn{2}{c}{{\it 7.6E+6}} &
   {\it 19.}&{\it 82} & {\it 19.}&{\it 34} & 0.&874 \\
41: ...\ $R_A \le 3 \; \bohr$ &
 141 & {\it 63.}&{\it 15} & {\it 53.}&{\it 84} &
   \multicolumn{2}{c}{{\it 2.4E+4}} &
   {\it 48.}&{\it 03} & {\it 46.}&{\it 88} & 2.&096 \\
42: ...\ $3 \; \bohr < R_A < 6 \; \bohr$ &
 201 & 0.&950 & 0.&602 &
   \multicolumn{2}{c}{{\it 8.9E+5}} &
   {\it 2.}&{\it 02} & {\it 1.}&{\it 63} & 0.&266 \\
43: ...\ $R_A \ge 6 \; \bohr$ &
 488 & {\it 0.}&{\it 184} & {\it 0.}&{\it 156} &
   \multicolumn{2}{c}{{\it 9.9E+6}} &
   {\it 0.}&{\it 0813} & 0.&0051 & 0.&0341 \\
\noalign{\smallskip}
\tableline
\noalign{\smallskip}
%
44: All fitted points; full-weight\tablenotemark[1] &
 $25\,065$ & {\it 329.}&{\it 02} & {\it 310.}&{\it 75} &
   \multicolumn{2}{c}{{\it 5.2E+7}} &
   {\it 28.}&{\it 70} & {\it 28.}&{\it 41} & 1.&383 \\
\end{tabular}
%
\smallskip

\tablenotetext[1]{
For energy-weighted rms errors,
deviations get only weight~$w_E$ from equation~(\ref{eq:wtE})
(unity for $E \le 0.2 \; \har$, reduced weight at higher energy),
as opposed to full-weight rms errors in last line of table (subset 44).
Note that the rms value for a surface is {\it italicized\/} for subsets
containing conformations in regions where that surface was not fitted.
}

\tablenotetext[2]{
WKB: Wilson, Kapral, \& Burns (Ref.~\onlinecite{wkb74}),
DR: Dove \& Raynor (Ref.~\onlinecite{dr78}),
SK: Schaefer \& K\"ohler (Ref.~\onlinecite{sk85}),
MR: Muchnick \& Russek (Ref.~\onlinecite{mr94}),
modMR: slightly modified Muchnick \& Russek surface of \S~\ref{sssec:modMuRu},
BMP: adopted surface of the present work.
}

\tablenotetext[3]{
Schaefer \& K\"ohler present 3 Legendre coefficients at 48
$R$-values for each of 5 $r$-values (namely, $r = 0.9$, 1.28, 1.449,
1.618, and~$2.0 \; \bohr$), plus formulae for very large~$R$.
}

\tablenotetext[4]{
$E < 0.174 \; \har$ corresponds to points lying below about twice the H$_2$
dissociation energy, relative to a H$_2$ molecule plus a distant He~atom.
}

\tablenotetext[5]{
$E < 0.0 \; \har$ corresponds to points lying below the H$_2$ dissociation
energy, relative to a H$_2$ molecule plus a distant He~atom.
}

\tablenotetext[6]{
These subsets 20 to~22 are also contained in subsets 1 to~3, respectively.
}

\tablenotetext[7]{
These subsets 23 to~25 were obtained using the same
probability distribution as subsets 20 to~22, but were {\it not\/} included
in the actual fitting process for the BMP surface of the present work.
}

\tablenotetext[8]{
The accuracy of these generated points is unknown.
They were intended only to yield ``reasonable'' values and prevent wild
excursions in the fitted surfaces, and were given relatively low weight
in the fit, resulting in the relatively large ``BMP'' rms for these points
at large~$r$.
The ``modMR'' surface was used to generate them, and thus by definition
has zero rms here (likewise, the very similar ``MR'' surface has a small
rms here); this is indicated by the square brackets enclosing their rms
values.
}

\tablenotetext[9]{
The ``SK'' and ``modMR'' surfaces were used to generate
these points, and thus they (and the ``MR'' surface) have small rms values
here (again indicated by square brackets).
}

\tablenotetext[10]{
These points comprise the \abinitio\ energies of Tao
(Ref.~\onlinecite{Tao94}) and points generated from them, and thus these
rms values provide a measure of the accuracy for {\it all\/} the surfaces
in the van der Waals well.
}

\end{table}

\clearpage

\clearpage

\clearpage

\clearpage

\clearpage

\clearpage

\clearpage

\clearpage

\clearpage

\clearpage

\clearpage

\clearpage

\clearpage

\end{document}